\documentclass[letterpaper,twocolumn,10pt]{article}
\usepackage{usenix-2020-09}

\usepackage{amsmath}
\usepackage{bm}
\usepackage{booktabs}   
\usepackage{tabularx}   
\usepackage{amsmath}    
\usepackage{amssymb}    
\usepackage{graphicx}
\usepackage{makecell}
\usepackage{multirow}
\usepackage{threeparttable}

\usepackage{colortbl} 
\usepackage{graphicx} 
\usepackage{diagbox}  

\usepackage{subfigure}

\usepackage{tikz}
\usetikzlibrary{arrows.meta,calc,positioning}

\usepackage{colortbl} 
\newcommand{\heading}[1]{{\vspace{3pt}\noindent{\textbf{#1}}}}

\usepackage{amsmath,amssymb}

\newenvironment{packeditemize}{
	\begin{list}{$\bullet$}{
			\setlength{\labelwidth}{4pt}
			\setlength{\itemsep}{0pt}
			\setlength{\leftmargin}{\labelwidth}
			\addtolength{\leftmargin}{\labelsep}
			\setlength{\parindent}{0pt}
			\setlength{\listparindent}{\parindent}
			\setlength{\parsep}{0pt}
			\setlength{\topsep}{1pt}}}{\end{list}}

\definecolor{light_cyan}{rgb}{0.53, 0.75, 0.77}
\definecolor{light_blue}{rgb}{0.466, 0.655, 0.94}
\definecolor{light_pink}{rgb}{0.98, 0.55, 0.565}
\definecolor{light_yellow}{rgb}{0.98, 0.83, 0.32}

\usepackage{color}
\usepackage{algorithm,algpseudocode} 

\usepackage[most]{tcolorbox}

\usepackage{bbding} 
\usepackage{pifont}
\newcommand{\cmark}{\textcolor{green!80!black}{\ding{51}}} 
\newcommand{\xmark}{\textcolor{red}{\ding{55}}} 
\newcommand{\pmark}{\textcolor{blue!90}{\ding{109}}}    

\AtBeginDocument{%
  }

\begin{document}

\date{}

\title{Why Neural Structural Obfuscation Can’t Kill \\White-Box Watermarks for Good!}

\author{{\rm Yanna Jiang$^{1,\textcolor{green}{*}}$, Guangsheng Yu$^{1,}\thanks{Contributed equally.}\,$,  Qingyuan Yu$^{4}$, Yi Chen$^{3}$, Qin Wang$^{1,2}$ }
\\ 
$^1$University of Technology Sydney $|$ $^2$CSIRO Data61 $|$ $^3$Tsinghua University $|$ $^4${Independent}  
}




\maketitle

\begin{abstract}

Neural Structural Obfuscation (NSO) (USENIX Security'23) is a family of ``zero cost'' structure-editing transforms (\texttt{nso\_zero}, \texttt{nso\_clique}, \texttt{nso\_split}) that inject dummy neurons.
By combining neuron permutation and parameter scaling, NSO makes a radical modification to the network structure and parameters while strictly preserving functional equivalence, thereby disrupting white-box watermark verification. This capability has been a fundamental challenge to the reliability of existing white-box watermarking schemes.

We rethink NSO and, for the first time, fully recover from the damage it has caused. 
We redefine NSO as a graph-consistent threat model within a \textit{producer--consumer} paradigm. This formulation posits that any obfuscation of a producer node necessitates a compatible layout update in all downstream consumers to maintain structural integrity.
Building on these consistency constraints on signal propagation, we present \textsc{Canon}, a recovery framework that probes the attacked model to identify redundancy/dummy channels and then \textit{globally} canonicalizes the network by rewriting \textit{all} downstream consumers by construction, synchronizing layouts across \texttt{fan-out}, \texttt{add}, and \texttt{cat}. 
Extensive experiments demonstrate that, even under strong composed and extended NSO attacks, \textsc{Canon} achieves \textbf{100\%} recovery success, restoring watermark verifiability while preserving task utility.

Our code is available at \url{https://anonymous.4open.science/r/anti-NSO-9874}.

\end{abstract}

\section{Introduction}
\label{sec:intro}

In USENIX Security'23. Yan et al. published \textit{Rethinking White-Box Watermarks on Deep Learning Models under Neural Structural Obfuscation} (NSO) that is a systematic study that claims many white-box watermarks can be ``removed at zero cost'' by injecting dummy neurons that preserve utility while disrupting verification~\cite{yan2023}. 
Their toolkit comprises three strategies: (i) zero injection (i.e., \texttt{nso\_zero}), which adds channels with all zero incoming or outgoing weights; (ii) clique cancellation (i.e., \texttt{nso\_clique}), which duplicates incoming weights and sets outgoing weights that sum to zero; and (iii) split preservation (i.e., \texttt{nso\_split}), which replaces a channel with $d\!+\!1$ substitutes whose outgoing weights sum to the original. 
The core observation underlying NSO is that mainstream white-box watermark verifiers implicitly bind to \textit{index-ordered parameters}: once channel indices are structurally perturbed, watermark extraction fails even though the network is functionally equivalent. 
This poses a fundamental threat to white-box watermark security~\cite{lu2024neural,pegoraro2024deepeclipse,zhong2025hardening,deng2025polo,yang2025effectiveness,hao2025gamc}, requiring a new approach to detect and recovery.

\begin{figure}
    \centering
    \includegraphics[width=\linewidth]{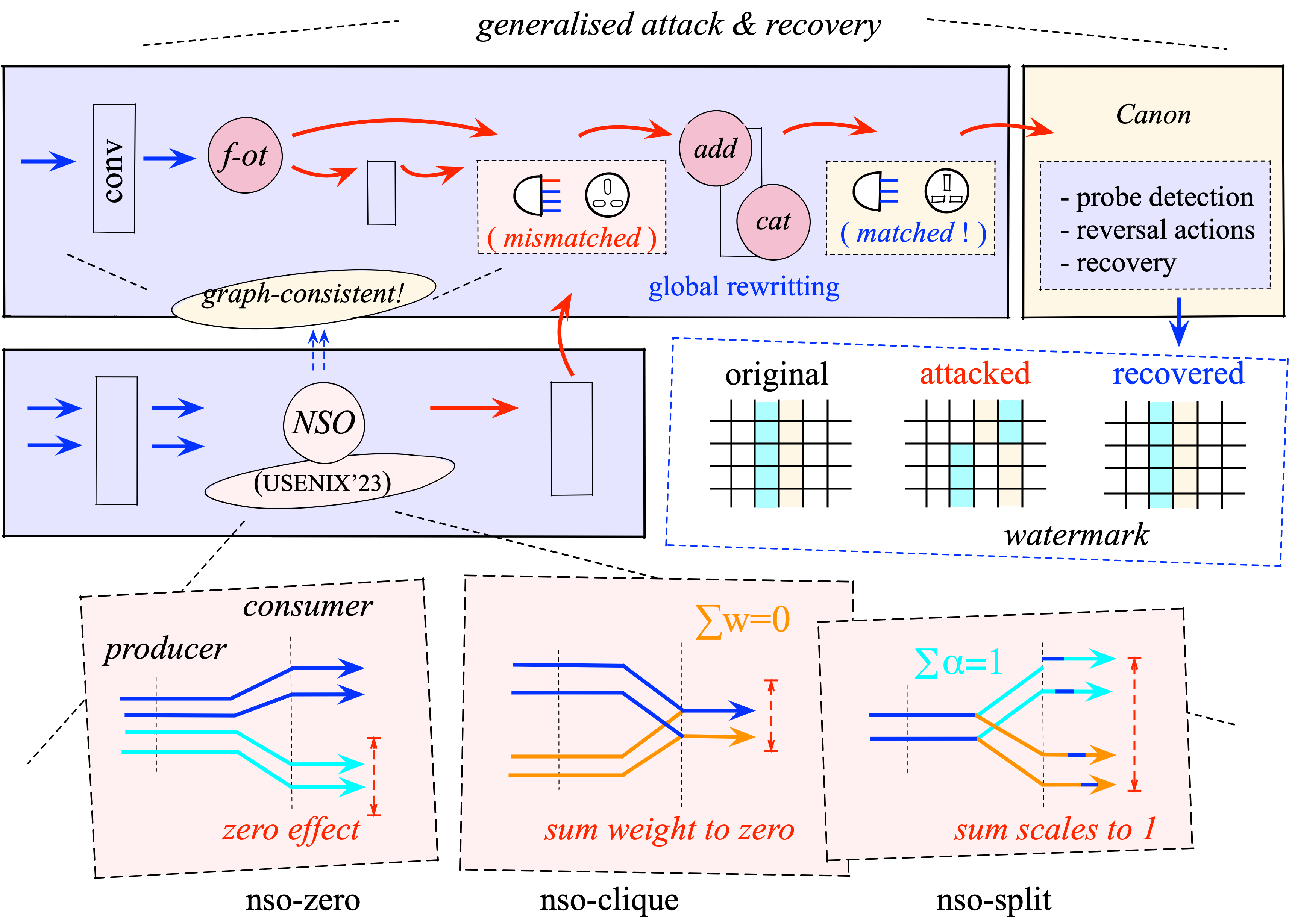}
    \vspace{-0.1in}
    \caption{\textsc{Canon} can recover a strengthened NSO variant that generalizes the original \textit{layer-local} edits into \textit{graph-aware} structural transformations that remain consistent across branches and merges, enabling attacks to be injected globally rather than only in isolated layers. The three NSO attacks correspond to different constrained forms of $M$ (zero injection, clique cancellation $\sum w=0$, and split preservation $\sum \alpha=1$).}
    \label{fig:Producer–Consumer}
\end{figure}


\smallskip
\noindent\textbf{We rethink NSO~\cite{yan2023} and show that it can be fully recovered, even under a stronger threat model, making embedding watermarks practical again.}
We recognize that existing NSO constructions, while effective, are largely confined to simple, sequential architectures and do not fully capture how structural obfuscation must behave in modern graph-structured networks. 
In contemporary models, such as ResNet~\cite{he2016deep}, EfficientNet~\cite{tan2019efficientnet}, InceptionV3~\cite{szegedy2016rethinking}, and DenseNet~\cite{huang2017densely}, structural edits are no longer layer-local. Instead, injected neurons must propagate their signals consistently through residual connections, multi-branch topologies, and shared downstream operators (e.g., \texttt{add}~\cite{he2016deep,tan2019efficientnet} and \texttt{cat}~\cite{szegedy2016rethinking,huang2017densely}) in order to preserve functional equivalence.

To make NSO representative of realistic deployment scenarios, we introduce a \textit{producer--consumer} perspective, as shown in Fig.\ref{fig:Producer–Consumer}, which explicitly models how injected neuron signals are produced and consumed across the computation graph. 
Under this view, NSO is not treated as a stronger attack, but as a completed and well-defined threat model that changes channel layout while remaining functionally equivalent and shape compatible in graph-consistent architectures. 

Building on this threat model, recovery and defense face a fundamentally harder challenge.
Once NSO is instantiated beyond simple sequential architectures, the attacker may arbitrarily combine multiple NSO primitives, place injections across layers and branches, and exploit residual merges, while the defender has no access to the attacker’s construction details. The central research problem becomes:

\begin{packeditemize}
\item[$\heartsuit$] \textit{How to distinguish attack-induced dummy neurones from benign structural components in a complex computation graph, when all edits are deliberately designed to remain functionally equivalent and appear structurally normal?}
\end{packeditemize}

\heading{Our approach: Functional equivalence enforces graph-consistent structure under NSO.}
Our key insight is that the indistinguishability introduced by NSO is only superficial.
To preserve functional equivalence, any NSO obfuscation \emph{must} induce a channel transform that is consistent across \textit{all} downstream consumers along every \textit{producer$\rightarrow$consumer} path of the computation graph: injected channels must satisfy the same cancellation/aggregation identities at \textit{every} downstream consumption site (fan-outs, residual merges, and concatenations), otherwise the obfuscated network would no longer be functionally equivalent. These constraints are not incidental artifacts of NSO but necessary conditions for its correctness, which we leverage to generalize NSO into a
graph-consistent producer-consumer threat model.

We thus design \textsc{Canon}, a graph-consistent recovery framework that identifies NSO redundancy and reverses structural obfuscation. Rather than viewing NSO as an irreversible removal of watermarks, \textsc{Canon} models NSO as a family of \textit{channel-basis rewrites} on the computation graph and recovers a canonical parameterization by (i) probing activations to infer redundancy and dummy channels and (ii) applying the induced ChannelTransform rewrites \textit{globally} to all downstream linear consumers, including \texttt{fan-out}, residual \texttt{add}, and channel \texttt{cat}. By synchronizing channel layouts across the graph, \textsc{Canon} eliminates NSO-induced index scrambling and restores a compact representation on which existing white-box watermark extractors can be directly applied. As a clean reference model is typically available in the white-box setting, we additionally provide a reference-equivalence certificate.

\heading{Why NSO fails under \textsc{Canon}.} NSO primitives preserve network functionality by construction: injected channels are either null, mutually cancel in cliques, or form proportional split duplicates whose aggregate contribution matches the original channel. Under \textsc{Canon}, these edits manifest as structured redundancy revealed by activation probes through consistent signatures and proportionality relations. Crucially, \textsc{Canon} operates \textit{graph-consistently}: once a ChannelTransform is inferred at a producer edge, it is propagated to \textit{all} downstream consumers, including residual merges and concatenations. This allows NSO edits to be removed via \textit{equivalence preserving compaction}, while downstream linear consumers are rewritten to the recovered channel basis and layouts are synchronized across the graph. The resulting parameters $\hat{\theta}$ (\textit{v.s.} original $\theta$) form a stable, compact representation on which existing white-box verifiers can be directly applied.

\smallskip
\noindent\textbf{Contributions.} We make the following contributions:

\begin{packeditemize}

 \item \textit{We generalise the threat model of NSO.} We reformulate NSO as a \emph{graph-consistent producer--consumer model} that captures fan-out, residual addition, and channel concatenation in modern architectures. This exposes the \emph{signal-consistency constraints} required for function-equivalent NSO, which are missing in prior layer-local formulations.

  \item \textit{We show that NSO-style obfuscation is fundamentally recoverable under graph consistency.} Contrary to the prevailing belief that NSO can ``remove'' white-box watermarks at zero cost, we prove constructively that NSO induces structured, globally constrained redundancy. These constraints make it possible to identify dummy channels and reverse layout obfuscation without access to the attacker’s construction details or the clean reference model. 
  
  \item \textit{We propose \textsc{Canon}, the first end-to-end canonicalization framework that defeats NSO by construction.} \textsc{CANON} recovers a graph-consistent channel basis using activation probes and enforces global layout synchronization across fan-out, residual merges, and concatenations, without requiring the original model.
  
  \item \textit{We empirically invalidate the ``zero-cost removal'' claim of NSO.}
  Under strong, compositional NSO attacks on modern CNNs, \textsc{Canon} restores watermark verifiability for all evaluated schemes with no accuracy loss, demonstrating that NSO fails as a permanent watermark removal strategy.

\end{packeditemize}

\section{Related Work}
\label{sec:related_work}

\subsection{Robustness of White-box Watermarking}

\heading{White-box watermarking schemes.}
White-box watermarking verifies model ownership with access to internal parameters and/or intermediate activations~\cite{uchida2017embedding,pan2023cracking}.
Existing schemes broadly include weight-based methods that encode messages into selected weights or subspaces~\cite{wang2021riga,ong2021protecting,liu2021watermarking,chen2021you}, activation-based methods that impose constraints on internal representations~\cite{darvish2019deepsigns,lim2022protect}, and passport-based methods that couple verification to secret keys embedded in dedicated modules or layer-wise passports~\cite{zhang2020passport,fan2021deepipr}.

\heading{Attacks and defenses in white-box watermarking.}
Most watermark robustness studies and defenses target value-level modifications that alter parameter values while keeping the architecture fixed, such as fine-tuning~\cite{11159508,cui2025ft,wang2021riga}, pruning~\cite{fan2019rethinking, zhang2018protecting,chen2021you}, compression~\cite{zhang2020passport}, parameter overwriting~\cite{ong2021protecting,yang2021robust}, distillation~\cite{yang2019effectiveness,wang2025toward}, and model extraction~\cite{lv2023robustness,tan2023deep}.
Correspondingly, many watermark constructions attempt to be robust to these edits by spreading the signal across weights~\cite{yang2024gaussian,zhang2021deep}, using redundancy encodings~\cite{luo2024wformer,pan2025robust}, adding regularizers~\cite{gan2023towards}, or designing verification rules that tolerate moderate weight perturbations~\cite{dai2025division}.
Some recent attacks further exploit invariances in verification, such as reparameterization~\cite{zhang2024remark,liu2024survey}, rescaling~\cite{kinakh2024evaluation,pan2023cracking}, or weight shifting style manipulations~\cite{yan2023,lukas2022sok,pan2023cracking} that preserve the function while disrupting the particular coordinates a verifier reads.
However, defenses for these invariance-based evasions are still rare, and most existing “robust watermarking” techniques still assume identifiable coordinates, motivating our focus on the recovery under structure/layout obfuscations.

\heading{Why these defenses do not cover NSO.}
NSO changes the problem at its root: it preserves utility by construction while performing structure edits, e.g., injecting/rearranging channels, that invalidate the index-ordered assumptions many white-box extractors rely on~\cite{yan2023}.
Under NSO, the watermark signal may remain present in a functional sense, yet its presumed host indices/locations are scrambled.
This breaks a common implicit premise behind ``robust watermarking'' against fine-tuning/pruning/overwriting: those defenses~\cite{cui2025ft,wang2021riga,fan2019rethinking,chen2021you,ong2021protecting,yang2021robust} typically do not provide a post attack recovery mechanism that canonically restores a compact and compatible parameterization after channel/layout rewriting.
As summarized in Table~\ref{tab:nso_defense_properties}, a robust NSO defense should provide (i) post attack recovery (Post-atk), (ii) attack independent (Atk-ind) without prior knowledge of the obfuscation, (iii) forward compatibility (Forw.) to support multiple watermark families, and (iv) robustness to compositional NSO (anti+), since attackers can mix primitives across layers/branches.
To our knowledge, prior watermark defenses do not offer a graph-consistent canonicalization procedure that can defeat NSO-style structure obfuscations while preserving functional equivalence across modern CNN graphs.

\begin{table}[t]
\centering
\caption{What a robust NSO defense must provide}
\label{tab:nso_defense_properties}
\renewcommand{\arraystretch}{1}
\resizebox{\linewidth}{!}{
\begin{threeparttable}
\begin{tabular}{l|cccc}
\midrule
\multicolumn{1}{c|}{\textbf{Representative line of work}} &
\rotatebox{90}{\textbf{Post-atk}} &
\rotatebox{90}{\textbf{Atk-ind}} &
\rotatebox{90}{\textbf{Forw.}} &
\rotatebox{90}{\textbf{anti+}} \\
\midrule
Value-robust watermarking~\cite{wang2021riga,fan2019rethinking,ong2021protecting,chen2021you,lv2023robustness,yang2019effectiveness} & \xmark & \cmark & \pmark & \xmark  \\
Invariant-based evasion~\cite{pan2023cracking,lukas2022sok} & \xmark & \xmark & \pmark & \xmark  \\ 
NSO preliminary defense (\S7.4)\cite{yan2023} & \pmark & \pmark & \pmark & \xmark  \\
\cmidrule{1-5}
\rowcolor{pink!25} \multicolumn{1}{c|}{\textsc{Canon} (ours)} & \cmark & \cmark & \cmark & \cmark \\
\cmidrule{1-5}
\end{tabular}
\begin{tablenotes}
\small
\item  Symbols: \cmark\ = yes; \pmark\ = partial; \xmark\ = no.
\item[$\bullet$] \textbf{Post-atk}: Post attack  recovery; \textbf{Atk-ind}: Attack independent;
\item[$\bullet$] \textbf{Forw.} (forward-compatible): Compatible with various watermarking schemes;
\item[$\bullet$] \textbf{anti+} (anti-compositinal NSO): Robustness to compositional NSO attack.
\end{tablenotes}
\end{threeparttable}
}
\end{table}

\subsection{Compared to Original NSO Design}
\label{sec:rw_vs_nso}

\heading{NSO and the verifier failure mode.}
Yan et al.~\cite{yan2023} introduce NSO as a \textit{function-equivalent} obfuscation that injects and rearranges internal neurons/channels while keeping task behavior intact.
By exploiting permutation invariances and other camouflaging steps, NSO preserves utility yet \textit{disrupts structure indexed assumptions}: a watermark signal may still exist in a functional sense, but the verifier’s presumed host locations no longer align, causing white-box extraction to fail.

\heading{Preliminary defense in the original NSO paper.}
The preliminary defense in \S7.4 of~\cite{yan2023} evaluates defenders under three knowledge levels (limited samples, full data distribution, and additionally the pre-attack reference model).
A key point is the gap between \textit{eliminating suspicious neurons} and \textit{recovering a parameterization compatible with the extractor}: without the  prior knowledge and reference model, the elimination procedure can remain non-restorative for watermark verification, while recovery becomes plausible mainly in the fully knowledgeable setting. Yan et al.~\cite{yan2023} also identify more effective dummy-neuron elimination and de-obfuscation.

\heading{Stronger attacker protocol and our recovery objective.}
Our evaluation targets a strictly stronger, compositional NSO regime than the single primitive settings used to evaluate \S7.4, which reflect graph constraints and maximize heterogeneity across residual \texttt{add} and channel \texttt{cat} structure.
Accordingly, we focus on recovery invoked during verification, without requiring prior knowledge of the attack procedure: \textsc{Canon} leverages signal-consistency constraints that NSO must satisfy along \textit{producer$\rightarrow$consumer paths} and performs \textit{global} consumer rewrites that remain consistent through \texttt{fan-out}, residual \texttt{add}, and channel \texttt{cat}.
This yields a compact, function-equivalent layout on which existing white-box watermark extractors become applicable again, without requiring the original model (used only for optional diagnostics/certification).
We provide additional details on the original defense settings and protocol mismatch in Appendix~\ref{app:nso_defense_settings}.
\section{System Overview and Architecture}
\label{sec:system-overview}

\subsection{Our Design in Nutshell}

\heading{Problem statement.} 
White-box watermark verifiers implicitly assume a \textit{fixed internal structure}: watermarked parameters reside in specific layers or channels and are accessed via \textit{index-ordered} extraction.
NSO violates this assumption without changing network functionality. By performing structural channel edits (i.e., dummy insertion, channel splitting, and scaling), NSO preserves functional equivalence while \textit{scrambling the channel
layout} expected by the verifier.

The challenge is that NSO is designed to be both \textit{function-equivalent} and \textit{structurally plausible}.
Once structural obfuscation is instantiated beyond simple sequential architectures, dummy channels propagate through fan-out, residual merges, and channel concatenations. As a result, attack redundancy becomes indistinguishable from benign model structure under local or layer-wise reasoning, rendering conventional defenses ineffective. The resulting problem is to recover a compact and compatible parameterization that preserves the original network function, given only white-box access to an NSO-obfuscated model and without relying on the attacker’s construction details or the clean reference model.

\heading{Our approach.} We address this problem by exploiting a fundamental property of graph structured neural networks: \textit{producer--consumer consistency}. To preserve functional equivalence, any channel layout change introduced at a producer must be absorbed by all downstream consumers along the producer$\rightarrow$consumer paths of the computation graph. This consistency requirement applies uniformly across fan-out, residual addition, and channel concatenation.

We present \textsc{Canon}, a graph-consistent recovery framework that models NSO edits as \textit{channel-basis transformations} and reverses structural obfuscation by construction. Rather than relying on local pruning or heuristic detection, \textsc{Canon} explicitly propagates layout corrections to all downstream linear consumers, ensuring global synchronization of channel representations while preserving end-to-end functionality.

\heading{Producer--consumer perspective.} 
We view NSO as an edit on \textit{producer edges} that changes the channel layout of intermediate tensors (e.g., from $C$ channels to $C{+}d$ channels via dummy injection). Any downstream operator that consumes this tensor is a \textit{consumer} and must remain compatible with the layout to preserve functional equivalence. This includes \texttt{fan-out} consumers, residual merges via elementwise \texttt{add}, and channel concatenation via \texttt{cat}.

Under function-equivalent obfuscation, any producer-side layout change must induce a \textit{consistent} rewrite across all downstream consumers along every producer$\rightarrow$consumer path. This graph-level consistency constraint is the key lever exploited by our recovery.

\subsection{System and Threat Model}

\heading{System model.} Let $f_{\theta}$ be a neural network with parameters $\theta$ and intermediate tensors flowing along a computation graph.
The model owner embeds a white-box watermark into $\theta$ using a specific embedding algorithm with the given scheme.
An adversary may apply NSO to produce an obfuscated parameter set $\theta'$ that preserves utility: $f_{\theta'} \approx f_{\theta}$. A verifier obtains white-box access to $\theta'$ (and the scheme's verification key/secret where applicable) and aims to verify the watermark originally embedded in $\theta$.

Our pipeline takes $\theta'$ as input and outputs recovered parameters $\hat{\theta}$ such that (i) $f_{\hat{\theta}} \approx f_{\theta'}$ ( hence also $f_{\hat{\theta}} \approx f_{\theta}$ under NSO's utility preserving promise), and (ii) watermark extraction on $\hat{\theta}$ yields high similarity according to the scheme's metric.
For \textit{evaluation} with a clean reference available, we report a reference-equivalence certificate that checks whether watermarked weights match the pre-attack clean model up to permissible channel equivalences (permutation and optional positive per-channel scaling) under a relative tolerance.

\heading{Threat model.} We consider a white-box adversary who has full knowledge of the victim model architecture and can modify internal parameters. The adversary performs NSO attacks by injecting dummy channels into Conv/Linear layers and instantiating these edits \textit{graph-consistently} in modern architectures with structural operators such as residual \texttt{add} and channel \texttt{cat}.
The adversary is allowed to apply standard NSO primitives, including zero-effect channels, mutually canceling channel groups, and redundant channel splits, as long as the resulting model remains shape compatible and functionally equivalent. The adversary’s goal is to preserve model utility ($f_{\theta'} \approx f_{\theta}$) while breaking structure-dependent watermark verification (e.g., index-ordered extraction), rather than destroying the watermark signal itself.

\heading{System assumptions.}
We assume that the model forward pass is sufficiently static to recover a computation graph (e.g., via \texttt{torch.fx}), allowing us to trace dataflow through structural operators such as residual \texttt{add} and channel \texttt{cat}. Conv/inear layers expose a well-defined channel that can be rewritten, while common nonlinear activations preserve channel count.

The defender is assumed to have probe access to the model and can evaluate it on a small set of probe inputs (synthetic or public) to collect activation signatures used for redundancy inference. Channel-layout rewrites are applied only to linear consumers (Conv/Linear); modules that explicitly mix channels (e.g., LayerNorm or attention) are treated as barriers and are not traversed by non-identity channel transformations.

When grouped or depthwise convolutions are present, we restrict rewrites to respect group boundaries. Channel transformations are constrained to be block-diagonal with blocks aligned to groups (i.e., a direct sum of per-group submatrices with zeros off the diagonal blocks~\cite{horn2012matrix}, so no mixing across groups is permitted.

\begin{figure*}[t]
    \centering
    \includegraphics[width=0.99\linewidth]{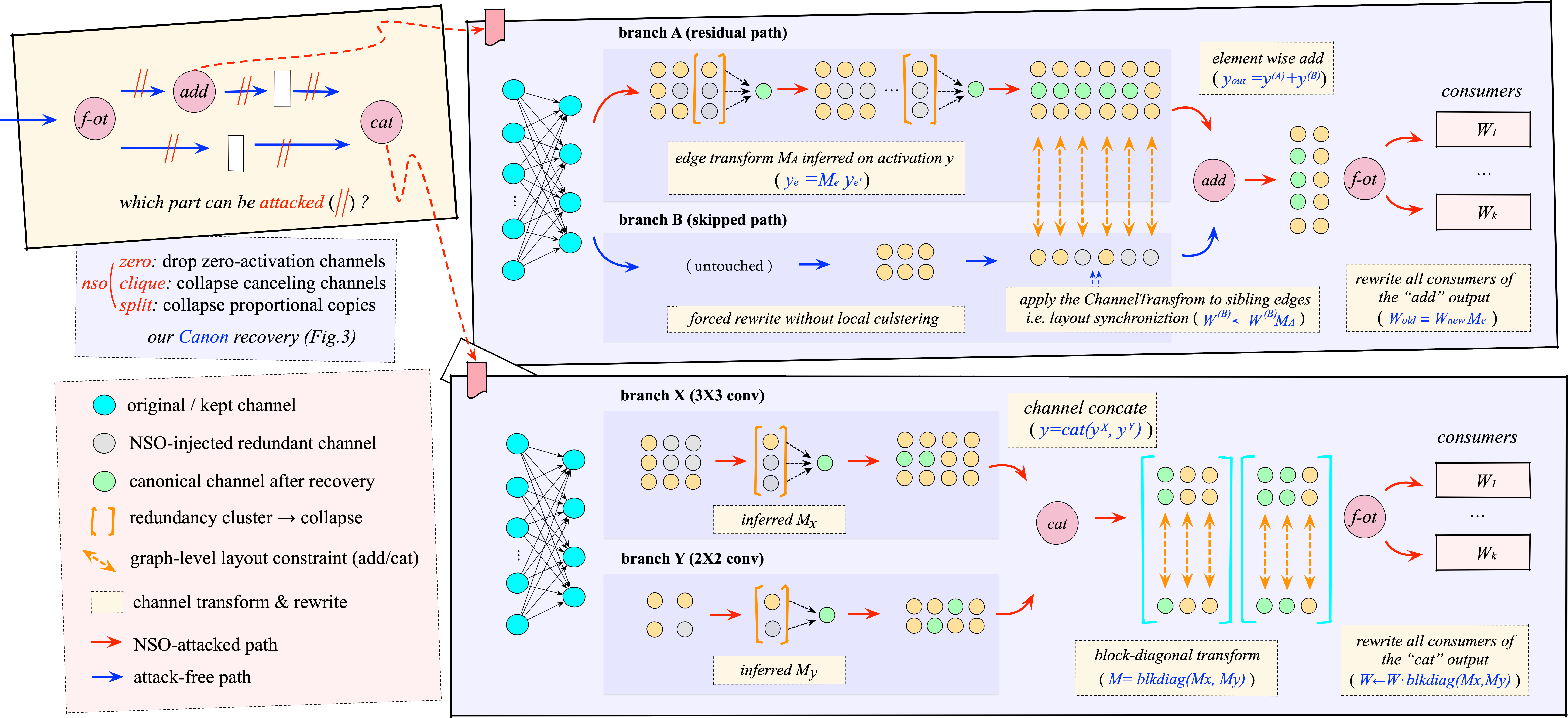}
    \caption{
    \textbf{Neuron-view global synchronization.} An adversary performs channel layout obfuscation by introducing a linear map on activations, writing $y_{\text{before}} = M_{\text{inj}}, y_{\text{after}}$ (with $M_{\text{inj}}$ typically sparse/structured and possibly non-square), and injecting redundant dummy channels (gray) via primitives such as \texttt{nso\_split} and \texttt{nso\_clique} while preserving end-to-end input–output behavior.  
    \textbf{(i)} For \textit{residual add (ResNet)}, From probe activations, we infer a channel transform $M_A$ on one branch. Since element-wise \texttt{add} requires layout compatibility, we enforce layout synchronization by applying a compatible rewrite on the sibling edge, even if it is attack-free. The unified add-output layout is then propagated by rewriting all downstream linear consumers (fan-out) via $W \leftarrow W\, M_A$, without requiring local rediscovery of redundancy. 
    \textbf{(ii)} For \textit{channel concatenation (Inception)}, each branch is compacted independently with transforms $(M_X, M_Y)$. At \texttt{cat}, these compose into a block-diagonal transform $\mathrm{blkdiag}(M_X, M_Y)$, which is propagated to all downstream consumers via $W \leftarrow W\times\mathrm{blkdiag}(M_X, M_Y)$.
}
    \label{fig:graph}
    \vspace{-0.05in}
\end{figure*}

\section{Graph-Consistent NSO Attacks}
\label{sec:graph-consistent}

We first introduce a \textit{ChannelTransform} abstraction that unifies NSO edits and our recovery actions.
We then formalize a \textit{graph-consistent} instantiation of NSO on modern architectures (\texttt{fan-out}, residual \texttt{add}, channel \texttt{cat}).

\subsection{ChannelTransform Abstraction}

\heading{``Before-from-after'' convention.}
We represent any channel layout rewrite on a \textit{producer edge} as a sparse linear transform on the channel axis:
\begin{equation}
  y_{\text{before}} = M\, y_{\text{after}},
  \quad
  M \in \mathbb{R}^{C_{\text{before}} \times C_{\text{after}}}.
  \label{eq:channeltransform}
\end{equation}
Here, ``before/after'' is \textit{local to a rewrite} on a given tensor edge; it does not assume ``before'' is the clean model globally.
For NSO injection, ``after'' is the expanded (obfuscated) representation; while for the recovery phase, ``after'' is typically the compacted representation.
This convention makes the consumer rewrite rule uniform across attack and recovery.

\heading{Consumer rewrite for dense Conv/Linear.}
Let a linear consumer produce $z = W\, y_{\text{before}}$ from the producer output $y_{\text{before}}$.
If we rewrite the producer interface using~\eqref{eq:channeltransform}, functional equivalence is preserved by updating the consumer parameters as:
\begin{equation}
\begin{aligned}
  z & = W\, y_{\text{before}} = W (M y_{\text{after}}) = (W M)\, y_{\text{after}} \\
 & \quad \Rightarrow \quad
  W_{\text{after}} = W_{\text{before}}\, M.
  \label{eq:consumer-rewrite}
\end{aligned}
\end{equation}
For Conv2D, the same rule applies by reshaping the convolution kernel as a matrix over the input-channel axis.

\heading{Consumer rewrite for grouped/depthwise Conv.}
For grouped convolution with $G$ groups, the input-channel axis is partitioned into $G$ disjoint blocks.
We restrict ChannelTransform to be block-diagonal~\cite{horn2012matrix} with respect to this partition and apply~\eqref{eq:consumer-rewrite} independently within each group, ensuring the rewritten operator remains a valid grouped convolution.
Depthwise convolution is the special case $G=C$. Because depthwise kernels do \textit{not} linearly mix channels, preserving the depthwise operator class requires a \textit{restricted} ChannelTransform: on depthwise consumers we allow only diagonal per-channel transforms (positive scaling, and optionally per-channel selection when pruning is enabled). We treat permutations as explicit channel reindexing operations that must be propagated across adjacent edges; if a nontrivial permutation cannot be propagated safely in a region, we treat the depthwise module as a barrier to that transform.

\begin{algorithm}[!t]
\footnotesize
\caption{Graph-consistent NSO injection}
\label{alg:nso-graph-consistent}
\begin{algorithmic}[1]
\Require model \textcolor{teal}{$f_\theta$}; injection ratio \textcolor{teal}{$\rho$}; split selector \textcolor{teal}{$p$};
primitive \textcolor{teal}{$\mathsf{op}\in\{\texttt{nso\_zero},\texttt{nso\_clique},\texttt{nso\_split}\}$}
\Ensure attacked model \textcolor{teal}{$f_{\theta'}$} with $f_{\theta'}\approx f_\theta$ and graph-consistent layouts

\State Trace forward graph; collect producer edges $\mathcal{E}$ (Conv/Linear outputs).
\State For each $e\in\mathcal{E}$, record downstream consumers: \texttt{fan-out}, \texttt{add}, \texttt{cat}.

\For{each $e\in\mathcal{E}$ with width $C_e$}
    \State $d_e \gets \lceil\rho\,C_e\rceil$
    \If{$\mathsf{op}=\texttt{nso\_split}$}
        \State $i_e \gets \min(C_e{-}1,\lfloor p\,C_e\rfloor)$
    \EndIf
\EndFor
\Comment{\textcolor{violet}{global injection plan (merge-safe)}}

\State For each residual \texttt{add} merge-group, enforce an identical expansion rule
\Statex \quad (same $d_e$ and channel ordering) across all incoming operands.
\State For each \texttt{cat}, record branch ordering for block-diagonal propagation. 
\Statex  \Comment{\textcolor{violet}{graph-consistency constraints}}

\State $\theta' \gets \theta$  \Comment{\textcolor{violet}{initialize recovery from attacked parameters}}

\For{each producer edge $e\in\mathcal{E}$ in topological order}
    \State $(\Delta\theta_e, M_e) \gets \textsc{Inject}(e,d_e,i_e,\mathsf{op})$
    \Comment{\textcolor{violet}{$y^{\mathrm{old}}_e = M_e\,y^{\mathrm{new}}_e$ (Eq.~\eqref{eq:channeltransform})}}

    \State Apply $\Delta\theta_e$ to update producer parameters in $\theta'$.

    \For{each downstream Conv/Linear consumer weight $W$ that consumes $y^{\mathrm{old}}_e$}
        \State $W \gets W\,M_e$ \Comment{\textcolor{violet}{consumer-consistent rewrite (Eq.~\eqref{eq:consumer-rewrite})}}
    \EndFor

    \State Synchronize across structural nodes:  \Comment{\textcolor{violet}{layout consistency}}
    \If{\texttt{add}}
        \State enforce layout compatibility at each residual merge (per merge-group).
    \ElsIf{\texttt{cat}}
        \State $M_{\textsf{cat}} \gets \mathrm{blkdiag}(M_1,\dots,M_k)$ on the \texttt{cat} output edge.
        \State rewrite its consumers as $W \gets W\,M_{\textsf{cat}}$.
    \EndIf
\EndFor

\State \Return $f_{\theta'}$
\end{algorithmic}
\end{algorithm}

\heading{Flatten/view boundary.}
For the common CNN pattern $y \in \mathbb{R}^{C\times H \times W}$ flattened to $\tilde{y}\in\mathbb{R}^{C H W}$ (channel major layout), the lifted transform is:
\begin{equation}
  \tilde{y}_{\text{before}} = (M \otimes I_{HW})\, \tilde{y}_{\text{after}},
  \label{eq:flatten-lift}
\end{equation}
where $I_{HW}$ is the $HW \times HW$ identity matrix and $\otimes$ denotes the Kronecker product~\cite{horn2012matrix}. So $(M\otimes I_{HW})$ applies the same channel transform $M$ independently to each of the $HW$ spatial locations under the channel major flattening.
So the corresponding linear consumer rewrite uses:
\begin{equation}
W_{\text{after}} = W_{\text{before}} (M \otimes I_{HW}).
\label{eq:flatten-lift-W}
\end{equation}

\heading{Permutation/scaling.}
Per-channel permutation and positive scaling are sparse \textit{ChannelTransforms}: 
permutation is a one-hot matrix $\Pi$ and scaling is a diagonal matrix $D$ with $D_{ii}>0$.
These correspond to $M=\Pi^\top D^{-1}$ as a sparse ChannelTransform, such that $y_{\text{before}} = M\,y_{\text{after}}$ (equivalently, $y_{\text{after}} = D\Pi\, y_{\text{before}}$).
Matching the threat model and our evaluation certificate, our recovery's proportionality estimates $\alpha_j$ explicitly target these scaling relationships, and our BN-aware capture is designed to remain correct when scaling is implemented with batch-normalization consistent compensation.

\subsection{Graph-consistent NSO}

\heading{Why ``graph-consistent'' matters.}
In sequential CNNs, NSO can be described as a layer-local channel edit plus a local consumer rewrite.
Modern models, however, are graph-structured~\cite{bhatti2023deep}: a single producer edge may feed \textit{multiple consumers} (\texttt{fan-out}), and structural operators impose global layout constraints.
Residual \texttt{add} requires its two operands to share an \textit{identical} channel layout, and channel \texttt{cat} requires consistent segment alignment across branches.
Therefore, to preserve functional equivalence, NSO must be instantiated as a \textit{graph-consistent} family of transforms plus \textit{consistent rewrites} across all downstream consumers along \textit{every} producer$\rightarrow$consumer path.

\heading{Graph-consistent injection policy (global plan).}
Following NSO~\cite{yan2023}, extending obfuscation to graph-structured architectures such as ResNet and Inception requires coordinated injection positions. We therefore adopt a \textit{graph-consistent injection policy}: a global injection plan is first determined over the traced computation graph, ensuring that every residual \texttt{add} merge observes compatible expanded layouts and that every \texttt{cat} node receives branches whose expanded layouts align correctly in the concatenated channel bus.

For each eligible producer edge $e$ with width $C_e$, we inject $d_e=\lceil \rho C_e \rceil$ dummy channels using standard NSO primitives: \texttt{nso\_zero} or \texttt{nso\_clique}, which append $d_e$ channels, or \texttt{nso\_split}, which replaces a baseline channel selected by $p$ with $d_e{+}1$ scaled duplicates. 
We additionally consider composed primitives (\texttt{mix-opseq}) and treat the resulting obfuscation as a composition of \textit{ChannelTransforms}.

\heading{Topological execution and immediate consumer rewrites.}
Algorithm~\ref{alg:nso-graph-consistent} executes the global injection plan in topological order over the computation graph. At each injected producer edge $e$, we construct a transform $M_e$ such that $y_e = M_e\, y'_e$, and immediately rewrite \textit{all} downstream linear consumers according to~\eqref{eq:consumer-rewrite}. This eager rewriting strategy ensures that channel layout changes introduced at the producer are consistently absorbed by every consumer along the producer$\rightarrow$consumer paths.

For residual merges implemented via \texttt{add}, we verify that both operands satisfy the required layout compatibility at the merge; otherwise, the injection placement is deemed unsupported. For channel concatenation via \texttt{cat}, we compose the per-branch transforms into a block-diagonal map and propagate the resulting transform to all downstream consumers.

\heading{Instantiating \textsf{Inject}.}
We instantiate \textsf{Inject} using NSO function-equivalence primitives. \texttt{nso\_zero} inserts null-effect channels, \texttt{nso\_split} duplicates a baseline channel into redundant copies whose aggregate contribution
matches the original, and \texttt{nso\_clique} introduces mutually canceling channels by ensuring that the corresponding consumers' weight columns sum to zero after rewriting. Each primitive is expressed as a \textit{ChannelTransform} paired with the consumer rewrite rule, which together preserve functional equivalence (up to numerical tolerance) by construction.

\section{End-to-End Recovery: \textsc{Canon}}
\label{subsec:canon-recovery}

\begin{figure}[!]
    \centering
    \includegraphics[width=\linewidth]{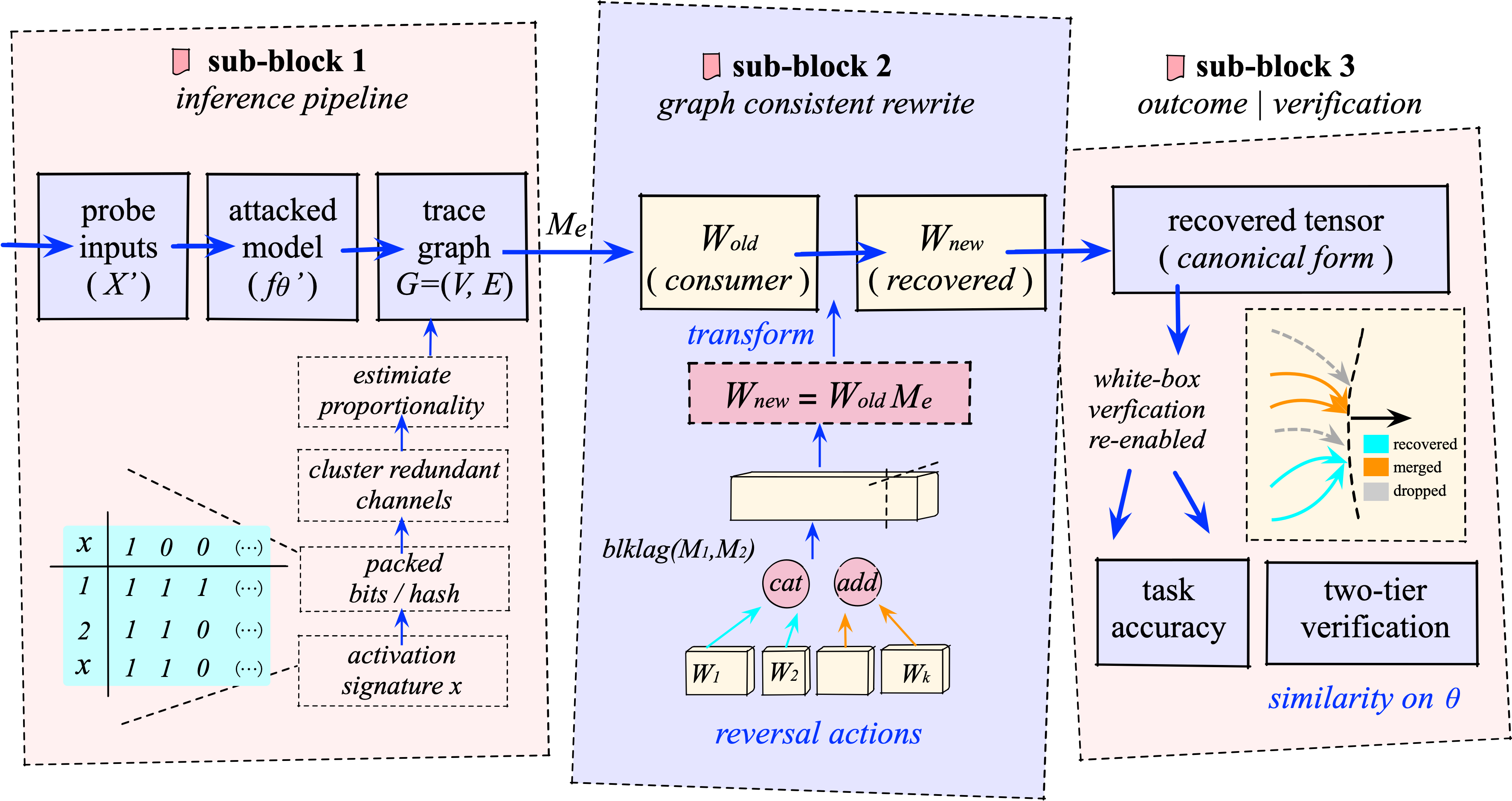}
    \vspace{-0.1in}
    \caption{\textbf{Recovery.} By clustering activation signatures from probe inputs, we infer a compacting transform $M_e$. Rather than local pruning, our \textit{graph-consistent policy} propagates $M_e$ to all downstream linear consumers via $W_{\text{new}} = W_{\text{old}}\, M_e$, enforcing consistent channel layouts across residual \texttt{add} and enabling block-diagonal handling of \texttt{cat}. This restores watermark verification on the recovered parameters $\hat{\theta}$.
    }
    \label{fig:attack_recovery}
\end{figure}

We present \textsc{Canon}, a recovery procedure (Fig.\ref{fig:attack_recovery}) that (i) infers channel redundancy from probe activations and (ii) propagates the resulting layout correction to \textit{all} downstream consumers \textit{by construction}, restoring a canonical, compact representation suitable for white-box watermark extraction.

\heading{Design principle.} \textsc{Canon} follows a \emph{ChannelTransform-first} design principle. Rather than requiring each downstream layer to independently rediscover redundancy, we infer a compacting transform at producer edges and rewrite all downstream linear consumers accordingly. This global rewriting strategy is essential under \texttt{fan-out} and structural merges, where local-only defences are brittle and inconsistent. Once a ChannelTransform is inferred, \textsc{Canon} enforces a synchronized channel basis across all producer$\rightarrow$consumer paths, including residual \texttt{add} and channel \texttt{cat}. As a result, the recovered model is graph-consistent by construction. 

We now present the details of our recovery steps.

\subsection{Inferring Channel Redundancy}

\subsubsection{Probe Capture}

\heading{Probe set and evaluation mode.}
We pre-generate a deterministic probe set $\mathcal{X}=\{x_t\}_{t=1}^{T}$ and run the attacked model $f_{\theta'}$ in evaluation mode (no dropout; BN uses running statistics).
For each eligible Conv/Linear producer edge $e$, we record per-channel activation summaries on $\mathcal{X}$.

\heading{BN-aware activation capture.}
To remain correct under BN-consistent scaling, we capture activations at a consistent semantic location:
if a BatchNorm immediately follows the producer edge before the next linear operator, we record the \textit{post-BN} tensor; otherwise we record the producer output.
More generally, we capture \textit{post-normalization} and \textit{before any cross-channel mixing}; if a channelwise nonlinearity follows (e.g., ReLU), we apply the same nonlinearity in the capture pipeline when forming state bits and summaries.

\subsubsection{Scalar Summaries and State Bits.}
For probe $t$, let the recorded tensor be $A^{(t)}_e$.
For each channel $i$, we compute a real-valued summary $u_{i,t}$ (e.g., spatial mean of $\mathrm{ReLU}(A^{(t)}_e[i])$) and an activity bit $b_{i,t}=\mathbb{I}[u_{i,t}>0]$.
This yields a per-channel bitstring $\mathbf{b}_i=(b_{i,1},\dots,b_{i,T})$.

\heading{Signature hashing (coarse bucketing).}
We pack $\mathbf{b}_i$ and hash it into a fixed-length signature:
\begin{equation}
\mathrm{sig}(i)=H\big(\mathrm{packbits}(\mathbf{b}_i)\big).
\end{equation}
Channels with identical signatures form coarse candidate buckets.
This is efficient and aligns with the NSO mechanism: injected redundant channels tend to share \textit{repeated activation state patterns} under diverse probes.

\subsubsection{Proportionality Refinement and Scale Estimation}

\heading{Why refinement is necessary.}
Signature buckets are coarse, where unrelated channels may collide, and true duplicates may coexist with channels that merely share the same active/inactive pattern.
We therefore refine each bucket using proportionality of real-valued summaries across probes.

\heading{Robust scale estimation.}
For channels $i,j$ in the same bucket, we estimate a scalar $\alpha_{j\leftarrow i}$ such that $u_{j,t}\approx \alpha_{j\leftarrow i}u_{i,t}$.
We compute a robust median ratio on probes where $|u_{i,t}|>\epsilon$ (and require at least $T_{\min}$ valid probes), and accept proportionality only if a symmetric relative error test is below a tolerance $\tau$.
Inside a bucket, we build a proportionality graph and take connected components as refined subclusters.

\subsection{Fan-out aware Merge/Drop Decisions}

\heading{Fan-out aware decision rule.}
Given a refined redundancy cluster $\mathcal{C}$ on a producer edge, we decide whether it is a pure dummy structure (\texttt{nso\_clique}) or represents a meaningful duplicated signal (\texttt{nso\_split}).
Crucially, because a producer edge may feed multiple consumers, this decision must hold across \textit{all} downstream linear consumers (\texttt{fan-out}).

\heading{Merged outgoing effect.}
For any consumer with input-channel columns $\{w_j\}$ on that edge, we form the merged representative:
\begin{equation}
  w_r' \;=\; \sum_{j\in\mathcal{C}} \alpha_j\, w_j.
  \label{eq:merged-column}
\end{equation}
We drop $\mathcal{C}$ as a \texttt{nso\_clique} dummy structure only if $\|w_r'\|$ is near zero for \textit{every} downstream linear consumer; otherwise, we keep a representative and merge the rest into it.

\subsection{Graph-consistent Rewrites}

\heading{Transform synthesis.}
From the merge/drop decisions, we synthesize a sparse edge transform $M_e$ such that $y_e = M_e\,y'_e$, where $y'_e$ is the compacted producer output.
This transform encodes kept representatives and merge coefficients.

\heading{Global consumer rewrites (by construction).}
We then rewrite \textit{every} downstream linear consumer that consumes $y_e$ by applying~\eqref{eq:consumer-rewrite}. This is the core \textit{ChannelTransform-first} step: we do not rely on consumers to rediscover redundancy locally.

\heading{Residual \texttt{add}: layout synchronization.}
For an elementwise merge $y=y^{(1)}+y^{(2)}$, both operands must share an identical recovered layout. We therefore enforce a shared recovered layout across the two branches: if one branch is compacted or reparameterized, the sibling branch is rewritten to a representation compatible with the recovered layout so the merge remains valid.
Equivalently, for each residual add merge group (a set of edges constrained by skip topology to be compatible with the layout), we enforce a single recovered channel basis and propagate it to all downstream consumers.

\heading{Channel concatenation: block-diagonal composition.}
For $y=\mathrm{cat}(y^{(1)},\dots,y^{(k)})$ along the channel axis, we compose per-branch transforms as:
\begin{equation}
  y_{\text{before}} = \mathrm{blkdiag}(M_1,\dots,M_k)\, y_{\text{after}}.
  \label{eq:blockdiag}
\end{equation}
This preserves branch semantics and enables downstream rewrites through the concatenation node.

\begin{algorithm}[!t]
\footnotesize
\caption{\textsc{Canon} Recovery}
\label{alg:recovery-graph-consistent}
\begin{algorithmic}[1]
\Require attacked model  \textcolor{teal}{$f_{\theta'}$}; probe set  \textcolor{teal}{$\mathcal{X}$};
thresholds \textcolor{teal}{$\epsilon,\tau$}; norms \textcolor{teal}{$\gamma_{\text{drop}}$ (drop)} and  \textcolor{teal}{$\gamma_{\text{keep}}$ (keep/merge)}
\Ensure recovered model \textcolor{teal}{$f_{\hat{\theta}}$}

\State Trace attacked forward graph.
\State For each channelized producer edge $e$, collect:
\Statex \quad downstream consumers (\texttt{fan-out}) and structural nodes (\texttt{add}, \texttt{cat}).
\State Run probes $x\in\mathcal{X}$ in eval mode.
\State For each eligible edge $e$, record BN-aware per-channel summaries $u_{i,t}$ and activity bits $b_{i,t}$.
\State $\hat{\theta} \gets \theta'$.  \Comment{\textcolor{violet}{initialize recovery from attacked parameters}}

\Statex \quad At each edge $e$, recover via $y^{\mathrm{atk}}_e = M_e\,y^{\mathrm{cmp}}_e$ and rewrite $W \leftarrow W M_e$.

\For{each eligible producer edge $e$ in topological order} 
    \Statex    \Comment{\textcolor{violet}{candidate channel groups for equivalence preserving pruning}}
    \State Compute channel signatures:
    \Statex \quad $\mathrm{sig}(i) \gets H(\mathrm{packbits}(\mathbf{b}_i))$. 
    \Comment{\textcolor{violet}{coarse bucketing}}

    \State Partition channels into buckets by identical $\mathrm{sig}(\cdot)$.
    \State Within each bucket, refine subclusters satisfying:
    \Statex \quad $u_{j,t} \approx \alpha\,u_{r,t}$ on probes with $|u_{r,t}|>\epsilon$ and relative error $\le\tau$.

    \For{each refined cluster $\mathcal{C}$ with representative $r$ and scales $\{\alpha_j\}_{j\in\mathcal{C}}$}
        \For{each downstream linear consumer $c$}
            \State $w^{(c)}_{\star} \gets \sum_{j\in\mathcal{C}} \alpha_j\, W^{(c)}_{:,j}$.
        \EndFor
        \If{$\max_c \|w^{(c)}_{\star}\| \le \gamma_{\text{drop}}$}
            \State mark $\mathcal{C}$ as \textbf{drop}.
        \Else
            \State merge $\mathcal{C}$ into $r$ and keep $r$.
        \EndIf
    \EndFor

    \State Construct sparse transform $M_e$ encoding keep/merge/drop
    \Statex \quad such that $y^{\mathrm{atk}}_e = M_e\,y^{\mathrm{cmp}}_e$.
    \Comment{\textcolor{violet}{synthesize  synthesize linear transform}}

    \State Compact producer parameters in $\hat{\theta}$ accordingly.

    \For{each consumer weight $W$ that previously consumes $y^{\mathrm{atk}}_e$}
        \State $W \gets W\,M_e$.
    \EndFor
    \Comment{\textcolor{violet}{propagate recovered layout to all consumers}}

    \State Synchronize across structural nodes: \Comment{\textcolor{violet}{layout consistency}}
    \If{\texttt{add}}
        \State enforce a shared recovered layout across both operands within each residual merge group.
    \ElsIf{\texttt{cat}}
        \State $M_{\textsf{cat}} \gets \mathrm{blkdiag}(M_1,\dots,M_k)$.
        \State rewrite consumers as $W \gets W\,M_{\textsf{cat}}$. 
    \EndIf
\EndFor

\State \Return $f_{\hat{\theta}}$
\end{algorithmic}
\end{algorithm}

\subsection{Verification and Evaluation}
\label{subsec:verification}

\heading{Two-tier evaluation protocol.}
Since we target \textbf{white-box} watermarking, a clean pre-attack reference model is typically available to the verifier/defender in practice.
Accordingly, \textsc{Canon} adopts a \textit{two-tier} evaluation protocol that leverages the reference when present for a strong certification, and otherwise falls back to a similarity-based check.
Concretely:
(i) with a clean reference available, we first attempt a strong reference-equivalence certificate on watermarked weights, aligned with the canonicalization goal; and
(ii) otherwise, or if the certificate does not pass, we fall back to an explicit attack-aware similarity PASS predicate based on the clean/attacked/recovered triplet.

\heading{Tier 1: reference-equivalence certificate.}
Given a clean pre-attack reference model, we certify that recovered watermarked weights match the reference up to output-channel permutation (and optional positive per-channel scaling) under a relative tolerance.
We flatten each output channel’s incoming weights into a vector $w_i$ (and optionally append bias) and form the pairwise $\ell_2$ residual matrix $R_{i,j}$.
We then perform an $\ell_2$-greedy one-to-one matching~\cite{munkres1957algorithms} by repeatedly selecting the smallest remaining $R_{i,j}$ while enforcing uniqueness, until a full matching $\pi$ is obtained.
Passing the certificate implies that the recovered watermarked tensors lie in the \textit{same symmetry-equivalence class} as the reference (permutation and, when enabled, BN-consistent positive scaling).
So, in this case, the verifier can treat the recovered parameters as a certified canonical representative that is reference-equivalent on watermarked weights, i.e., verification behaves as if performed on the reference modulo the allowed symmetries. Watermark similarity is then reported for completeness.

\heading{Tier 2: attack-aware similarity PASS.}
If Tier 1 is unavailable or does not pass, we apply an empirical PASS predicate based on each scheme's native similarity score.
Let $c=\mathrm{Sim}(\theta)$, $a=\mathrm{Sim}(\theta')$, and $r=\mathrm{Sim}(\hat{\theta})$ denote the similarity on the clean, attacked, and recovered models, respectively.
Define the similarity drop caused by the attack as $\Delta \triangleq \max\{0,\,c-a\}$.
We declare recovery \textsc{PASS} under Tier 2 if $r-a \;\ge\; \lambda\,\Delta \;-\; \delta$, where $\lambda\in(0,1]$ specifies the required fraction of the lost similarity to be recovered, and $\delta\ge 0$ is a tolerance term that absorbs extractor noise / numerical drift.

\heading{Reporting metrics.}
We report post-recovery task accuracy and watermark similarity computed by the extractor on $\hat{\theta}$.

\section{Experiments}
\label{sec:exp}

\begin{table*}[t]
    \centering
    \caption{Watermark similarity restoration for \textit{ResNet-18} and \textit{EfficientNet} under NSO \textcolor{magenta}{\textbf{\texttt{add-ops}}} attacks (\textcolor{magenta}{\textbf{$0.2$}} attack ratio). 
    Tier-1 reference-equivalence verification passes (\cmark) in all runs. We thus use Tier-1 watermark similarity of the recovered models as the primary outcome. Tier-2 similarity is included only as a conservative fallback for completeness rather than an invoked decision.
    }
    \label{tab:nso_recovery_suite_add_02}
    \vspace{5pt}
    \renewcommand{\arraystretch}{1.1}
    \resizebox{\linewidth}{!}{
    \begin{threeparttable}
    \begin{tabular}{c | c | ccc | c | ccccc }
    \toprule
    & \multirow{3}{*}{\textbf{Watermark}} &
    \multicolumn{3}{c|}{\textbf{Clean / Functional Drift}}  &
    \multirow{3}{*}{\textbf{\makecell{Tier-1\\Pass}}} &
    \multicolumn{5}{c}{\textbf{Attack Sim.$\rightarrow$ Recovery Tier-1 Sim. {\color{gray!80}(Tier-2 Sim.)}}}\\

    \cmidrule(lr){3-5} 
    \cmidrule(lr){7-11}
     & &
     \makecell{\textbf{Clean}\\ \textbf{Sim.}} &
     \makecell{$\boldsymbol{\Delta_{\mathrm{acc}}}$\\ \textbf{(att.)}} &
     \makecell{$\boldsymbol{\Delta_{\mathrm{acc}}}$\\ \textbf{(rec.)}} & &
     \makecell{\textbf{\texttt{nso\_zero}}} &
     \makecell{\textbf{\texttt{nso\_clique}}} &
     \multicolumn{1}{c}{\textbf{\texttt{nso\_split}}} &
     \makecell{\textbf{\texttt{mix-opseq}}} &
     \makecell{\textbf{\texttt{mix-opseq}}\\\textbf{\texttt{(per-merge-group)}}} 
     \\
    \midrule

    \multirow{9}{*}{\rotatebox{90}{\makecell{ResNet-18}}}
    & uchida17~\cite{uchida2017embedding} &\cellcolor{blue!9} 1.0000 & \cellcolor{blue!10} 0.0000 & \cellcolor{blue!10} 0.0000 & \cmark
      & \cellcolor{pink!25} 0.5703$\rightarrow$1.0000 {\color{gray!80}(0.5781)} & \cellcolor{pink!25} 0.5312$\rightarrow$1.0000 {\color{gray!80}(0.5781)} & \cellcolor{pink!25} 0.4844$\rightarrow$1.0000 {\color{gray!80}(0.4297)} & \cellcolor{pink!27} 0.4531$\rightarrow$1.0000 {\color{gray!80}(0.6016)} & \cellcolor{pink!27} 0.6094$\rightarrow$1.0000 {\color{gray!80}(0.4844)}  \\
      
    & greedy~\cite{liu2021watermarking} &\cellcolor{blue!9} 1.0000& \cellcolor{blue!10} 0.0000 & \cellcolor{blue!10} 0.0000 & \cmark
      & \cellcolor{pink!25} 0.5000$\rightarrow$1.0000 {\color{gray!80}(0.5000)}& \cellcolor{pink!25} 0.4844$\rightarrow$1.0000 {\color{gray!80}(0.5000)}& \cellcolor{pink!25} 0.4609$\rightarrow$1.0000 {\color{gray!80}(0.5234)} & \cellcolor{pink!27}  0.5703$\rightarrow$1.0000 {\color{gray!80}(0.5000)} & \cellcolor{pink!27} 0.5000$\rightarrow$1.0000 {\color{gray!80}(0.5156)}\\
      
    & lottery~\cite{chen2021you} &\cellcolor{blue!9} 1.0000& \cellcolor{blue!10} 0.0000 & \cellcolor{blue!10} 0.0000 & \cmark
      & \cellcolor{pink!25} 0.5547$\rightarrow$1.0000 {\color{gray!80}(0.4766)} & \cellcolor{pink!25} 0.5000$\rightarrow$1.0000 {\color{gray!80}(0.4766)} & \cellcolor{pink!25} 0.5312$\rightarrow$1.0000 {\color{gray!80}(0.5000)} & \cellcolor{pink!27} 0.4688$\rightarrow$1.0000 {\color{gray!80}(0.4844)} & \cellcolor{pink!27} 0.4609$\rightarrow$1.0000 {\color{gray!80}(0.4766)} \\
      
    & ipr\_gan~\cite{ong2021protecting} &\cellcolor{blue!9} 1.0000 & \cellcolor{blue!10} 0.0000 & \cellcolor{blue!10} 0.0000 & \cmark
      & \cellcolor{pink!25} 0.5703$\rightarrow$1.0000 {\color{gray!80}(0.5156)} & \cellcolor{pink!25} 0.5703$\rightarrow$1.0000 {\color{gray!80}(0.5156)} & \cellcolor{pink!25} 0.5625$\rightarrow$1.0000 {\color{gray!80}(0.5000)} & \cellcolor{pink!27} 0.5234$\rightarrow$1.0000 {\color{gray!80}(0.5156)} & \cellcolor{pink!27} 0.5469$\rightarrow$1.0000 {\color{gray!80}(0.4844)} \\
      
    & riga21~\cite{wang2021riga} & \cellcolor{blue!9} 0.9766 & \cellcolor{blue!10} 0.0000 & \cellcolor{blue!10} 0.0000 & \cmark
      & \cellcolor{pink!25} 0.4531$\rightarrow$0.9766 {\color{gray!80}(0.9766)} & \cellcolor{pink!25} 0.4531$\rightarrow$0.9766 {\color{gray!80}(0.9766)} & \cellcolor{pink!25} 0.4531$\rightarrow$0.9766 {\color{gray!80}(0.9922)} & \cellcolor{pink!27} 0.4531$\rightarrow$0.9766 {\color{gray!80}(0.9922)} & \cellcolor{pink!27} 0.4531$\rightarrow$0.9766 {\color{gray!80}(0.9922)} \\
      
    & ipr\_ic~\cite{lim2022protect} &\cellcolor{blue!9} 1.0000 & \cellcolor{blue!10} 0.0000 & \cellcolor{blue!10} 0.0000 & \cmark
      & \cellcolor{pink!25} 0.5156$\rightarrow$1.0000 {\color{gray!80}(0.5312)} & \cellcolor{pink!25} 0.4766$\rightarrow$1.0000 {\color{gray!80}(0.5312)} & \cellcolor{pink!25} 0.5156$\rightarrow$1.0000 {\color{gray!80}(0.5078)} & \cellcolor{pink!27} 0.5234$\rightarrow$1.0000 {\color{gray!80}(0.5312)} & \cellcolor{pink!27} 0.4141$\rightarrow$1.0000 {\color{gray!80}(0.4609)} \\
      
    & deepsigns~\cite{darvish2019deepsigns} &\cellcolor{blue!9} 1.0000 & \cellcolor{blue!10} 0.0000 & \cellcolor{blue!10} 0.0000 & \cmark
      & \cellcolor{pink!25} 0.5547$\rightarrow$1.0000 {\color{gray!80}(0.4844)} & \cellcolor{pink!25} 0.5547$\rightarrow$1.0000 {\color{gray!80}(0.4844)} & \cellcolor{pink!25} 0.5547$\rightarrow$1.0000 {\color{gray!80}(0.5000)} & \cellcolor{pink!27} 0.4844$\rightarrow$1.0000 {\color{gray!80}(0.5000)} & \cellcolor{pink!27} 0.5391$\rightarrow$1.0000 {\color{gray!80}(0.4531)} \\
      
    & deepipr~\cite{fan2021deepipr} &\cellcolor{blue!9} 1.0000 & \cellcolor{blue!10} 0.0000 & \cellcolor{blue!10} 0.0000 & \cmark
      & \cellcolor{pink!25} 0.4062$\rightarrow$1.0000 {\color{gray!80}(0.5312)} & \cellcolor{pink!25} 0.4062$\rightarrow$1.0000 {\color{gray!80}(0.5312)} & \cellcolor{pink!25} 0.4062$\rightarrow$1.0000 {\color{gray!80}(0.3594)} & \cellcolor{pink!27} 0.4062$\rightarrow$1.0000 {\color{gray!80}(0.6094)} & \cellcolor{pink!27} 0.4062$\rightarrow$1.0000 {\color{gray!80}(0.6094)} \\
      
    & passport\_aware~\cite{zhang2020passport} &\cellcolor{blue!9} 1.0000 & \cellcolor{blue!10} 0.0000 & \cellcolor{blue!10} 0.0000 & \cmark
      & \cellcolor{pink!25} 0.5703$\rightarrow$1.0000 {\color{gray!80}(0.5156)} & \cellcolor{pink!25} 0.5703$\rightarrow$1.0000 {\color{gray!80}(0.5156)} & \cellcolor{pink!25} 0.5625$\rightarrow$1.0000 {\color{gray!80}(0.5391)} & \cellcolor{pink!27} 0.5312$\rightarrow$1.0000 {\color{gray!80}(0.6094)} & \cellcolor{pink!27} 0.5000$\rightarrow$1.0000 {\color{gray!80}(0.4609)} \\

    \cmidrule{1-7}

    \multirow{9}{*}{\rotatebox{90}{\makecell{EfficientNet}}}
    & uchida17~\cite{uchida2017embedding} &\cellcolor{blue!9} 1.0000 & \cellcolor{blue!10} 0.0000 & \cellcolor{blue!10} 0.0000 & \cmark
      & \cellcolor{pink!25} 0.5547$\rightarrow$1.0000 {\color{gray!80}(0.4375)} & \cellcolor{pink!25} 0.4688$\rightarrow$1.0000 {\color{gray!80}(0.4375)} & \cellcolor{pink!25} 0.4766$\rightarrow$1.0000 {\color{gray!80}(0.4531)} & \cellcolor{pink!27} 0.4609$\rightarrow$1.0000 {\color{gray!80}(0.5234)} & \cellcolor{pink!27} 0.4688$\rightarrow$1.0000 {\color{gray!80}(0.4609)} \\
      
    & greedy~\cite{liu2021watermarking} &\cellcolor{blue!9} 1.0000& \cellcolor{blue!10} 0.0000 & \cellcolor{blue!10} 0.0000 & \cmark
      & \cellcolor{pink!25} 0.4844$\rightarrow$1.0000 {\color{gray!80}(0.4531)} & \cellcolor{pink!25} 0.4453$\rightarrow$1.0000 {\color{gray!80}(0.4531)} & \cellcolor{pink!25} 0.4922$\rightarrow$1.0000 {\color{gray!80}(0.5156)} & \cellcolor{pink!27} 0.5859$\rightarrow$1.0000 {\color{gray!80}(0.5312)} & \cellcolor{pink!27} 0.5078$\rightarrow$1.0000 {\color{gray!80}(0.5469)} \\
      
    & lottery~\cite{chen2021you} &\cellcolor{blue!9} 1.0000& \cellcolor{blue!10} 0.0000 & \cellcolor{blue!10} 0.0000 & \cmark
      & \cellcolor{pink!25} 0.5391$\rightarrow$1.0000 {\color{gray!80}(0.4531)} & \cellcolor{pink!25} 0.4688$\rightarrow$1.0000 {\color{gray!80}(0.4531)} & \cellcolor{pink!25} 0.4609$\rightarrow$1.0000 {\color{gray!80}(0.4688)}& \cellcolor{pink!27} 0.5078$\rightarrow$1.0000 {\color{gray!80}(0.4609)}& \cellcolor{pink!27} 0.5156$\rightarrow$1.0000 {\color{gray!80}(0.4531)}\\
      
    & ipr\_gan~\cite{ong2021protecting} &\cellcolor{blue!9} 1.0000 & \cellcolor{blue!10} 0.0000 & \cellcolor{blue!10} 0.0000 & \cmark
      & \cellcolor{pink!25} 0.4844$\rightarrow$1.0000 {\color{gray!80}(0.5391)} & \cellcolor{pink!25} 0.5547$\rightarrow$1.0000 {\color{gray!80}(0.5391)} & \cellcolor{pink!25} 0.4844$\rightarrow$1.0000 {\color{gray!80}(0.5156)} & \cellcolor{pink!27} 0.4375$\rightarrow$1.0000 {\color{gray!80}(0.5312)} & \cellcolor{pink!27} 0.5234$\rightarrow$1.0000 {\color{gray!80}(0.5469)} \\
      
    & riga21~\cite{wang2021riga} & \cellcolor{blue!9} 0.9062 & \cellcolor{blue!10} 0.0000 & \cellcolor{blue!10} 0.0000 & \cmark
      & \cellcolor{pink!25} 0.6172$\rightarrow$0.9062 {\color{gray!80}(0.8984)} & \cellcolor{pink!25} 0.6172$\rightarrow$0.9062 {\color{gray!80}(0.8984)} & \cellcolor{pink!25} 0.6172$\rightarrow$0.9062 {\color{gray!80}(0.8984)} & \cellcolor{pink!27} 0.6172$\rightarrow$0.9062 {\color{gray!80}(0.9141)} & \cellcolor{pink!27} 0.6172$\rightarrow$0.9062 {\color{gray!80}(0.9219)}  \\
      
    & ipr\_ic~\cite{lim2022protect} &\cellcolor{blue!9} 1.0000 & \cellcolor{blue!10} 0.0000 & \cellcolor{blue!10} 0.0000 & \cmark
      & \cellcolor{pink!25} 0.5000$\rightarrow$1.0000 {\color{gray!80}(0.4766)} & \cellcolor{pink!25} 0.4922$\rightarrow$1.0000 {\color{gray!80}(0.4766)} & \cellcolor{pink!25} 0.3750$\rightarrow$1.0000 {\color{gray!80}(0.4922)} & \cellcolor{pink!27} 0.4766$\rightarrow$1.0000 {\color{gray!80}(0.5078)} & \cellcolor{pink!27} 0.4766$\rightarrow$1.0000 {\color{gray!80}(0.5156)} \\
      
    & deepsigns~\cite{darvish2019deepsigns} &\cellcolor{blue!9} 1.0000 & \cellcolor{blue!10} 0.0000 & \cellcolor{blue!10} 0.0000 & \cmark
      & \cellcolor{pink!25} 0.4766$\rightarrow$1.0000 {\color{gray!80}(0.5234)} & \cellcolor{pink!25} 0.5312$\rightarrow$1.0000 {\color{gray!80}(0.5234)} & \cellcolor{pink!25} 0.4922$\rightarrow$1.0000 {\color{gray!80}(0.5000)} & \cellcolor{pink!27} 0.4297$\rightarrow$1.0000 {\color{gray!80}(0.5000)} & \cellcolor{pink!27} 0.5156$\rightarrow$1.0000 {\color{gray!80}(0.5312)} \\
      
    & deepipr~\cite{fan2021deepipr} &\cellcolor{blue!9} 1.0000 & \cellcolor{blue!10} 0.0000 & \cellcolor{blue!10} 0.0000 & \cmark
      & \cellcolor{pink!25} 0.4062$\rightarrow$1.0000 {\color{gray!80}(0.5469)} & \cellcolor{pink!25} 0.4062$\rightarrow$1.0000 {\color{gray!80}(0.5469)} & \cellcolor{pink!25} 0.4062$\rightarrow$1.0000 {\color{gray!80}(0.5469} & \cellcolor{pink!27} 0.4062$\rightarrow$1.0000 {\color{gray!80}(0.4219)} & \cellcolor{pink!27} 0.4062$\rightarrow$1.0000 {\color{gray!80}(0.5156)} \\
      
    & passport\_aware~\cite{zhang2020passport} &\cellcolor{blue!9} 1.0000 & \cellcolor{blue!10} 0.0000 & \cellcolor{blue!10} 0.0000 & \cmark
      & \cellcolor{pink!25} 0.4922$\rightarrow$1.0000 {\color{gray!80}(0.4922)} & \cellcolor{pink!25} 0.4844$\rightarrow$1.0000 {\color{gray!80}(0.4922)} & \cellcolor{pink!25} 0.5625$\rightarrow$1.0000 {\color{gray!80}(0.5469)} & \cellcolor{pink!27} 0.5391$\rightarrow$1.0000 {\color{gray!80}(0.4531)} & \cellcolor{pink!27} 0.5312$\rightarrow$1.0000 {\color{gray!80}(0.4766)} \\

    \bottomrule
    \end{tabular}

    \begin{tablenotes}
    \item \textbf{$\Delta_{\mathrm{acc}}$($\mathrm{clean},\mathrm{\textbf{att}acked})$} and \textbf{$\Delta_{\mathrm{acc}}(\mathrm{clean},\mathrm{\textbf{rec}overed})$} report the \emph{worst-case} absolute accuracy gap over the five listed attacks for each watermark.
    \item \textbf{Tier-1 pass} indicates that the Tier-1 reference-equivalence certificate over watermarked layers passes for all runs in the suite for that model/watermark. We use \cmark to denote a 100\% match rate.
    \end{tablenotes}

    \end{threeparttable}
    }
\end{table*}
\begin{table*}[t]
    \centering
    \caption{Watermark similarity restoration for \textit{InceptionV3} and \textit{DenseNet} under NSO \textcolor{magenta}{\textbf{\texttt{cat-ops}}} attacks (\textcolor{magenta}{\textbf{$0.2$}} attack ratio). Tier-1  verification passes (\cmark) in all runs and is used as the primary outcome, while Tier-2 serves as a conservative fallback.}
    \label{tab:nso_recovery_suite_cat_02}
    \vspace{5pt}

    \renewcommand{\arraystretch}{1.1}
    \resizebox{\linewidth}{!}{
    \begin{threeparttable}
    \begin{tabular}{c |c | ccc |c| ccccc}
    \toprule
    &
    \multirow{3}{*}{\textbf{Watermark}} &
    \multicolumn{3}{c|}{\textbf{Clean / Functional Drift}} &
    \multirow{3}{*}{\textbf{\makecell{Tier-1\\Pass}}} &
    \multicolumn{5}{c}{\textbf{Attack Sim.$\rightarrow$ Recovery Tier-1 Sim. {\color{gray!80}(Tier-2 Sim.)}}}
    \\

    \cmidrule(lr){3-5} 
    \cmidrule(lr){7-11}
     & &
     \makecell{\textbf{Clean}\\\textbf{Sim.}} &
     \makecell{$\boldsymbol{\Delta_{\mathrm{acc}}}$\\\textbf{(att.)}} &
     \makecell{$\boldsymbol{\Delta_{\mathrm{acc}}}$\\\textbf{(rec.)}} &
     &
     \makecell{\textbf{\texttt{zero}}} &
     \makecell{\textbf{\texttt{clique}}} &
     \makecell{\textbf{\texttt{split}}} &
     \makecell{\textbf{\texttt{mix-opseq}}} &
     \makecell{\textbf{\texttt{mix-opseq}}\\\textbf{\texttt{(per-merge-group)}}} 
     \\
    \midrule

    \multirow{9}{*}{\rotatebox{90}{\makecell{InceptionV3}}}
    & uchida17~\cite{uchida2017embedding} &\cellcolor{yellow!13} 1.0000 & \cellcolor{yellow!15} 0.0000 & \cellcolor{yellow!15} 0.0000 & \cmark
      & \cellcolor{blue!9}  0.3984$\rightarrow$1.0000 {\color{gray!80}(0.5391)} & \cellcolor{blue!9} 0.4766$\rightarrow$1.0000 {\color{gray!80}(0.5391)} & \cellcolor{blue!9}  0.5625$\rightarrow$1.0000 {\color{gray!80}(0.5938)} & \cellcolor{blue!10} 0.5000$\rightarrow$1.0000 {\color{gray!80}(0.5547)} & \cellcolor{blue!10} 0.5000$\rightarrow$1.0000 {\color{gray!80}(0.5547)} \\
      
    & greedy~\cite{liu2021watermarking} &\cellcolor{yellow!13} 1.0000& \cellcolor{yellow!15} 0.0000 & \cellcolor{yellow!15} 0.0000 & \cmark
      & \cellcolor{blue!9} 0.5938$\rightarrow$1.0000 {\color{gray!80}(0.5703)}& \cellcolor{blue!9}  0.5781$\rightarrow$1.0000 {\color{gray!80}(0.5703)}& \cellcolor{blue!9} 0.5156$\rightarrow$1.0000 {\color{gray!80}(0.5156)}& \cellcolor{blue!10} 0.5078$\rightarrow$1.0000 {\color{gray!80}(0.5234)}& \cellcolor{blue!10} 0.4609$\rightarrow$1.0000 {\color{gray!80}(0.4453)}  \\
      
    & lottery~\cite{chen2021you} &\cellcolor{yellow!13} 1.0000& \cellcolor{yellow!15} 0.0000 & \cellcolor{yellow!15} 0.0000 & \cmark
      & \cellcolor{blue!9}  0.4844$\rightarrow$1.0000 {\color{gray!80}(0.4766)}& \cellcolor{blue!9}  0.4766$\rightarrow$1.0000 {\color{gray!80}(0.4766)}& \cellcolor{blue!9}  0.4688$\rightarrow$1.0000 {\color{gray!80}(0.4609)}& \cellcolor{blue!10} 0.4688$\rightarrow$1.0000 {\color{gray!80}(0.4844)}& \cellcolor{blue!10} 0.3906$\rightarrow$1.0000 {\color{gray!80}(0.4531)}  \\
      
    & ipr\_gan~\cite{ong2021protecting} &\cellcolor{yellow!13} 1.0000 & \cellcolor{yellow!15} 0.0000 & \cellcolor{yellow!15} 0.0000 & \cmark
      & \cellcolor{blue!9}  0.5078$\rightarrow$1.0000 {\color{gray!80}(0.5000)} & \cellcolor{blue!9}  0.5469$\rightarrow$1.0000 {\color{gray!80}(0.5000)} & \cellcolor{blue!9}  0.5156$\rightarrow$1.0000 {\color{gray!80}(0.5234)}  & \cellcolor{blue!10} 0.5000$\rightarrow$1.0000 {\color{gray!80}(0.5391)} & \cellcolor{blue!10} 0.5000$\rightarrow$1.0000 {\color{gray!80}(0.5391)}  \\
      
    & riga21~\cite{wang2021riga} & \cellcolor{yellow!13} 0.9688 & \cellcolor{yellow!15} 0.0000 & \cellcolor{yellow!15} 0.0000 & \cmark
      & \cellcolor{blue!9}  0.4453$\rightarrow$0.9688 {\color{gray!80}(0.9688)} & \cellcolor{blue!9}  0.4453$\rightarrow$0.9688 {\color{gray!80}(0.9688)} & \cellcolor{blue!9}  0.4453$\rightarrow$0.9688 {\color{gray!80}(0.9766)} & \cellcolor{blue!10} 0.4453$\rightarrow$0.9688 {\color{gray!80}(0.9766)} & \cellcolor{blue!10} 0.4453$\rightarrow$0.9688 {\color{gray!80}(0.9766)}  \\
      
    & ipr\_ic~\cite{lim2022protect} &\cellcolor{yellow!13} 1.0000 & \cellcolor{yellow!15} 0.0000 & \cellcolor{yellow!15} 0.0000 & \cmark
      & \cellcolor{blue!9} 0.5547$\rightarrow$1.0000 {\color{gray!80}(0.5156)} & \cellcolor{blue!9}  0.5391$\rightarrow$1.0000 {\color{gray!80}(0.5156)} & \cellcolor{blue!9}  0.4766$\rightarrow$1.0000 {\color{gray!80}(0.4531)} & \cellcolor{blue!10} 0.5938$\rightarrow$1.0000 {\color{gray!80}(0.4453)} & \cellcolor{blue!10} 0.5938$\rightarrow$1.0000 {\color{gray!80}(0.4453)}  \\
      
    & deepsigns~\cite{darvish2019deepsigns} &\cellcolor{yellow!13} 1.0000 & \cellcolor{yellow!15} 0.0000 & \cellcolor{yellow!15} 0.0000 & \cmark
      & \cellcolor{blue!9}  0.5781$\rightarrow$1.0000 {\color{gray!80}(0.4531)} & \cellcolor{blue!9}  0.6094$\rightarrow$1.0000 {\color{gray!80}(0.4531)} & \cellcolor{blue!9}  0.5547$\rightarrow$1.0000 {\color{gray!80}(0.4688)} &  \cellcolor{blue!10}  0.5391$\rightarrow$1.0000 {\color{gray!80}(0.4844)} &  \cellcolor{blue!10}  0.5391$\rightarrow$1.0000 {\color{gray!80}(0.4844)}  \\
      
    & deepipr~\cite{fan2021deepipr} &\cellcolor{yellow!13} 1.0000 & \cellcolor{yellow!15} 0.0000 & \cellcolor{yellow!15} 0.0000 & \cmark
      & \cellcolor{blue!9}  0.4062$\rightarrow$1.0000 {\color{gray!80}(0.4688)} & \cellcolor{blue!9}  0.4062$\rightarrow$1.0000 {\color{gray!80}(0.4688)} & \cellcolor{blue!9}  0.4062$\rightarrow$1.0000 {\color{gray!80}(0.4219)} &  \cellcolor{blue!10} 0.4062$\rightarrow$1.0000 {\color{gray!80}(0.3281)} &  \cellcolor{blue!10}  0.4062$\rightarrow$1.0000 {\color{gray!80}(0.3281)}  \\
      
    & passport\_aware~\cite{zhang2020passport} &\cellcolor{yellow!13} 1.0000 & \cellcolor{yellow!15} 0.0000 & \cellcolor{yellow!15} 0.0000 & \cmark
      & \cellcolor{blue!9}  0.4844$\rightarrow$1.0000 {\color{gray!80}(0.4922)} & \cellcolor{blue!9}  0.4766$\rightarrow$1.0000 {\color{gray!80}(0.4922)} & \cellcolor{blue!9}  0.5391$\rightarrow$1.0000 {\color{gray!80}(0.4453)} &  \cellcolor{blue!10}  0.4453$\rightarrow$1.0000 {\color{gray!80}(0.5000)} &  \cellcolor{blue!10}  0.4453$\rightarrow$1.0000 {\color{gray!80}(0.5000)}  \\

    \cmidrule{1-9}

    \multirow{9}{*}{\rotatebox{90}{\makecell{DenseNet}}}
    & uchida17~\cite{uchida2017embedding} &\cellcolor{yellow!13} 1.0000 & \cellcolor{yellow!15} 0.0000 & \cellcolor{yellow!15} 0.0000 & \cmark
      & \cellcolor{blue!9}  0.5078$\rightarrow$1.0000 {\color{gray!80}(0.4844)} & \cellcolor{blue!9}  0.4531$\rightarrow$1.0000 {\color{gray!80}(0.4844)} & \cellcolor{blue!9}  0.5000$\rightarrow$1.0000 {\color{gray!80}(0.5703)} & \cellcolor{blue!10} 0.4453$\rightarrow$1.0000 {\color{gray!80}(0.5859)} & \cellcolor{blue!10} 0.4453$\rightarrow$1.0000 {\color{gray!80}(0.4375)} \\
      
    & greedy~\cite{liu2021watermarking} &\cellcolor{yellow!13} 1.0000& \cellcolor{yellow!15} 0.0000 & \cellcolor{yellow!15} 0.0000 & \cmark
      & \cellcolor{blue!9}  0.4531$\rightarrow$1.0000 {\color{gray!80}(0.4609)} & \cellcolor{blue!9} 0.5000$\rightarrow$1.0000 {\color{gray!80}(0.4609)}& \cellcolor{blue!9}  0.4766$\rightarrow$1.0000 {\color{gray!80}(0.5391)}& \cellcolor{blue!10} 0.4531$\rightarrow$1.0000 {\color{gray!80}(0.3984)}& \cellcolor{blue!10} 0.6328$\rightarrow$1.0000 {\color{gray!80}(0.4922)}  \\
      
    & lottery~\cite{chen2021you} &\cellcolor{yellow!13} 1.0000& \cellcolor{yellow!15} 0.0000 & \cellcolor{yellow!15} 0.0000 & \cmark
      & \cellcolor{blue!9}  0.5078$\rightarrow$1.0000 {\color{gray!80}(0.5078)} & \cellcolor{blue!9}  0.5156$\rightarrow$1.0000 {\color{gray!80}(0.5078)} & \cellcolor{blue!9}  0.5000$\rightarrow$1.0000 {\color{gray!80}(0.5078)} & \cellcolor{blue!10} 0.4922$\rightarrow$1.0000 {\color{gray!80}(0.5000)}& \cellcolor{blue!10} 0.5469$\rightarrow$1.0000 {\color{gray!80}(0.5000)} \\
      
    & ipr\_gan~\cite{ong2021protecting} &\cellcolor{yellow!13} 1.0000 & \cellcolor{yellow!15} 0.0000 & \cellcolor{yellow!15} 0.0000 & \cmark
      & \cellcolor{blue!9}  0.4922$\rightarrow$1.0000 {\color{gray!80}(0.4922)} & \cellcolor{blue!9}  0.5000$\rightarrow$1.0000 {\color{gray!80}(0.4922)} & \cellcolor{blue!9}  0.4922$\rightarrow$1.0000 {\color{gray!80}(0.5000)} & \cellcolor{blue!10} 0.4766$\rightarrow$1.0000 {\color{gray!80}(0.4688)} & \cellcolor{blue!10} 0.5000$\rightarrow$1.0000 {\color{gray!80}(0.5312)}  \\
      
    & riga21~\cite{wang2021riga} &\cellcolor{yellow!13} 1.0000 & \cellcolor{yellow!15} 0.0000 & \cellcolor{yellow!15} 0.0000 & \cmark
      & \cellcolor{blue!9} 0.4922$\rightarrow$1.0000 {\color{gray!80}(0.9531)} & \cellcolor{blue!9} 0.4922$\rightarrow$1.0000 {\color{gray!80}(0.9531)} & \cellcolor{blue!9} 0.4922$\rightarrow$1.0000 {\color{gray!80}(0.9531)} & \cellcolor{blue!10} 0.4922$\rightarrow$1.0000 {\color{gray!80}(0.9531)} & \cellcolor{blue!10} 0.4922$\rightarrow$1.0000 {\color{gray!80}(0.9531)}  \\
      
    & ipr\_ic~\cite{lim2022protect} & \cellcolor{yellow!13} 0.9922 & \cellcolor{yellow!15} 0.0000 & \cellcolor{yellow!15} 0.0000 & \cmark
      & \cellcolor{blue!9} 0.4531$\rightarrow$0.9922 {\color{gray!80}(0.5625)} & \cellcolor{blue!9}  0.5469$\rightarrow$0.9922 {\color{gray!80}(0.5625)} & \cellcolor{blue!9} 0.5391$\rightarrow$0.9922 {\color{gray!80}(0.5000)} & \cellcolor{blue!10} 0.4844$\rightarrow$0.9922 {\color{gray!80}(0.3906)} & \cellcolor{blue!10} 0.5312$\rightarrow$0.9922 {\color{gray!80}(0.4922)} \\
      
    & deepsigns~\cite{darvish2019deepsigns} &\cellcolor{yellow!13} 1.0000 & \cellcolor{yellow!15} 0.0000 & \cellcolor{yellow!15} 0.0000 & \cmark
      & \cellcolor{blue!9} 0.4844$\rightarrow$1.0000 {\color{gray!80}(0.5078)} & \cellcolor{blue!9}  0.4922$\rightarrow$1.0000 {\color{gray!80}(0.5078)} & \cellcolor{blue!9} 0.4922$\rightarrow$1.0000 {\color{gray!80}(0.5312)} & \cellcolor{blue!10} 0.5078$\rightarrow$1.0000 {\color{gray!80}(0.4844)} & \cellcolor{blue!10} 0.5078$\rightarrow$1.0000 {\color{gray!80}(0.5625)} \\
      
    & deepipr~\cite{fan2021deepipr} & \cellcolor{yellow!13}\cellcolor{yellow!13} 1.0000 & \cellcolor{yellow!15} \cellcolor{yellow!15} 0.0000 &  \cellcolor{yellow!15} \cellcolor{yellow!15} 0.0000 & \cmark
      &  \cellcolor{blue!9} 0.4062$\rightarrow$1.0000 {\color{gray!80}(0.4688)} & \cellcolor{blue!9} 0.4062$\rightarrow$1.0000 {\color{gray!80}(0.4688)} & \cellcolor{blue!9} 0.4062$\rightarrow$1.0000 {\color{gray!80}(0.5625)} & \cellcolor{blue!10} 0.4062$\rightarrow$1.0000 {\color{gray!80}(0.5469)} & \cellcolor{blue!10} 0.4062$\rightarrow$1.0000 {\color{gray!80}(0.5312)}  \\
      
    & passport\_aware~\cite{zhang2020passport} & \cellcolor{yellow!13}\cellcolor{yellow!13} 1.0000 &  \cellcolor{yellow!15} \cellcolor{yellow!15} 0.0000 &  \cellcolor{yellow!15} \cellcolor{yellow!15} 0.0000 & \cmark
      & \cellcolor{blue!9} 0.4844$\rightarrow$1.0000 {\color{gray!80}(0.3984)} & \cellcolor{blue!9} 0.4453$\rightarrow$1.0000 {\color{gray!80}(0.3984)} & \cellcolor{blue!9} 0.5156$\rightarrow$1.0000 {\color{gray!80}(0.4297)} & \cellcolor{blue!10} 0.4922$\rightarrow$1.0000 {\color{gray!80}(0.4844)} & \cellcolor{blue!10}  0.5469$\rightarrow$1.0000 {\color{gray!80}(0.5156)}  \\

    \bottomrule
    \end{tabular}

    \begin{tablenotes}
    \item $\Delta_{\mathrm{acc}}(\mathrm{clean},\mathrm{\textbf{att}acked})$ and $\Delta_{\mathrm{acc}}(\mathrm{clean},\mathrm{\textbf{rec}overed})$ report the \emph{worst-case} absolute accuracy gap over the five listed attacks for each watermark.
    \item \textbf{Tier-1 pass} indicates that the Tier-1 reference-equivalence certificate over watermarked layers passes for all runs in the suite for that model/watermark (\cmark means 100\% match rate).
    \end{tablenotes}

    \end{threeparttable}
    }
\end{table*}

We evaluate \textsc{Canon} for its practicality. We focus on whether \textsc{Canon} recovers watermark similarity across various white-box schemes and whether graph-consistent rewrites remain correct on nontrivial graphs (\texttt{fan-out}, residual \texttt{add}, channel \texttt{cat}) under strong, compositional NSO obfuscations.

For each model/dataset/watermark configuration, we run the full end-to-end pipeline (train+embed $\rightarrow$ NSO $\rightarrow$ recover) and report accuracy and watermark similarity on the clean, attacked, and recovered models. 
We additionally evaluate recovery with our two-tier verifier: first applying the reference-equivalence certificate on watermarked weights, and falling back to the attack-aware similarity PASS criterion when the certificate does not hold.

\subsection{Experimental Settings}

To capture a broad range of practical settings, we evaluate across multiple architectures and datasets. All experiments are run on a dedicated server environment to ensure a consistent and controlled comparison.

\smallskip
\noindent\textbf{Configurations.}
Our testbed consists of a dual-socket Intel Xeon Gold 6126 system (12 cores per CPU) with 192\,GB of 2666\,MHz ECC DDR4 memory (six-channel configuration) and 2$\times$1.2\,TB 10,000\,RPM SAS II drives in RAID~1.
We use two NVIDIA Tesla V100 GPUs (5,120 CUDA cores, 640 Tensor cores, 32\,GB memory each) on 64-bit Ubuntu 18.04.

\smallskip
\noindent\textbf{Datasets.}
As a \textit{structure-level} defense against NSO, \textsc{Canon} operates on computation graphs and channel-wise activation signatures, and is thus largely agnostic to semantic data distributions.
Accordingly, experiments are conducted on MNIST~\cite{cohen2017emnist} as a \textit{controlled} testbed to isolate the impact of structural obfuscation and graph-consistent recovery from confounders that dominate larger benchmarks (e.g., heavy augmentation, long training schedules, and dataset-specific inductive biases).
This choice enables exhaustive evaluation across all watermark schemes and NSO variants with multiple random seeds under a fixed compute budget, yielding stable conclusions about recovery correctness and verifiability.
MNIST images are adapted by repeating the grayscale channel to three channels and resizing $28\times 28$ to $32\times 32$.

\smallskip
\noindent\textbf{Model architectures.}
We evaluate four canonical CNN backbones, grouped by \textit{dominant graph motif} that governs NSO behavior and determines the requirements of \textit{graph-consistent} ChannelTransform inference and consumer rewriting.
\begin{packeditemize} 
\item \textit{Residual \textit{add} dominant.} These models enforce strict channel alignment at residual merges, requiring any
channel or layout transform to propagate consistently across all branches and downstream consumers.
    We consider \textit{ResNet-18}~\cite{he2016deep}, a minimal and widely adopted residual architecture that provides a clean baseline for testing layout synchronization under residual addition, and \textit{EfficientNet}~\cite{tan2019efficientnet}, which retains residual \texttt{add} while introducing MBConv-style heterogeneity (e.g., squeeze-and-excitation and compound scaling), thereby stressing ChannelTransform inference beyond simple residual blocks.
    
\item \textit{Concat dominant graphs.} These models rely heavily on channel concatenation across branches or over depth,
requiring recovery to preserve branch-structured channel partitions and maintain consistency under extensive feature reuse.
We evaluate \textit{InceptionV3}~\cite{szegedy2016rethinking}, which exercises multi-branch modules composed via \texttt{cat} and directly tests block-structured rewrites, and \textit{DenseNet}~\cite{huang2017densely}, which repeatedly applies \texttt{cat} along depth with heavy \texttt{fan-out}, stressing global consistency when a single producer feeds many downstream consumers.
\end{packeditemize}


\smallskip
\noindent\textbf{Watermarking schemes.}
We implement nine representative white-box watermarking schemes spanning three
mainstream design families. (i) \emph{weight-based} schemes embed ownership evidence directly into model parameters, such as selected weights, subspaces, or weight-derived structures, and verify ownership by extracting a message or checking parameter-level similarity. This category includes \textit{uchida17}~\cite{uchida2017embedding}, \textit{greedy}~\cite{liu2021watermarking}, \textit{lottery}~\cite{chen2021you}, \textit{ipr\_gan}~\cite{ong2021protecting}, and \textit{riga21}~\cite{wang2021riga}.
(ii) \emph{activation-based} schemes bind the watermark to internal representations rather than fixed weight indices, and verify ownership via activation- or feature-level signals measured on probe inputs. We evaluate two representative methods in this family, \textit{ipr\_ic}~\cite{lim2022protect} and \textit{deepsigns}~\cite{darvish2019deepsigns}, which are explicitly designed to be more robust to parameter reindexing and structural perturbations.
(iii) \emph{passport-based} schemes couple watermark verification to secret keys and dedicated passport mechanisms embedded in the model, such that correct key-dependent behavior is required for successful verification. We consider \textit{deepipr}~\cite{fan2021deepipr} and \textit{passport\_aware}~\cite{zhang2020passport}, which implement passport-aware normalization or explicit passport modules to enforce ownership control.

\heading{Attack settings.}
Assuming that an attacker has white-box access to the model parameters and can apply NSO-style, function-equivalent channel edits to conv layers. 
We evaluate three NSO primitives and two compositional variants:
\begin{packeditemize}
  \item \texttt{nso\_zero}: injects $d = \lceil \rho \cdot C \rceil$ dummy output channels with zero-initialized weights and biases into each eligible convolutional or linear layer, where $\rho\in[0,1]$ is the attack ratio and $C$ is the original channel count. 
  \item \texttt{nso\_clique}: injects $d$ dummy channels whose incoming weights are duplicated from a randomly sampled Gaussian filter $\mathbf{w}_{\text{base}} \sim \mathcal{N}(\mu, \sigma^2)$, where $\mu$ and $\sigma$ are computed from the original weight distribution, ensuring that the weighted sum of contributions from all clique members equals zero.
  \item \texttt{nso\_split}: replaces one channel by $k=d{+}1$ scaled duplicates, each weighted by coefficient $1/k$ such that their sum reconstructs the original channel's contribution.
  \item \texttt{mix-opseq}: applies a sampled global sequence of primitives sequentially within each eligible layer.In this mode, the injected channel count is recomputed at each step $t$ as $d^{(t)}=\lceil \rho\,C^{(t)}\rceil$ from the current width. 
  \item \texttt{mix-opseq (per-merge-group)}: samples sequences per residual merge group while allowing independence across channel concatenation branches, reflecting graph-consistency constraints and maximizing heterogeneity. It is an adaptive variant of \texttt{mix-opseq}.
\end{packeditemize}

The attacker injects a global fraction of dummy channels (default $\rho = 0.2$) into every eligible Conv/Linear output using a fixed per-layer placement rule that ensures shape compatibility across residual \texttt{add} and channel \texttt{cat}. We use interior placement rather than tail appends: after applying \texttt{nso\_zero}/\texttt{\_clique} (which create new channels) and \textit{\_split} (which selects a baseline channel at relative position $p \in [0,1]$, default $p=1.0$), we apply a merge-safe permutation so injected channels need not be tail-only. This may introduce tiny numerical drift from floating point summation order changes without breaking functionality.
We additionally apply random channel permutation and positive per-channel scaling (uniformly sampled in $[0.6,1.4]$ with batch-normalization consistent compensation) at a selected anchor location to stress test \textsc{Canon} under the strongest NSO setting.
A variant that disables permutation and scaling is reported in Appendix~\ref{sec:tab_nso_without_permute_scale}.

\heading{Evaluation metrics.}
We report four metrics to characterize NSO attacks and \textsc{Canon} recovery: 
(i) \textit{task utility}, measured as the accuracy gaps $\Delta_{\mathrm{acc}}(\mathrm{clean},\mathrm{attacked})$ and $\Delta_{\mathrm{acc}}(\mathrm{clean},\mathrm{recovered})$; (ii) \textit{watermark evidence}, measured by each scheme’s native similarity score $\mathrm{Sim}$ under our two-tier protocol, where Tier-1 reports $\mathrm{Sim}$ on the recovered model when the reference-equivalence certificate passes, and Tier-2 provides a conservative, attack-aware fallback score when certification is unavailable or fails; (iii) \textit{recovery cost}, measured as end-to-end wall-clock runtime of \textsc{Canon}; and (iv) \textit{false positives on clean models}, quantified by unnecessary structural edits when running \textsc{Canon} on non-attacked watermarked models (fraction of parameters removed and fraction of Conv/Linear layers whose output shapes change). Ideally, \textsc{Canon} preserves utility and watermark evidence while performing no structural edits on clean models, i.e., $\Delta_{\mathrm{acc}}(\mathrm{clean},\mathrm{recovered})=0$, $\mathrm{Sim}_{\mathrm{clean}}=\mathrm{Sim}_{\mathrm{recovered}}$, and both quantities in (iv) are zero.

\subsection{Experimental Results}
\label{sec:exp_results}

We examine whether \textsc{Canon} can restore white-box watermark verifiability while preserving task utility (\S\ref{subsubsec-utility}). We then analyze the cost of recovery and scalability under
different attack strengths and probing configurations (\S\ref{subsubsec-cost}). We assess the conservativeness of \textsc{Canon} by measuring false positives when it is applied to non-attacked models (\S\ref{subsubsec-clean}).

\subsubsection{Recovery Utility and Verification}\label{subsubsec-utility}

We evaluate \textsc{Canon}'s \textit{end-to-end effectiveness}: whether it can neutralize NSO's structure/layout obfuscations while preserving model utility and restoring white-box verifiability across watermark families and graph motifs.

We summarize the full attack-and-recovery suite at attack ratio $\rho=0.2$ in two tables: Table~\ref{tab:nso_recovery_suite_add_02} (add-dominant: ResNet-18 and EfficientNet) and Table~\ref{tab:nso_recovery_suite_cat_02} (cat-dominant: InceptionV3 and DenseNet).
Each row corresponds to a watermarking scheme; the left block reports clean similarity and main task accuracy gap, and the right block reports watermark similarity restoration (attack-time similarity $\rightarrow$ recovered similarity) together with the outcome of our two-tier verification.

Experimental results support the central claim of this work: NSO-style channel/layout obfuscations can preserve the model function while breaking structure-indexed white-box verification, and \textsc{Canon} can \textit{fully restore} a compact, graph-consistent parameterization that re-enables verification.
Across all four model architectures and nine watermarking schemes, we observe \textit{zero accuracy drift} under attack and after recovery (i.e., $\Delta_{\text{acc}}(\text{clean},\text{attacked}) = 0$ and $\Delta_{\text{acc}}(\text{clean},\text{recovered}) = 0$ in the reported suites).
This confirms that the attacks operate in the intended regime of \textit{structural obfuscation without utility loss}, and that any degradation in watermark similarity is isolated to the \textit{verification interface} (index/layout dependence), rather than to changes in learnt functionality.
The fact that \textsc{Canon} succeeds across both motifs in these two tables indicates that it is enforcing the global constraints induced by \texttt{fan-out}, residual \texttt{add}, and channel \texttt{cat}, highlighting why defending against NSO requires \textit{graph-consistent} recovery rather than layer-local heuristics.

Importantly, our reported recovery uses the \textit{two-tier verifier} described earlier.
Across this experimental suite, recovery succeeds at Tier-1 (reference-equivalence certification) for \textit{all} model--watermark pairs, demonstrating that the recovered watermarked weights match the pre-attack reference up to allowed channel symmetries (permutation and BN-consistent positive scaling).
This certificate provides scheme-agnostic evidence that recovery reconstructs a \textit{compact, reference-equivalence} parameterization on the attacked regions, rather than merely increasing a heuristic similarity score. Thus, watermark verifiability is \textit{fully recovered} across diverse watermark families while maintaining functional equivalence.

\begin{figure*}[!t]
    \centering
    \begin{minipage}[c]{1\textwidth}
      \centering
      \includegraphics[width=7in]{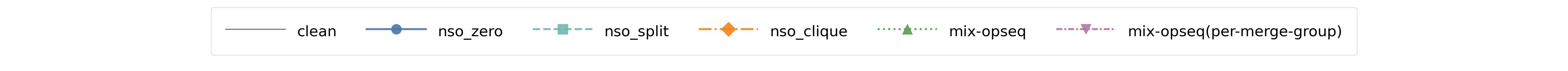}
    \end{minipage}
    \vspace{-15pt}
    \\
    \subfigure[\shortstack{\small probe batch $=32$,\\[-1pt]\small attack ratio $=0.2$}]{
    \begin{minipage}[t]{0.235\textwidth}
    \centering
    \includegraphics[width=1.8in]{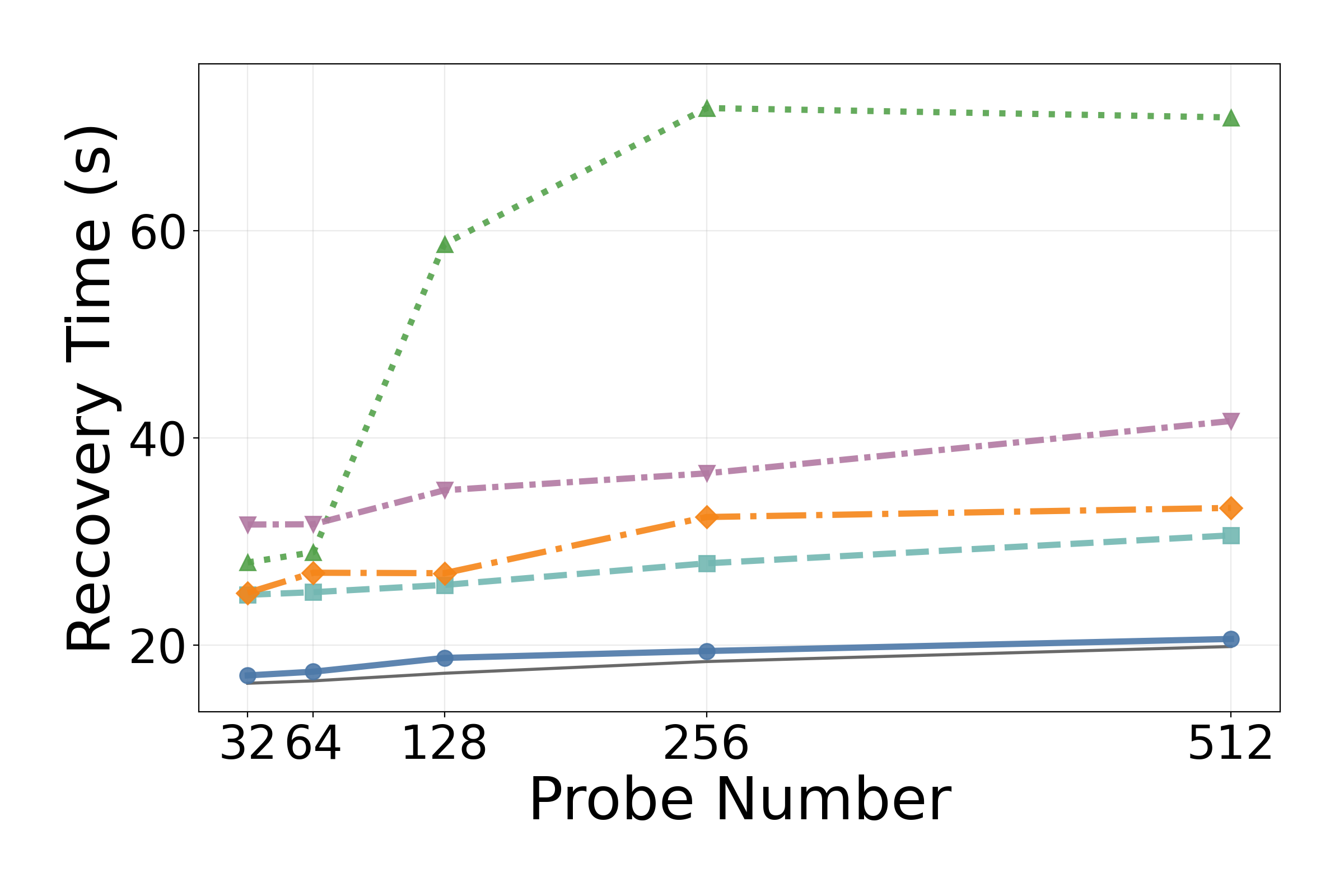}
    \end{minipage}
    \label{recover_time_batch32_ratio02}
    }
    \subfigure[\shortstack{\small probe batch $=32$,\\[-1pt]\small attack ratio $=0.5$}]{
    \begin{minipage}[t]{0.235\textwidth}
    \centering
    \includegraphics[width=1.8in]{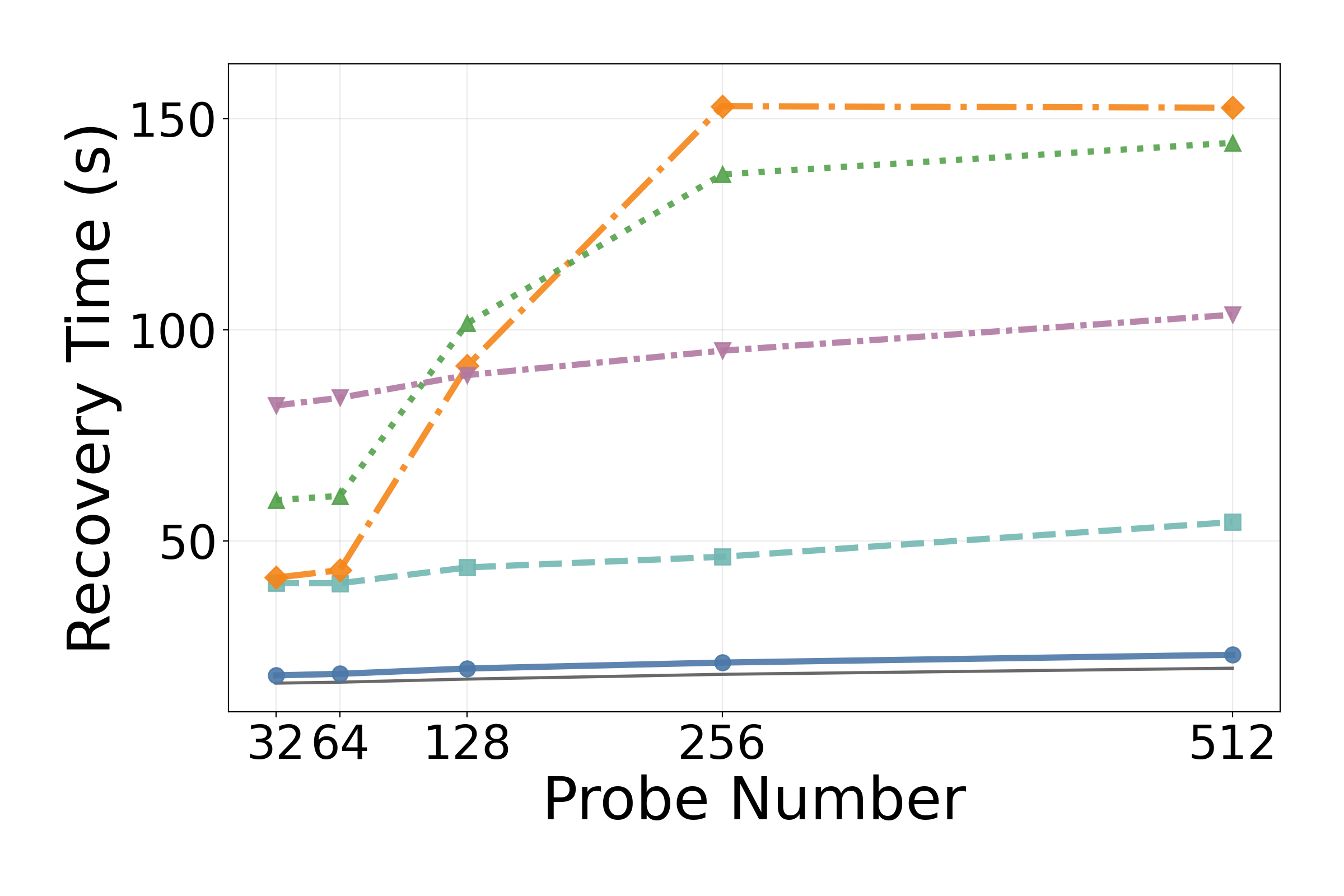}
    \end{minipage}
    \label{recover_time_batch32_ratio05}
    }
    \subfigure[\shortstack{\small probe batch $=64$,\\[-1pt]\small attack ratio $=0.2$}]{
    \begin{minipage}[t]{0.235\textwidth}
    \centering
    \includegraphics[width=1.8in]{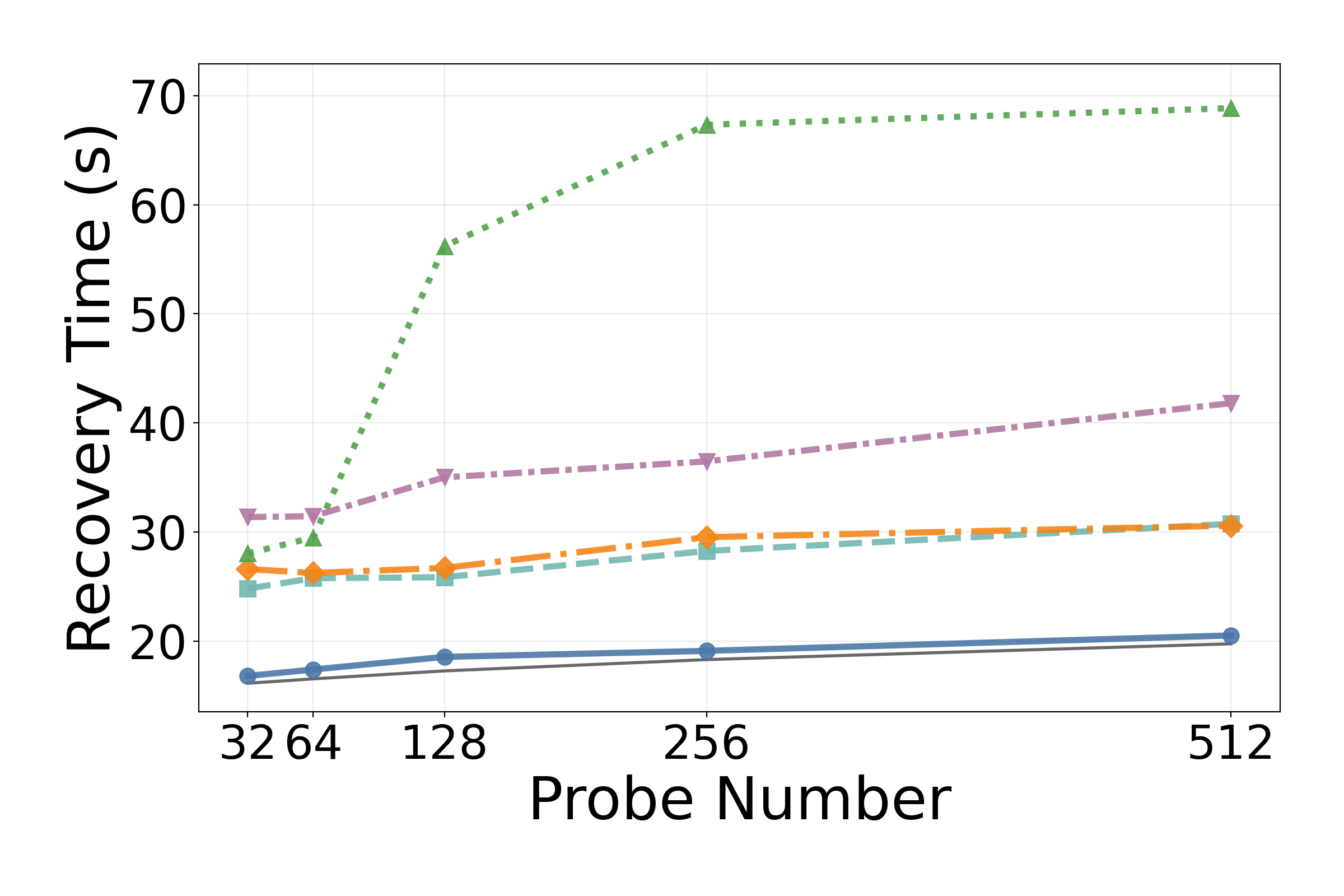}
    \end{minipage}
    \label{recover_time_batch64_ratio02}
    }
    \subfigure[\shortstack{\small probe batch $=64$,\\[-1pt]\small attack ratio $=0.5$}]{
    \begin{minipage}[t]{0.235\textwidth}
    \centering
    \includegraphics[width=1.8in]{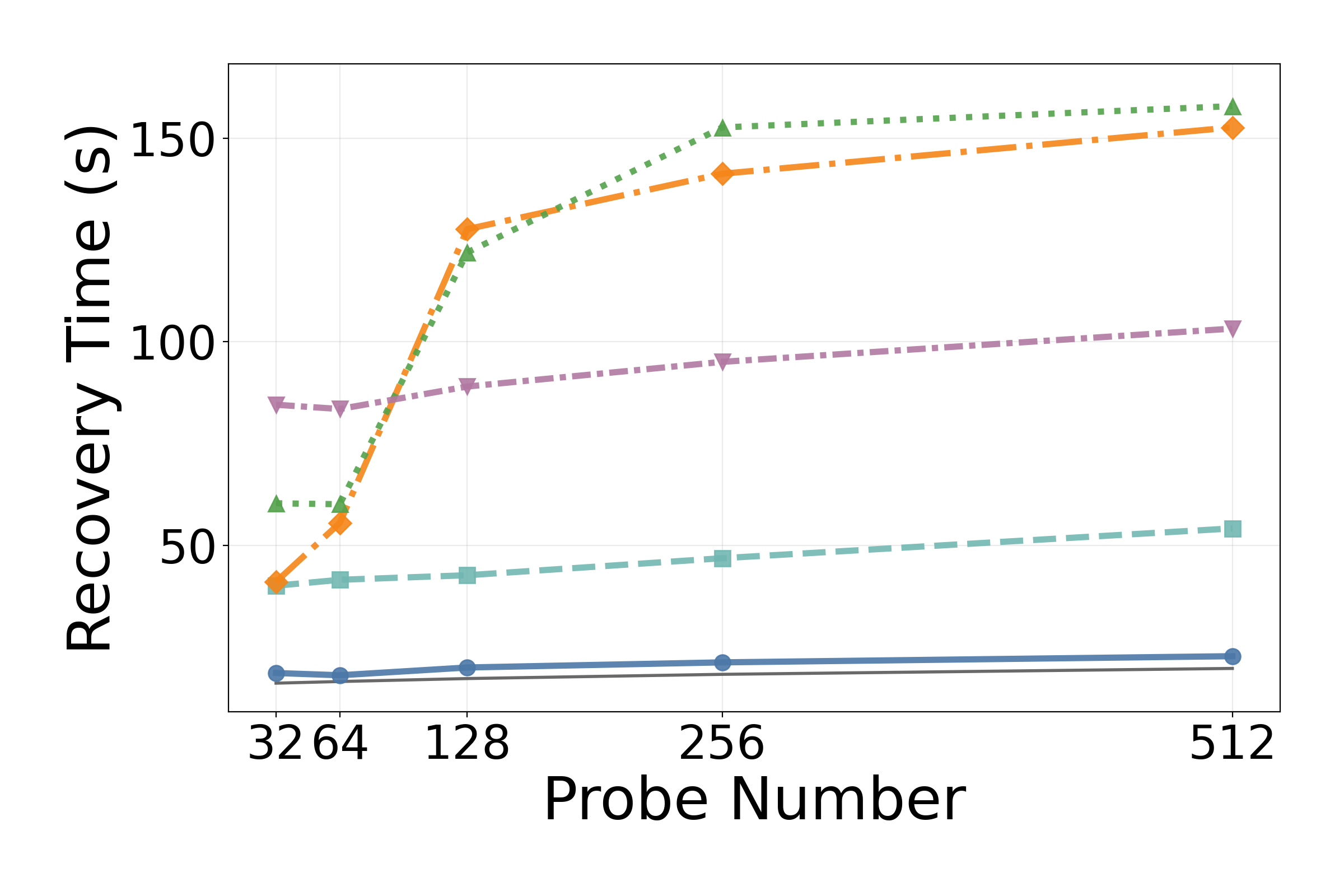}
    \end{minipage}
    \label{recover_time_batch64_ratio05}
    }
    \caption{\textsc{Canon} recovery time on ResNet-18 versus the number of probe inputs under different probe batch sizes and attack ratios.
    Each plot includes the clean baseline and the attacked cases (\texttt{nso\_zero}, \texttt{nso\_split}, \texttt{nso\_clique}, \texttt{mix-opseq}, \texttt{mix-opseq (per-merge-group)}).
    Increasing the probe number generally increases runtime due to additional activation collection and clustering, while higher injection ratios amplify the cost of redundancy detection and graph-consistent consumer rewrites. The legend is shown only in the first subgraph, as it applies to all subgraphs.}
    \label{fig:runtime_resnet18_probeT}
\end{figure*}

\begin{figure*}[!t]
    \centering
    \begin{minipage}[c]{1\textwidth}
      \centering
      \includegraphics[width=7in]{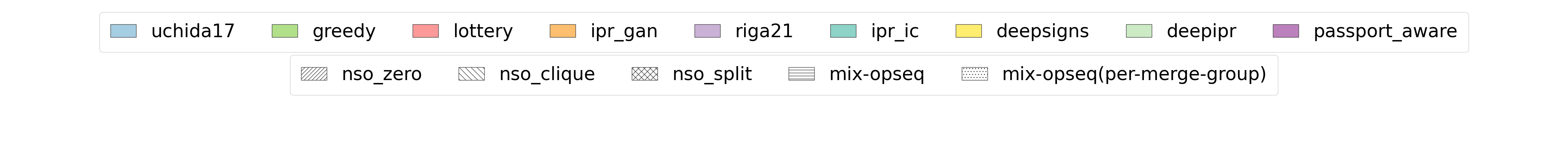}
    \end{minipage}
    \vspace{-20pt}
    \\
    \subfigure[\shortstack{\small ResNet-18,\\[-1pt]\small attack ratio $=0.2$}]{
    \begin{minipage}[t]{0.235\textwidth}
    \centering
    \includegraphics[width=1.74in]{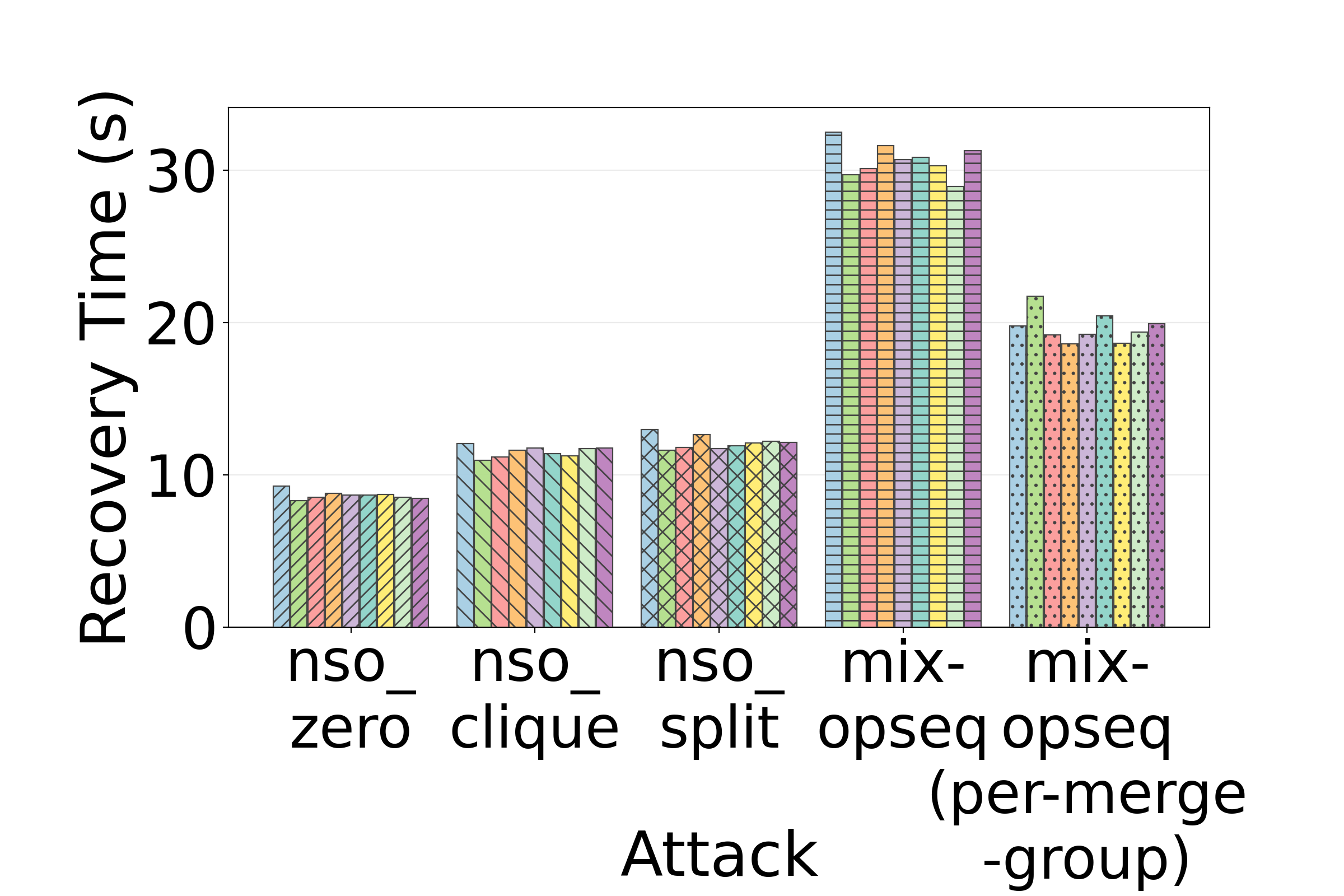}
    \end{minipage}
    \label{attack_recover_time_resnet_ratio02}
    }
    \subfigure[\shortstack{\small EfficientNet,\\[-1pt]\small attack ratio $=0.2$}]{
    \begin{minipage}[t]{0.235\textwidth}
    \centering
    \includegraphics[width=1.74in]{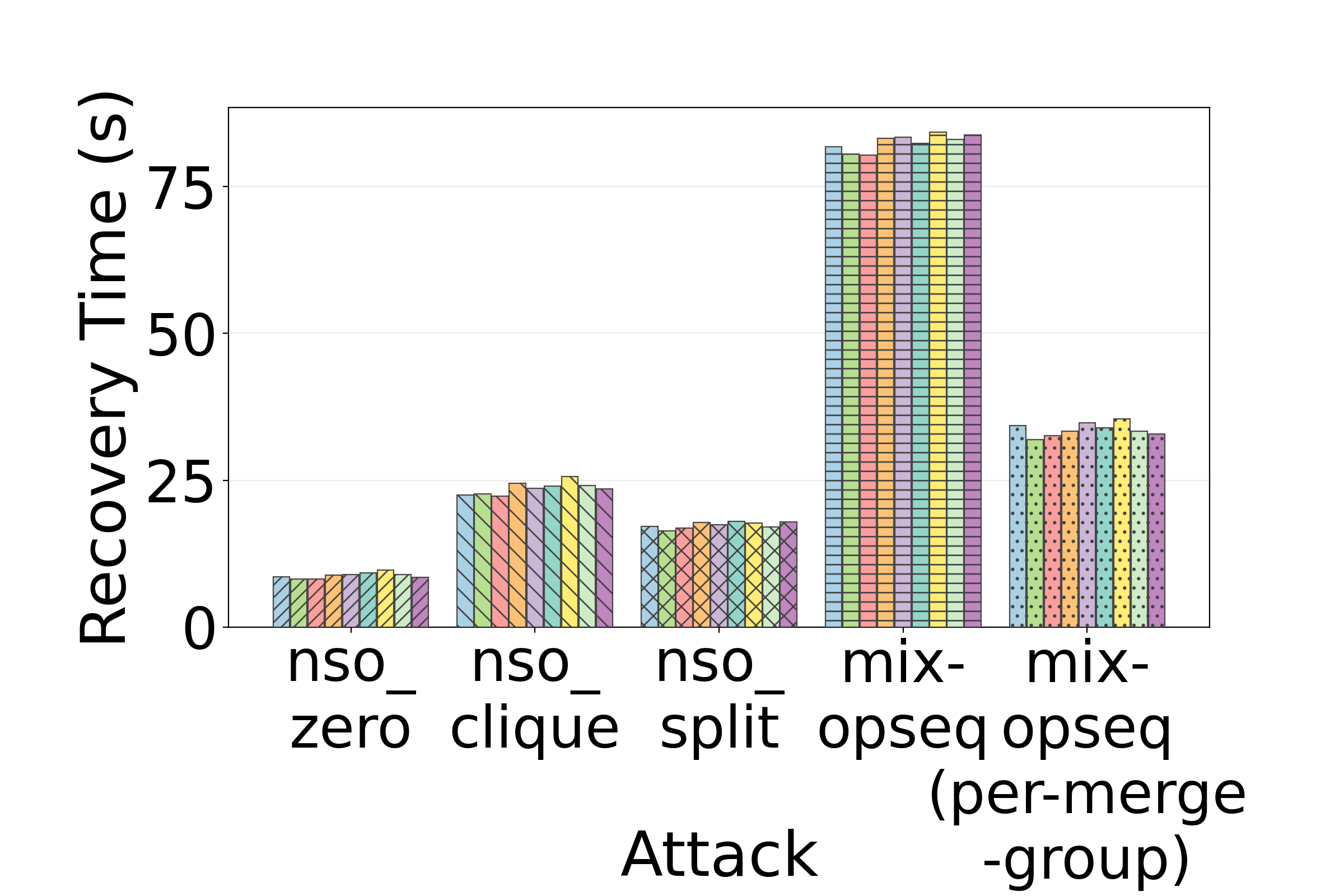}
    \end{minipage}
    \label{attack_recover_time_efficientnet_ratio02}
    }
    \subfigure[\shortstack{\small InceptionV3,\\[-1pt]\small attack ratio $=0.2$}]{
    \begin{minipage}[t]{0.235\textwidth}
    \centering
    \includegraphics[width=1.74in]{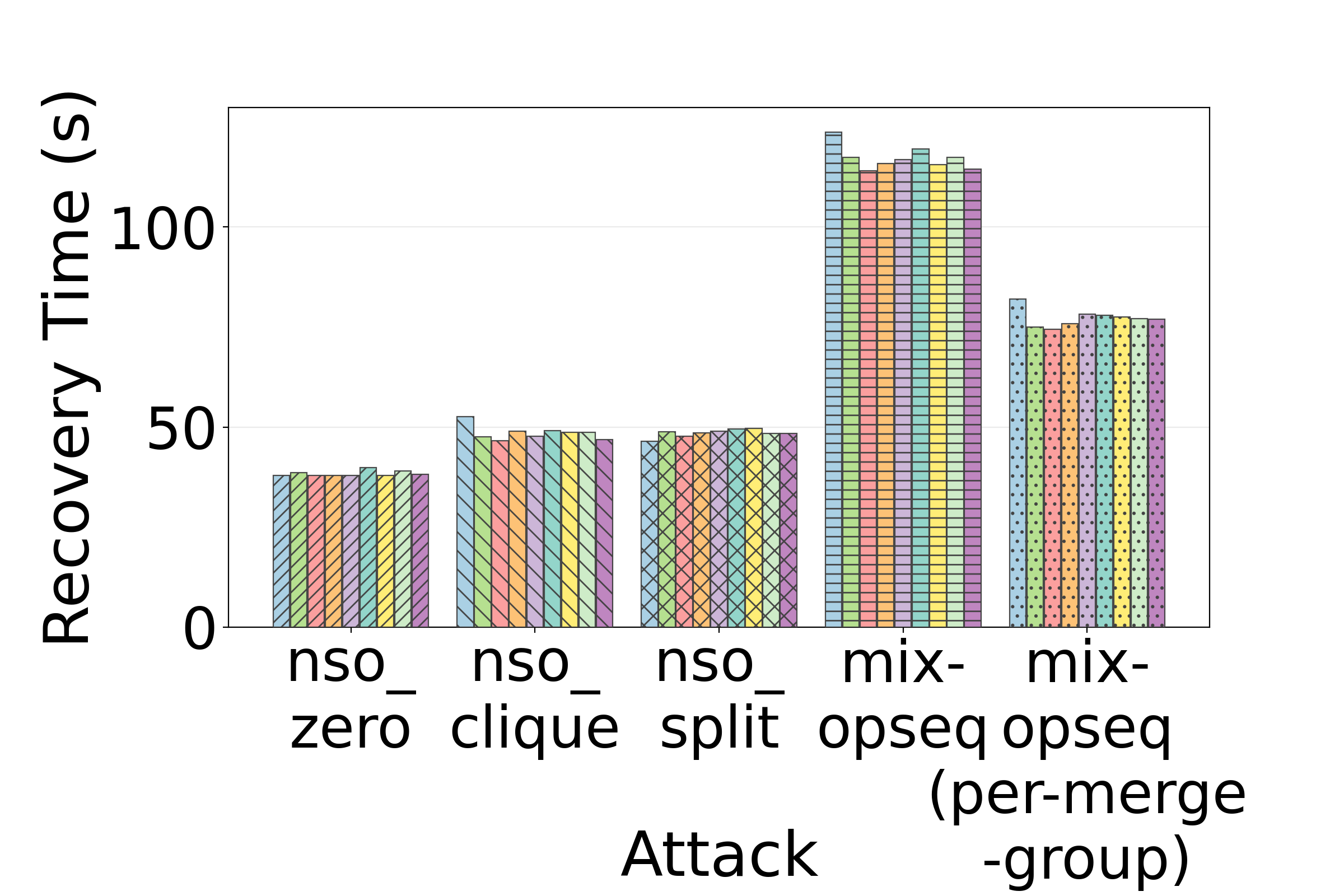}
    \end{minipage}
    \label{attack_recover_time_inceptionv3_ratio02}
    }
    \subfigure[\shortstack{\small DenseNet,\\[-1pt]\small attack ratio $=0.2$}]{
    \begin{minipage}[t]{0.235\textwidth}
    \centering
    \includegraphics[width=1.74in]{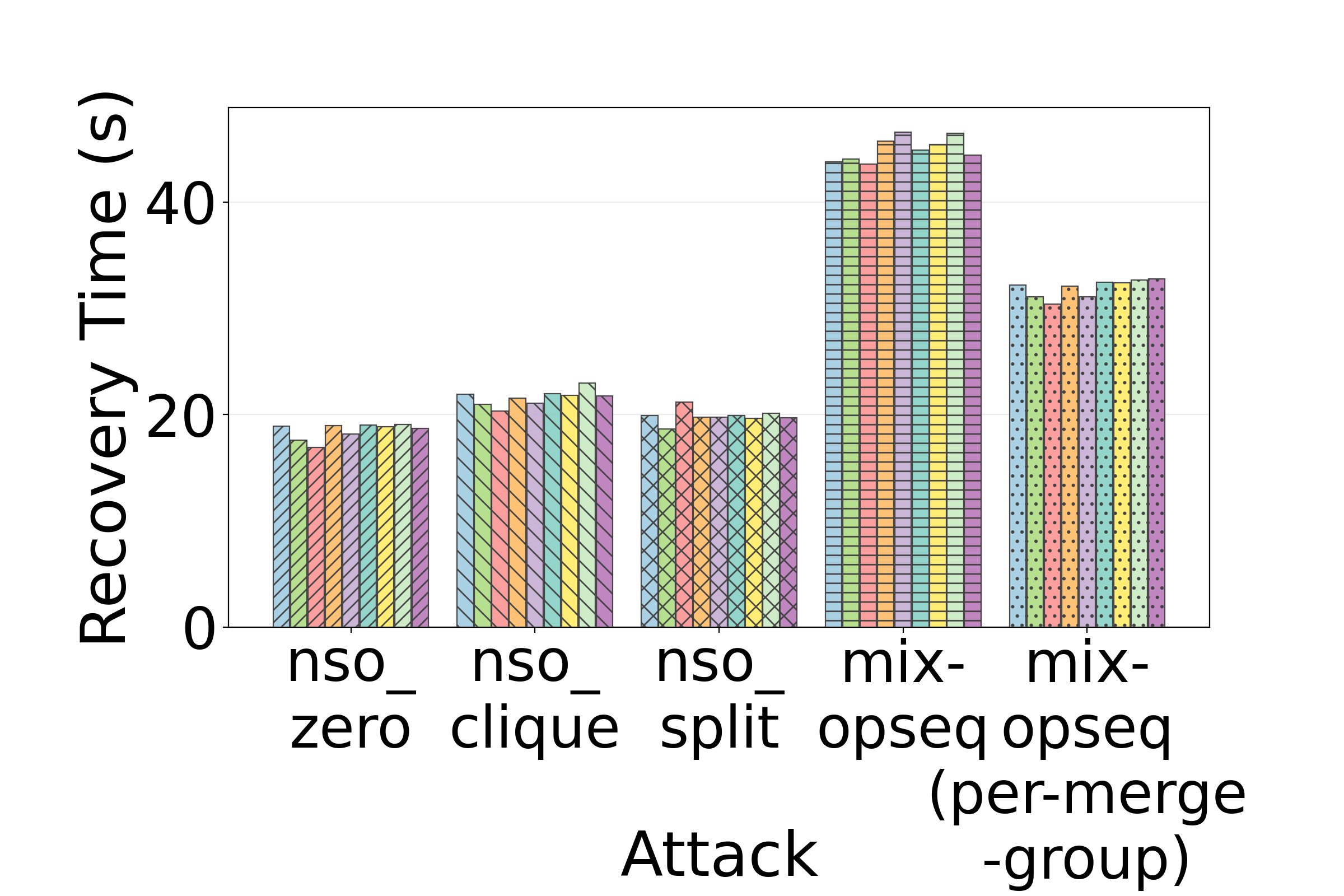}
    \end{minipage}
    \label{attack_recover_time_densenet_ratio02}
    }
    \\
    \subfigure[\shortstack{\small ResNet-18,\\[-1pt]\small attack ratio $=0.5$}]{
    \begin{minipage}[t]{0.235\textwidth}
    \centering
    \includegraphics[width=1.74in]{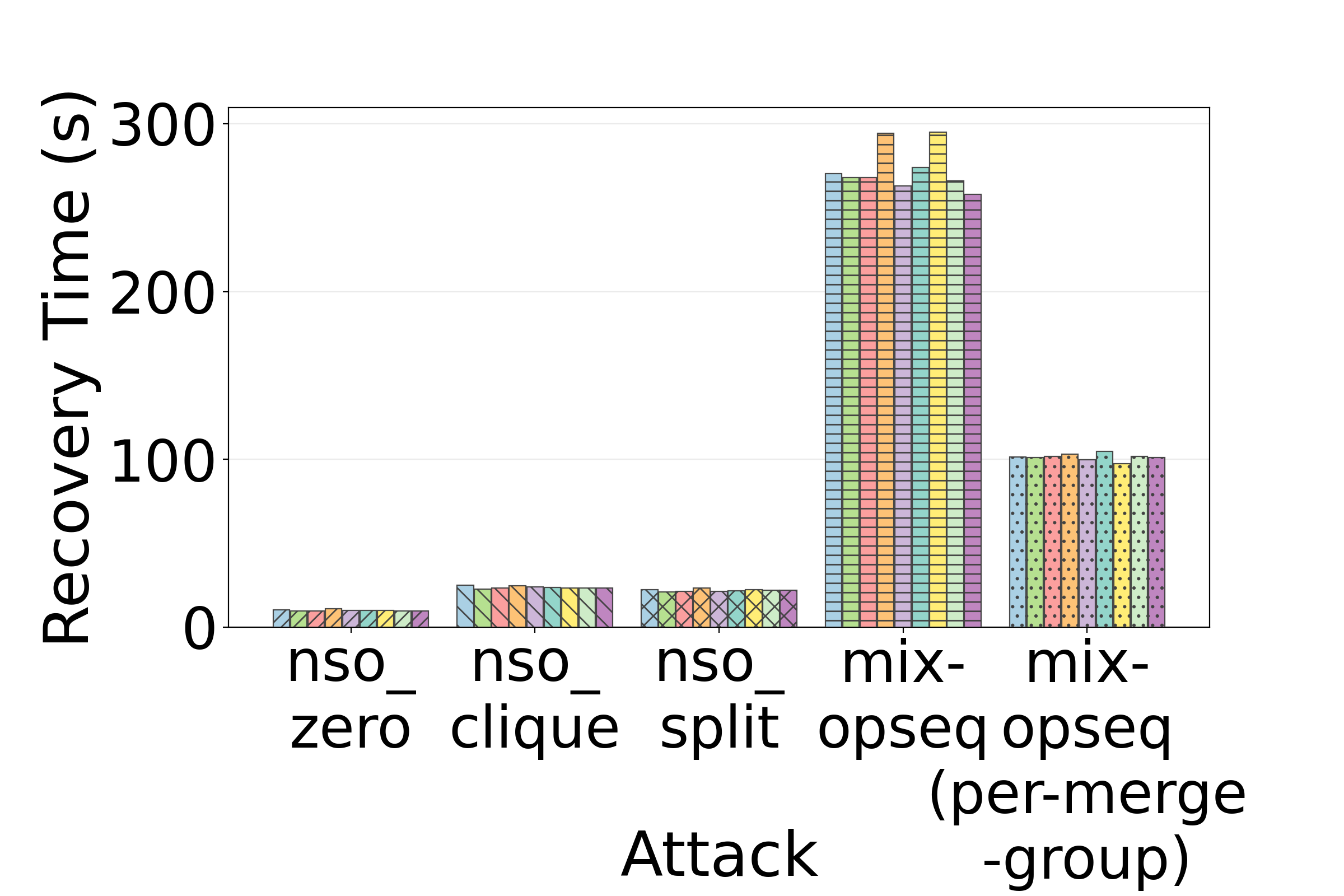}
    \end{minipage}
    \label{attack_recover_time_resnet_ratio05}
    }
    \subfigure[\shortstack{\small EfficientNet,\\[-1pt]\small attack ratio $=0.5$}]{
    \begin{minipage}[t]{0.235\textwidth}
    \centering
    \includegraphics[width=1.74in]{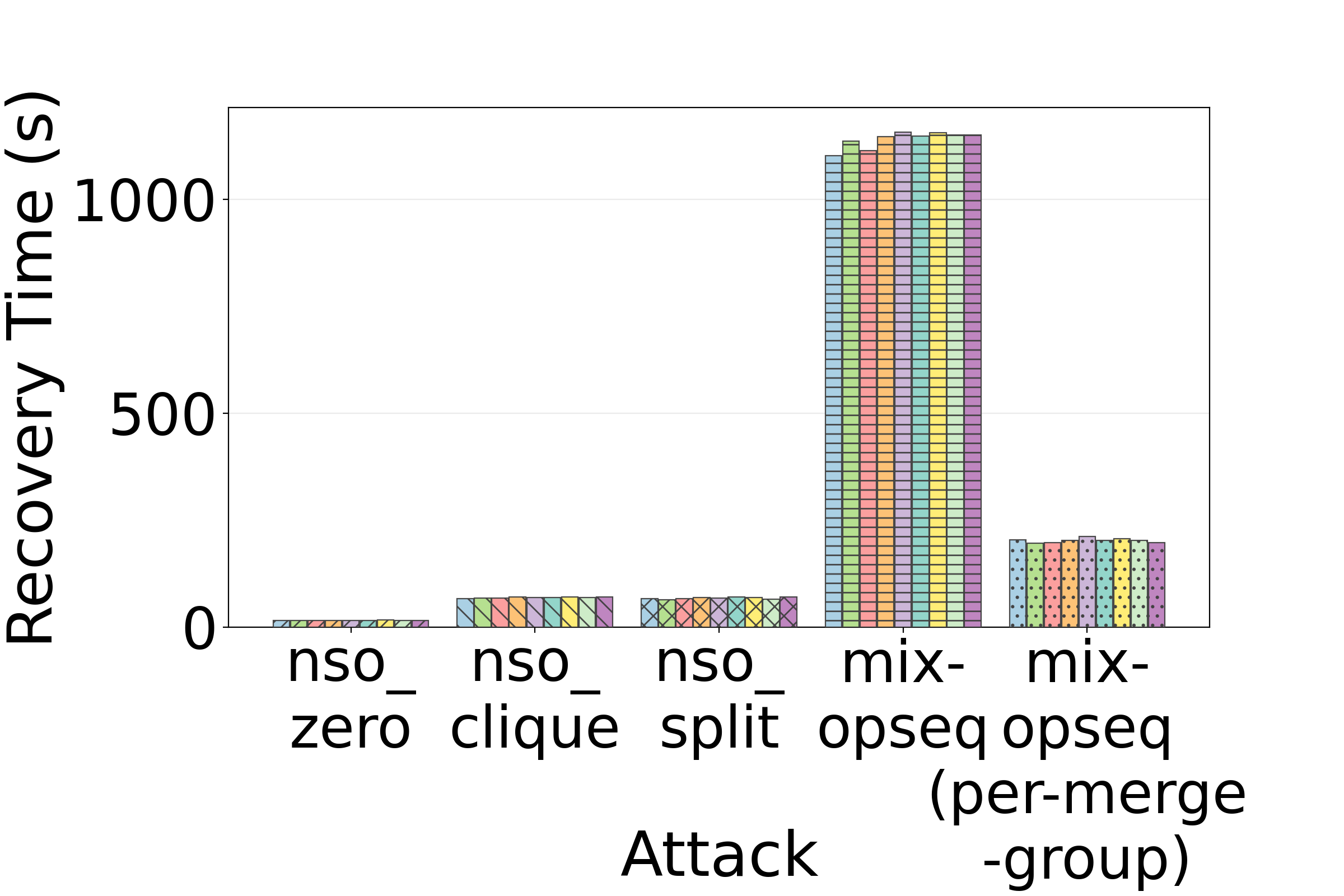}
    \end{minipage}
    \label{attack_recover_time_efficientnet_ratio05}
    }
    \subfigure[\shortstack{\small InceptionV3,\\[-1pt]\small attack ratio $=0.5$}]{
    \begin{minipage}[t]{0.235\textwidth}
    \centering
    \includegraphics[width=1.74in]{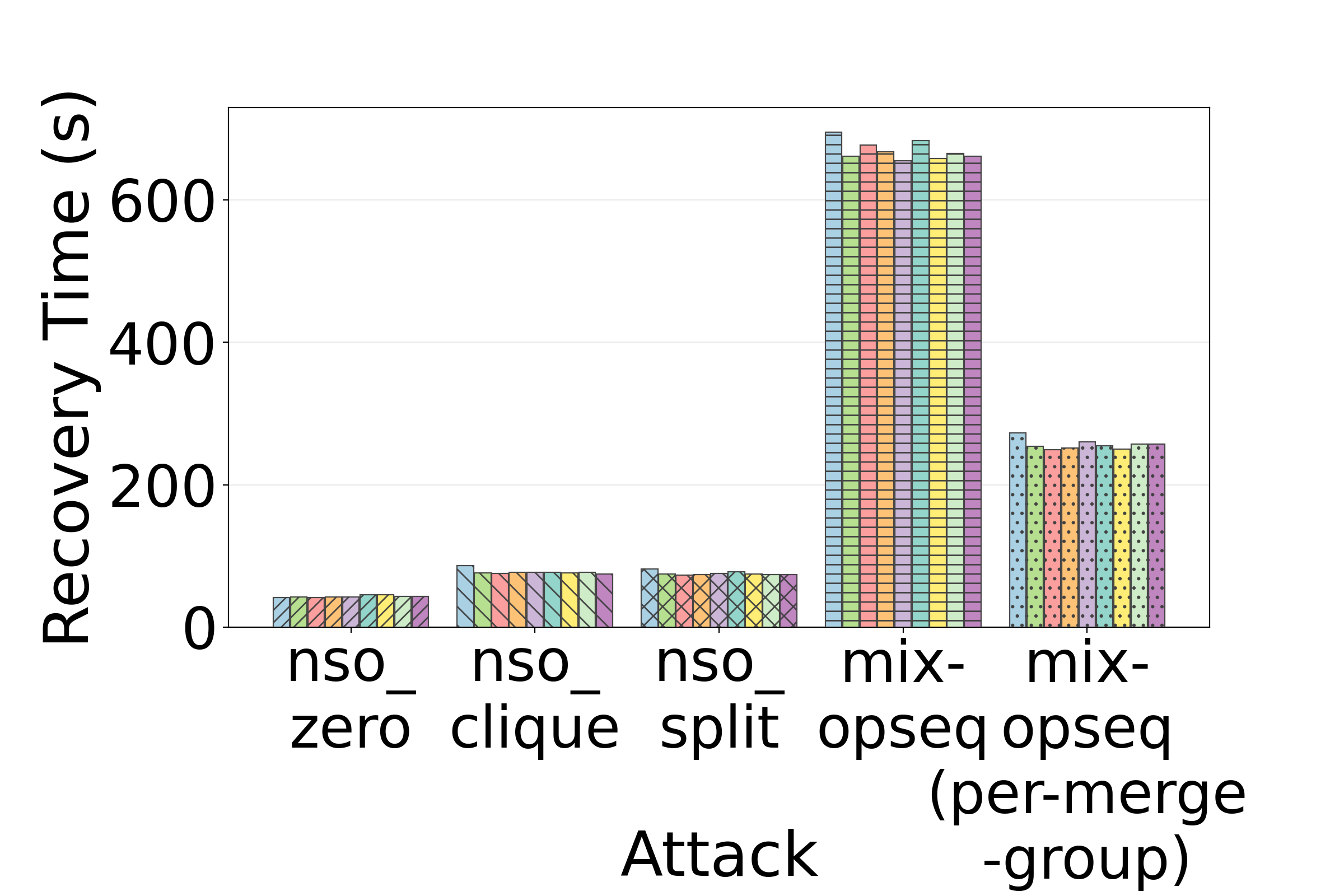}
    \end{minipage}
    \label{attack_recover_time_inceptionv3_ratio05}
    }
    \subfigure[\shortstack{\small DenseNet,\\[-1pt]\small attack ratio $=0.5$}]{
    \begin{minipage}[t]{0.235\textwidth}
    \centering
    \includegraphics[width=1.74in]{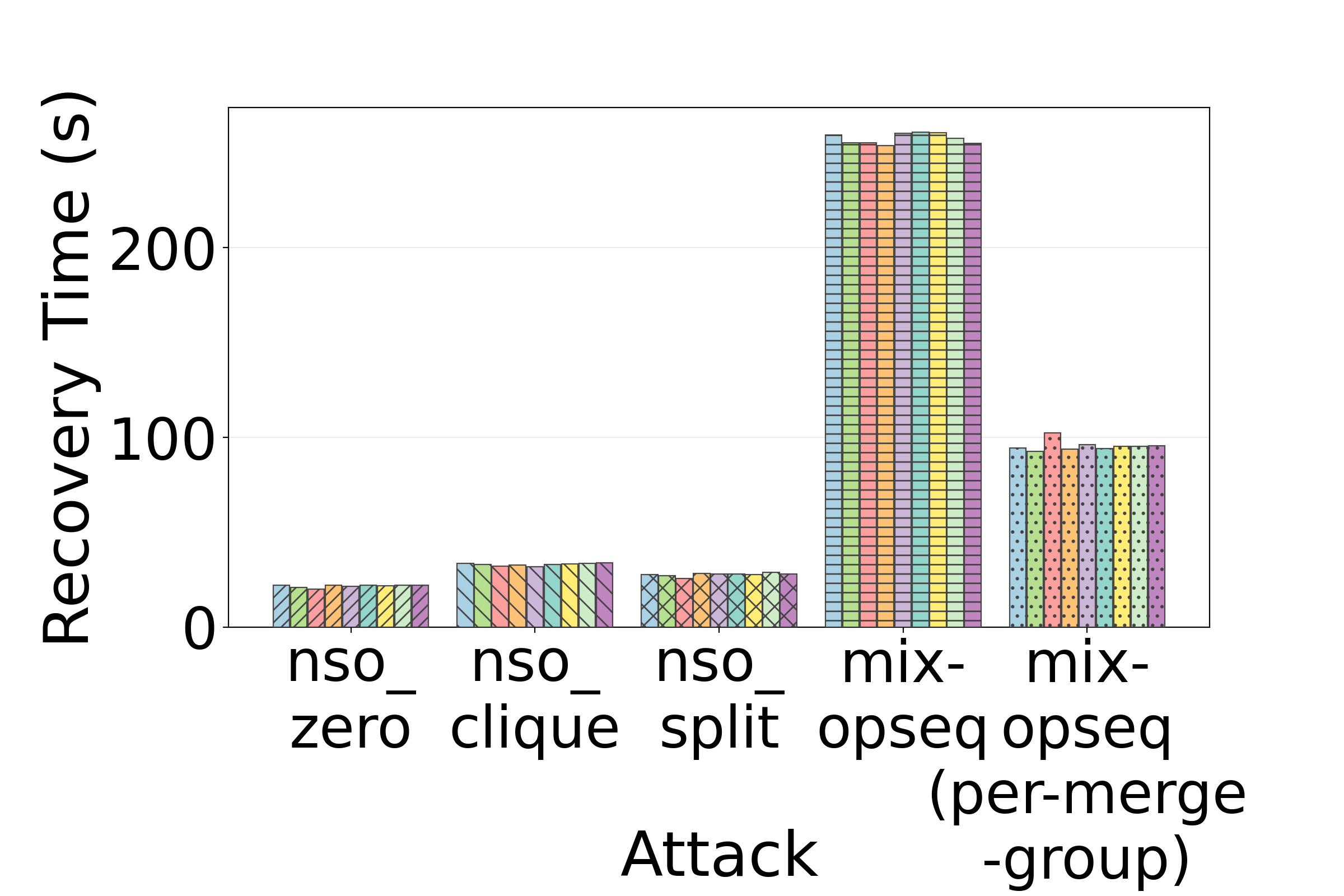}
    \end{minipage}
    \label{attack_recover_time_densenet_ratio05}
    }
    \caption{\textsc{Canon} recovery time under five NSO attack variants for ResNet-18 and DenseNet across multiple white-box watermarking schemes. Overall, recovery cost is attack-dependent and consistently highest for the composed mix-opseq setting, with a pronounced increase at $\rho=0.5$, while simpler attacks (\texttt{nso\_zero} / \texttt{nso\_split} / \texttt{nso\_clique}) remain comparatively inexpensive and stable across watermark methods.}
    \label{fig:runtime_watermark}
\end{figure*}

For completeness, we also report the Tier-2 similarity scores produced by our two-tier verifier as a \textit{fallback reference}. Although this fallback is not exercised in our reported runs as \textit{all configurations pass at Tier-1}, it provides a conservative view of what verification would look like \textit{in the worst case} where certification fails.
In this hypothetical setting, the recovered similarity typically improves over the attacked similarity across schemes, indicating that canonicalization still makes the extractor’s coordinates substantially more compatible.
Notably, for \textit{riga21}, the similarity can rise dramatically, e.g., from $45.31\%$ to $99.22\%$, suggesting that even when exact reference-equivalence cannot be established, \textsc{Canon} can still restore a highly usable level of ownership evidence. We emphasize that these Tier-2 numbers are provided only as contextual evidence of robustness; the main conclusion remains that Tier-1 certification succeeds throughout.

\smallskip
\noindent\textbf{Results at larger attack ratio.}
We defer the corresponding tables for attack ratio $\rho=0.5$ to Appendix~\ref{sec:tab_nso_0.5}, to reduce redundancy while showing that the above conclusions remain stable under heavier structural obfuscation.

\begin{center}
\fbox{%
\begin{minipage}{0.9\linewidth}
\textbf{Takeaway (full recovery under function-equivalent NSO).} NSO preserves accuracy yet breaks structure-indexed verification, whereas \textsc{Canon} fully restores a compact, graph-consistent layout and achieves complete verification recovery with Tier-1 reference-equivalence certification passing throughout.
\end{minipage}
}
\end{center}

\begin{table}[t]
\centering
\caption{\textbf{False positive evaluation.} ResNet-18 recovery on clean (non-attacked) watermarked models. All test cases show zero structural false positives (no parameter pruning, no layer shape changes), perfect accuracy preservation ($\Delta_{\mathrm{acc}} = 0$), and 100\%  Tier-1 reference-equivalence certificate pass rate. Watermark similarity gaps are minimal (within noise tolerance), with one case showing slight improvement.}

\label{tab:fp_resnet}
\vspace{5pt}
\begin{threeparttable}

\resizebox{\linewidth}{!}{
\begin{tabular}{c|cc|cc|l}
\toprule
\multirow{2}{*}{\textbf{Watermark}} & \multirow{2}{*}{\textbf{P-FPR}} & \multirow{2}{*}{\textbf{L-FP}} &  \multirow{2}{*}{\textbf{\textbf{$\Delta_{\mathrm{sim}}$}}} &  \multirow{2}{*}{\textbf{$\Delta_{\mathrm{acc}}$}} & \multicolumn{1}{c}{\textbf{ Tier-1}} \\

& &  &  &  & \multicolumn{1}{c}{\textbf{Pass}} \\

\midrule
uchida17    & \cellcolor{blue!10} $0.0000$ & \cellcolor{blue!10} $0/21$ & \cellcolor{pink!25} $0.0000$ & \cellcolor{pink!25} $0.0000$  & \cmark\,   ($2/2$) \\
greedy       & \cellcolor{blue!10} $0.0000$ & \cellcolor{blue!10} $0/21$ & \cellcolor{pink!25} $0.0000$ & \cellcolor{pink!25} $0.0000$  & \cmark\,   ($4/4$) \\
lottery  & \cellcolor{blue!10} $0.0000$ & \cellcolor{blue!10} $0/21$ & \cellcolor{pink!25} $0.0000$ & \cellcolor{pink!25} $0.0000$  & \cmark\,   ($18/18$) \\
ipr\_gan      & \cellcolor{blue!10} $0.0000$ & \cellcolor{blue!10} $0/21$ & \cellcolor{pink!25} $0.0000$ & \cellcolor{pink!25} $0.0000$  & \cmark\,   ($2/2$) \\
riga21 & \cellcolor{blue!10} $0.0000$ & \cellcolor{blue!10} $0/21$ & \cellcolor{pink!25} $0.0000$ & \cellcolor{pink!25} $0.0078$ & \cmark\,   ($2/2$) \\
ipr\_ic & \cellcolor{blue!10} $0.0000$ & \cellcolor{blue!10} $0/21$ & \cellcolor{pink!25} $0.0000$ & \cellcolor{pink!25} $0.0000$  & \cmark\,   ($1/1$) \\
deepsigns   & \cellcolor{blue!10} $0.0000$ & \cellcolor{blue!10} $0/21$ & \cellcolor{pink!25} $0.0000$ & \cellcolor{pink!25} $0.0000$  & \cmark\,   ($2/2$) \\
deepipr & \cellcolor{blue!10} $0.0000$ & \cellcolor{blue!10} $0/21$ & \cellcolor{pink!25} $0.0000$ & \cellcolor{pink!25} $0.0000$  & \cmark\,   ($8/8$) \\
passport\_aware & \cellcolor{blue!10} $0.0000$ & \cellcolor{blue!10} $0/21$ & \cellcolor{pink!25} $0.0000$ &  \cellcolor{pink!25} $0.0000$  & \cmark\,   ($21/21$) \\
\bottomrule
\end{tabular}
}
    \begin{tablenotes}
    \scriptsize
    \item \textbf{P-FPR}: parameter false positive rate;
    \item \textbf{L-FP}: layer false positive, listed in \textit{changed / total};
    \item \textbf{$\Delta_{\mathrm{acc}}$} for accuracy gap while \textbf{$\Delta_{\mathrm{sim}}$} for watermark similarity gap.
    \end{tablenotes}
\end{threeparttable}
\end{table}

\subsubsection{Recovery Runtime and Scalability}\label{subsubsec-cost}

We characterize the \textit{practical cost} of recovery and how it scales with attack complexity and probing configuration.

We first present Fig.\ref{fig:runtime_watermark}, which aggregates \textsc{Canon} recovery time across all nine watermarking schemes for four architectures and two injection ratios under the five NSO variants.
We then zoom in on ResNet-18 in Fig.\ref{fig:runtime_resnet18_probeT} to isolate how runtime scales with the number of probe inputs and probe batch size, and to attribute costs to the activation collection and clustering procedures that underpin ChannelTransform inference.

Fig.\ref{fig:runtime_watermark} shows that \textit{runtime is primarily attack-dependent}, while variation across watermarking schemes is comparatively minor.
This is expected because \textsc{Canon} operates on the attacked computation graph and channel signatures, and its dominant costs arise from redundancy detection/clustering and graph-consistent consumer rewrites, both of which are determined mainly by the \textit{structural complexity} introduced by NSO, rather than by the downstream watermark extractor.
Across backbones, single primitive attacks (e.g., \texttt{nso\_zero}, \texttt{nso\_split}) remain inexpensive and stable, whereas the compositional setting \texttt{mix-opseq} is consistently the most costly.
This gap is more pronounced at higher injection ratios, reflecting the intended hardness of compositional NSO: sequentially applying multiple primitives entangles redundancy patterns and enlarges the scope of downstream rewrites required to maintain global consistency.
Fig.\ref{fig:runtime_watermark} highlights a key practical point that the recovery cost tracks the attacker’s structural aggressiveness, making the most damaging NSO variants also the most expensive to apply and to undo.

Motivated by the weak dependence on watermark type and the consistent trends across architectures in Fig.\ref{fig:runtime_watermark}, we next focus on ResNet-18 to study probe-driven scaling in greater detail.
Fig.\ref{fig:runtime_resnet18_probeT} confirms that \textit{runtime increases with the probe number} because more probes imply more activation collection and more stable clustering statistics, and that it also increases with the injection ratio because heavier obfuscation amplifies redundancy detection and consumer rewrite cost.
Notably, the \texttt{nso\_clique} family exhibits a more pronounced sensitivity at $\rho=0.5$.
A plausible explanation is that \texttt{nso\_clique} obfuscation induces larger and denser equivalence structures at higher injection ratios, which increases clustering complexity and expands the set of consumers that must be rewritten consistently to preserve cancellation constraints.
In contrast, \texttt{nso\_zero} and \texttt{nso\_split} exhibit smoother scaling because their redundancy signatures are typically easier to separate and their downstream rewrites are less entangled.

These results suggest that \textsc{Canon} offers a \textit{tunable cost} profile: probe budgets can be adjusted to trade runtime for robustness margin, while the dominant driver remains the underlying NSO complexity.
Crucially, even under strong graph-consistent NSO stressors, \textsc{Canon} typically runs at the \textit{10-second} scale and in most cases stays well below \textit{one minute}.
Given that NSO can \textit{silently} invalidate white-box watermark verification without harming model utility, this one-time overhead during verification is \textit{practically acceptable}, because recovery is invoked only when a model is challenged for ownership, and it is usually performed only once per disputed artifact.
In this sense, \textsc{Canon} introduces a modest, controllable verification cost in exchange for restoring verifiability against a highly destructive class of structure obfuscations.

\begin{center}
\fbox{%
\begin{minipage}{0.9\linewidth}
\textbf{Takeaway (acceptable one-time verification cost).} \textsc{Canon} runtime is dominated by NSO complexity: single primitive attacks are inexpensive, while compositional \texttt{mix-opseq} is the most costly. This overhead is incurred only at verification time once and remains practically acceptable, while probe budgets provide a direct knob to trade runtime for robustness margin.
\end{minipage}
}
\end{center}

\subsubsection{False Positives on Clean Models}\label{subsubsec-clean}

We evaluate \textsc{Canon}'s \textit{conservativeness} when no NSO is present, i.e., whether the recovery pipeline introduces functional artifacts on clean watermarked models.
Table~\ref{tab:fp_resnet} measures structural changes (parameter pruning and layer shape changes), utility drift, and watermark similarity drift when \textsc{Canon} is applied directly to clean (non-attacked) models.
Therefore, this table tests a ``do no harm'' requirement for our defenses so that meaningful rewrites should be triggered only when structural obfuscation is detected.

Table~\ref{tab:fp_resnet} confirms \textit{zero} false positives.
All cases show zero structural changes (no parameter pruning or Conv/Linear output shape changes) and perfect accuracy preservation (i.e., $\Delta_{\text{acc}}(\text{clean},\text{recovered}) = 0$).
Watermark similarity gaps on clean models are also zero.

We observe one minor special case for \textit{riga21}, where recovery yields a slightly higher similarity than the clean baseline (i.e., one additional bit matches for a 128-bit message).
This behavior is consistent with benign numerical variability. Some extractors depend on thresholding or floating-point-sensitive computations, and equivalence preserving rewrites can shift a borderline decision across a threshold without indicating any functional damage. \textsc{Canon} is safe to apply as a pre-processing step during verification even when NSO is absent.


\begin{center}
\fbox{%
\begin{minipage}{0.9\linewidth}
\textbf{Takeaway (zero false positives, safe for clean model).} On clean (non-attacked) models, \textsc{Canon} makes no structural changes, preserves accuracy exactly, and leaves watermark similarity unchanged, indicating a safe defense during verification.
\end{minipage}
}
\end{center}
\section{Conclusion}
\label{sec:conclusion}


We demonstrate that NSO's \textit{zero-cost} structural obfuscation does \textbf{NOT} always break white-box watermarking. By modeling NSO as a graph-consistent producer–consumer threat and enforcing global layout consistency, we substantially strengthen the attack model and expose its recoverability. This transformation converts previously unrecoverable structural edits into a canonical, verifiable form. We present \textsc{Canon} to reliably recover obfuscated models. Across strong composed NSO attacks, \textsc{Canon} achieves 100\% recovery, incurs no utility loss, and introduces no false positives. Our recovery fully restores watermark verification without losing task accuracy.


\bibliographystyle{unsrt}
\bibliography{bib.bib}

\appendix
\renewcommand{\thesubsection}{\arabic{subsection}}


\section{Notation (Table~\ref{tab:notation})}
\label{appendix_notation}

\begin{table}[t]
\caption{Notations}
\label{tab:notation}
\centering
\vspace{5pt}
\begin{threeparttable}
\resizebox{\linewidth}{!}{%
\begin{tabular}{c|l}
\toprule
\textbf{Symbol} & \multicolumn{1}{c}{\textbf{Meaning}} \\

\midrule

\cellcolor{pink!25} $y$ & Channelized activation on a producer edge (Conv output). \\
\cellcolor{pink!25} $C$ & Channel width. \\
\cellcolor{pink!25} $y_{\text{before}},\,y_{\text{after}}$ & Activation before/after a local channel rewrite. \\
\cellcolor{pink!25} $M \in \mathbb{R}^{C_{\text{before}}\times C_{\text{after}}}$  & Channel transform with $y_{\text{before}} = M\,y_{\text{after}}$. \\
\cellcolor{pink!25} $W$ & Linear consumer weights; rewritten as $W_{\text{after}} = W_{\text{before}}\,M$. \\
\cellcolor{pink!25} $\mathcal{X}=\{x_t\}_{t=1}^{T}$ & Probe set used by \textsc{Canon}. \\
\cellcolor{pink!25} $T$ & Number of probes. \\
\cellcolor{pink!25} $u_{i,t}$ & BN-aware scalar summary of channel $i$ on probe $t$. \\
\cellcolor{pink!25} $b_{i,t}$ & Activity bit; $\mathbf{b}_i=(b_{i,1},\dots,b_{i,T})$. \\
\cellcolor{pink!25} $\mathrm{sig}(i)$ & Hashed activity signature for coarse bucketing. \\
\cellcolor{pink!25} $\epsilon$ & Magnitude threshold for stable ratio estimation. \\
\cellcolor{pink!25} $T_{\min}$ & Minimum probes for ratio/scale estimation. \\
\cellcolor{pink!25} $\tau$ & Proportionality tolerance. \\
\cellcolor{pink!25} $\mathcal{C}$ & Candidate redundancy cluster of channels. \\
\cellcolor{pink!25} $r$ & Representative channel index. \\
\cellcolor{pink!25} $\alpha_j$ & Estimated scale of channel $j$ relative to $r$. \\
\cellcolor{pink!25} $\gamma_{\text{drop}},\,\gamma_{\text{keep}}$ & Norm thresholds for drop vs.\ keep/merge (fan-out aware). \\
\cellcolor{pink!25} $\epsilon_{\mathrm{cert}}$ & Tolerance for Tier-1 reference-equivalence certificate. \\
\cellcolor{pink!25} $\eta$ & Stabilizer in certificate denominator. \\
\cellcolor{pink!25} $\lambda,\,\delta$ & Tier-2 PASS parameters (recovery fraction, noise slack). \\
\cellcolor{pink!25} $\rho$ & NSO injection ratio, $\rho\in[0,1]$. \\
\cellcolor{pink!25} $d=\lceil \rho C\rceil$ & Injected dummy channels (single-step). \\
\cellcolor{pink!25} $p$ & Split-baseline selector in $[0,1]$. \\
\cellcolor{pink!25} $S$ & Operation-sequence length for \texttt{mix-opseq}. \\

\midrule

\cellcolor{yellow!10} $G=(V,E)$ & FX-traced forward DAG with shape/dataflow metadata. \\
\cellcolor{yellow!10} $\mathcal{P}\subseteq E$ & Eligible producer edges (Conv/Linear outputs). \\
\cellcolor{yellow!10} $C_\ell$ & Output width of producer edge $\ell$. \\
\cellcolor{yellow!10} $d_\ell=\lceil \rho C_\ell\rceil$ & Injected channels for $\ell$; $C'_\ell=C_\ell+d_\ell$. \\
\cellcolor{yellow!10} $C_\ell^{(t)}$ & Width of $\ell$ after $t$ \texttt{mix-opseq} steps. \\
\cellcolor{yellow!10} $d_\ell^{(t)}=\lceil \rho C_\ell^{(t)}\rceil$ & Per-step injection amount. \\
\cellcolor{yellow!10} $\mathrm{Cons}(\ell)$ & Downstream linear consumers of $\ell$. \\
\cellcolor{yellow!10} $W_c$ & Weight tensor of consumer $c$. \\
\cellcolor{yellow!10} $|W_c|$ & Number of parameters in $W_c$. \\
\cellcolor{yellow!10} $M,\,M'$ & Clean and attacked models. \\
\cellcolor{yellow!10} $\mathrm{FLOPs}(M)$ & Forward-pass compute cost of $M$. \\

\bottomrule
\end{tabular}%
}
    \begin{tablenotes}
    \scriptsize
    \item the upper part lists notations appearing in the \textit{main text} (\S\ref{sec:graph-consistent});
    \item the lower part covers notations used for the \textit{complexity analysis} (Appendix~\ref{sec:complexity}).
    \end{tablenotes}
 \end{threeparttable}
\end{table}

\section{Complexity Analysis}
\label{sec:complexity}

This appendix characterizes the time and memory cost of (i) graph-consistent NSO attacks and (ii) \textsc{Canon} recovery.
We use standard big-$O$ notation and measure costs in terms of (a) forward-pass compute and (b) the number of scalar parameters / activation elements read or written.
Constants hidden in $O(\cdot)$ include fixed kernel areas (e.g., $k^2$ for $k{\times}k$ convolutions) and implementation-level constants.


\subsection{Complexity of graph-consistent NSO}\label{app:complexity:nso}

\paragraph{Graph extraction and shape bookkeeping.}
Graph-consistent NSO requires a one-time graph extraction and shape bookkeeping pass.
FX tracing is $O(|V|{+}|E|)$ time and $O(|V|{+}|E|)$ memory.
If tensor shapes are inferred via a \emph{real} example forward pass (rather than meta propagation), this adds an additional $O(\mathrm{FLOPs}(M))$ time term.
In total, a safe upper bound is
\[
\texttt{time}_{\mathrm{fx}} = O(|V|+|E|+\mathrm{FLOPs}(M)),
\quad
\texttt{mem}_{\mathrm{fx}} = O(|V|+|E|).
\]

\paragraph{Single-step primitives (\texttt{nso\_zero} / \texttt{nso\_split} / \texttt{nso\_clique}).}
For each attacked producer $\ell\in\mathcal{P}_{\mathrm{atk}}$, NSO (i) materializes the widened producer weights, and (ii) rewrites every downstream linear consumer $c\in\mathrm{Cons}(\ell)$ to preserve functional equivalence.

Let $M$ denote the induced channel mapping (a \emph{ChannelTransform}) between pre- and post-attack channel coordinates on edge $\ell$, so that consumer weights are updated as $W_c \leftarrow W_c M$.
In general, a sparse multiply costs $O(\mathrm{nnz}(M)\cdot C_{\mathrm{out}} \cdot k^2)$ for a conv consumer (or $O(\mathrm{nnz}(M)\cdot C_{\mathrm{out}})$ for linear), because each nonzero corresponds to copying/scaling/adding one input-channel slice across all output channels.
For NSO primitives, $M$ is \emph{highly structured} (permutation/diagonal plus $O(1)$-fan-in injections), so $\mathrm{nnz}(M)=O(C'_\ell)$ and the update can be implemented as gather/scale/merge with cost linear in the written tensor size.
Thus, per attacked producer $\ell$,
\[
\texttt{time}_{\mathrm{single}}(\ell)
=
O(|W'_\ell|)
\;+\;
\sum_{c\in \mathrm{Cons}(\ell)} O(|W'_c|),
\]
where $W'_\ell$ is the widened producer tensor and each $W'_c$ is the rewritten consumer tensor whose input width increases from $C_\ell$ to $C'_\ell$.
Any additional constraint enforcement for \texttt{split}/\texttt{clique} touches only affected slices and remains linear in the same tensors, so it does not change the asymptotic order.

\paragraph{Memory footprint.}
After injection, per-edge widths scale as $C'_\ell \approx (1{+}\rho)C_\ell$ (ignoring rounding).
For \emph{interior} dense regions where consecutive layers widen both input and output widths, weight tensors can scale quadratically in the width factor because $|W|$ depends on $C_{\mathrm{out}}C_{\mathrm{in}}$; e.g., a conv weight can scale as $(1{+}\rho)^2$ in a widened block.
Peak temporary memory during rewriting is dominated by holding an old and a new tensor simultaneously and is therefore
\[
\texttt{peak-mem}_{\mathrm{single}}
=
O\!\Bigl(\max_{c \text{ rewritten}} |W'_c|\Bigr),
\]
in addition to the resident parameters.

\paragraph{\texttt{mix-opseq} (length $S$).}
If an operation sequence repeatedly injects $d_\ell^{(t)}=\lceil \rho\, C_\ell^{(t)}\rceil$ channels at step $t$, then
\[
C_\ell^{(t+1)} = C_\ell^{(t)} + d_\ell^{(t)} \approx (1+\rho)\,C_\ell^{(t)}
\quad\Rightarrow\quad
C_\ell^{(S)} \approx (1+\rho)^S C_\ell,
\]
up to rounding.
Per attacked producer $\ell$, the total rewrite work across $S$ steps forms a geometric series and is dominated by the last step:
\begin{equation}
\begin{aligned}
\texttt{time}_{\mathrm{opseq}}(\ell)
&=
\sum_{t=0}^{S-1}
\left(
O(|W_\ell^{(t+1)}|)
+
\sum_{c\in\mathrm{Cons}(\ell)} O(|W_c^{(t+1)}|)
\right) \\
&=
O(|W_\ell^{(S)}|)
+
\sum_{c\in\mathrm{Cons}(\ell)} O(|W_c^{(S)}|).
\end{aligned}
\end{equation}
Consequently, \texttt{mix-opseq} can induce geometric cost increases in $S$ through the widened widths; in dense regions, weight sizes may scale as $\Theta((1+\rho)^{2S})$.

\paragraph{\texttt{mix-opseq} (per-merge-group).}
Per-merge-group sampling changes only which edges are attacked together; it does not change the dominant tensor-materialization and rewrite terms above.
It adds a near-linear preprocessing step (e.g., union-find over the FX graph) to compute merge groups:
\[
\texttt{time}_{\mathrm{group}} = O((|V|+|E|)\,\alpha(|V|+|E|)),
\quad
\texttt{mem}_{\mathrm{group}} = O(|V|),
\]
where $\alpha(\cdot)$ is the inverse-Ackermann function.
Optional interior-placement bookkeeping (when enabled) is linear in the affected widths and does not change the dominant rewrite cost.

\subsection{Complexity of \textsc{Canon} recovery}\label{app:complexity:canon}

\textsc{Canon} recovery consists of (i) probe capture, (ii) per-layer redundancy discovery (bucketing + proportionality refinement), and (iii) graph-consistent rewrites that propagate a recovered channel layout through fan-out and merge nodes.

Let $M'$ be the attacked model and let $\mathcal{P}'$ be its eligible producer edges with attacked widths $C'_\ell$.

\paragraph{(i) Probe forward passes and activation summaries.}
Running $T$ probes requires $T$ forward passes:
\[
\texttt{time}_{\mathrm{probe}} = O(T\cdot \mathrm{FLOPs}(M')).
\]
In addition, \textsc{Canon} computes per-channel summaries/bits from probed tensors.
If an edge $e$ has spatial dimensions $H_e\times W_e$, computing channel-wise reductions costs
\[
\texttt{time}_{\mathrm{summ}} = O\!\Bigl(T\sum_{e\in\mathcal{P}'} C'_e H_e W_e\Bigr),
\]
which is typically lower order than the forward pass but is included here for completeness.
Because only per-channel scalars/bits are stored (not full activations), memory scales as
\[
\texttt{mem}_{\mathrm{probe}} = O\!\Bigl(T\sum_{\ell\in\mathcal{P}'} C'_\ell\Bigr),
\]
up to constant factors depending on whether summaries are stored as floats or bit-packed.

\paragraph{(ii) Signature bucketing and proportionality refinement.}
For a producer edge $\ell$, computing / hashing a $T$-length signature per channel is $O(C'_\ell T)$.
If channels are partitioned into signature buckets $\mathcal{B}_\ell$ and a bucket $b$ contains $|b|$ channels, pairwise proportionality checks within $b$ cost $O(|b|^2 T)$.
Thus,
\[
\texttt{time}_{\mathrm{cluster}}(\ell)
=
O(C'_\ell T)
+
O\!\Bigl(T\sum_{b\in\mathcal{B}_\ell} |b|^2\Bigr).
\]
The worst case (all channels collide into one bucket) is $O((C'_\ell)^2T)$, while typical NSO-induced redundancy yields smaller buckets and substantially reduces the quadratic term.
Auxiliary memory is upper bounded by the largest bucket if an explicit adjacency structure is materialized:
\[
\texttt{mem}_{\mathrm{cluster}} = O\!\Bigl(\max_{\ell}\max_{b\in\mathcal{B}_\ell} |b|^2\Bigr),
\]
and can be reduced in practice by streaming pairwise checks.

\paragraph{(iii) Graph-consistent consumer rewrites.}
Once a channel layout map is inferred for an edge, \textsc{Canon} rewrites all downstream linear consumers (fan-out) and synchronizes layouts across merge nodes (\texttt{add}/\texttt{cat}) by propagating composed (still sparse/structured) transforms.
Each rewrite is implemented as gather/scale/merge over weight slices and is therefore linear in the tensor size being written.
Let $\mathcal{W}_{\mathrm{rw}}$ denote the multiset of weight tensors rewritten during recovery; then
\begin{equation}
\begin{aligned}
&\texttt{time}_{\mathrm{rewrite}}
=
O\!\Bigl(\sum_{W\in\mathcal{W}_{\mathrm{rw}}} |W|\Bigr),\\
& \texttt{peak-mem}_{\mathrm{rewrite}}
=
O\!\Bigl(\max_{W\in\mathcal{W}_{\mathrm{rw}}} |W|\Bigr).
\end{aligned}
\end{equation}
In rare corner cases, an optional second synchronization pass may re-touch some tensors, multiplying the rewrite term by a small constant (e.g., $K\in\{1,2\}$).

\paragraph{Overall recovery cost and dependence on $\rho$.}
Combining the above terms yields
\begin{equation}
\begin{aligned}
\texttt{time}_{\mathrm{recover}}
= &
O(T\cdot \mathrm{FLOPs}(M'))
+
O\!\Bigl(T\sum_{e\in\mathcal{P}'} C'_e H_e W_e\Bigr)\\
& +
\sum_{\ell\in\mathcal{P}'}
O\!\Bigl(C'_\ell T + T\sum_{b\in\mathcal{B}_\ell} |b|^2\Bigr)
+
O\!\Bigl(\sum_{W\in\mathcal{W}_{\mathrm{rw}}} |W|\Bigr),
\end{aligned}
\end{equation}
and
\begin{equation}
\begin{aligned}
\texttt{mem}_{\mathrm{recover}}
=&
O\!\Bigl(T\sum_{\ell\in\mathcal{P}'} C'_\ell\Bigr)
+
O\!\Bigl(\max_{\ell}\max_{b\in\mathcal{B}_\ell} |b|^2\Bigr) \\
&+
O\!\Bigl(\max_{W\in\mathcal{W}_{\mathrm{rw}}} |W|\Bigr)
+
O(|V|+|E|).
\end{aligned}
\end{equation}
Crucially, the injection ratio $\rho$ affects recovery \emph{indirectly} through attacked widths $C'_\ell$: larger $\rho$ increases $\mathrm{FLOPs}(M')$ (probe time), stored probe summaries, and the worst-case clustering term.
For \texttt{mix-opseq} of length $S$, widths can scale as $C'_\ell \approx (1+\rho)^S C_\ell$, yielding geometric blow-ups in both attack and recovery cost when $\rho$ and/or $S$ are large.

\begin{table*}[t]
    \centering
    \caption{Watermark similarity restoration for \textit{ResNet-18} and \textit{EfficientNet} under NSO \textcolor{magenta}{\textbf{\texttt{add-ops}}} attacks (\textcolor{magenta}{\textbf{$0.5$}} attack ratio). Tier-1  verification passes (\cmark) in all runs and is used as the primary outcome, while Tier-2 serves as a conservative fallback.}
    \label{tab:nso_recovery_suite_add_05}
    \vspace{5pt}
    
    \renewcommand{\arraystretch}{1.1}
    \resizebox{\linewidth}{!}{
    \begin{threeparttable}
    \begin{tabular}{c |c | ccc |c| ccccc}
    \toprule
    &
    \multirow{3}{*}{\textbf{Watermark}} &
    \multicolumn{3}{c|}{\textbf{Clean / Functional Drift}} &
    \multirow{3}{*}{\textbf{\makecell{Tier-1\\Pass}}} &
    \multicolumn{5}{c}{\textbf{Attack Sim.$\rightarrow$ Recovery Tier-1 Sim. {\color{gray!80}(Tier-2 Sim.)}}}
    \\

    \cmidrule(lr){3-5} 
    \cmidrule(lr){7-11}
     & &
     \makecell{\textbf{Clean}\\\textbf{Sim.}} &
     \makecell{$\boldsymbol{\Delta_{\mathrm{acc}}}$\\\textbf{(att.)}} &
     \makecell{$\boldsymbol{\Delta_{\mathrm{acc}}}$\\\textbf{(rec.)}} &
     &
     \makecell{\textbf{\texttt{zero}}} &
     \makecell{\textbf{\texttt{clique}}} &
     \makecell{\textbf{\texttt{split}}} &
     \makecell{\textbf{\texttt{mix-opseq}}} &
     \makecell{\textbf{\texttt{mix-opseq}}\\\textbf{\texttt{(per-merge-group)}}} 
     \\
    \midrule

        \multirow{9}{*}{\rotatebox{90}{\makecell{ResNet-18}}}
    & uchida17~\cite{uchida2017embedding} &\cellcolor{blue!9} 1.0000 &\cellcolor{blue!10} 0.0000 &\cellcolor{blue!10} 0.0000 & \cmark
      & \cellcolor{pink!25} 0.4844$\rightarrow$1.0000 {\color{gray!80}(0.4844)} & \cellcolor{pink!25} 0.4141$\rightarrow$1.0000 {\color{gray!80}(0.4844)} & \cellcolor{pink!25} 0.5547$\rightarrow$1.0000 {\color{gray!80}(0.4922)} & \cellcolor{pink!27} 0.5000$\rightarrow$1.0000 {\color{gray!80}(0.4219)} & \cellcolor{pink!27} 0.5000$\rightarrow$1.0000 {\color{gray!80}(0.4766)} \\
      
    & greedy~\cite{liu2021watermarking} &\cellcolor{blue!9} 1.0000&\cellcolor{blue!10} 0.0000 &\cellcolor{blue!10} 0.0000 & \cmark
      & \cellcolor{pink!25} 0.6406$\rightarrow$1.0000 {\color{gray!80}(0.5781)}& \cellcolor{pink!25} 0.5469$\rightarrow$1.0000 {\color{gray!80}(0.5781)}& \cellcolor{pink!25} 0.5312$\rightarrow$1.0000 {\color{gray!80}(0.5703)}& \cellcolor{pink!27} 0.5078$\rightarrow$1.0000 {\color{gray!80}(0.5469)}& \cellcolor{pink!27} 0.5391$\rightarrow$1.0000 {\color{gray!80}(0.5234)} \\
      
    & lottery~\cite{chen2021you} &\cellcolor{blue!9} 1.0000&\cellcolor{blue!10} 0.0000 &\cellcolor{blue!10} 0.0000 & \cmark
      & \cellcolor{pink!25} 0.5703$\rightarrow$1.0000 {\color{gray!80}(0.4766)} & \cellcolor{pink!25} 0.4922$\rightarrow$1.0000 {\color{gray!80}(0.4766)} & \cellcolor{pink!25} 0.5391$\rightarrow$1.0000 {\color{gray!80}(0.4844)} & \cellcolor{pink!27} 0.5156$\rightarrow$1.0000 {\color{gray!80}(0.4922)} & \cellcolor{pink!27} 0.5469$\rightarrow$1.0000 {\color{gray!80}(0.5000)}   \\
      
    & ipr\_gan~\cite{ong2021protecting} &\cellcolor{blue!9} 1.0000 &\cellcolor{blue!10} 0.0000 &\cellcolor{blue!10} 0.0000 & \cmark
      & \cellcolor{pink!25} 0.5078$\rightarrow$1.0000 {\color{gray!80}(0.5156)} & \cellcolor{pink!25} 0.5391$\rightarrow$1.0000 {\color{gray!80}(0.5156)} & \cellcolor{pink!25} 0.5391$\rightarrow$1.0000 {\color{gray!80}(0.5469)} & \cellcolor{pink!27} 0.4531$\rightarrow$1.0000 {\color{gray!80}(0.5312)} & \cellcolor{pink!27} 0.4922$\rightarrow$1.0000 {\color{gray!80}(0.4531)}   \\
      
    & riga21~\cite{wang2021riga} & \cellcolor{blue!9} 0.9766 &\cellcolor{blue!10} 0.0000 &\cellcolor{blue!10} 0.0000 & \cmark
      & \cellcolor{pink!25} 0.4531$\rightarrow$0.9766 {\color{gray!80}(0.9922)} & \cellcolor{pink!25} 0.4531$\rightarrow$0.9766 {\color{gray!80}(0.9922)} & \cellcolor{pink!25} 0.4531$\rightarrow$0.9766 {\color{gray!80}(0.9844)} & \cellcolor{pink!27} 0.4531$\rightarrow$0.9766 {\color{gray!80}(0.9922)} & \cellcolor{pink!27} 0.4531$\rightarrow$0.9766 {\color{gray!80}(1.0000)} \\
      
    & ipr\_ic~\cite{lim2022protect} &\cellcolor{blue!9} 1.0000 &\cellcolor{blue!10} 0.0000 &\cellcolor{blue!10} 0.0000 & \cmark
      & \cellcolor{pink!25} 0.5000$\rightarrow$1.0000 {\color{gray!80}(0.5156)} & \cellcolor{pink!25} 0.4609$\rightarrow$1.0000 {\color{gray!80}(0.5156)} & \cellcolor{pink!25} 0.4375$\rightarrow$1.0000 {\color{gray!80}(0.5156)} & \cellcolor{pink!27} 0.5312$\rightarrow$1.0000 {\color{gray!80}(0.5078)} & \cellcolor{pink!27} 0.4375$\rightarrow$1.0000 {\color{gray!80}(0.5000)}  \\
      
    & deepsigns~\cite{darvish2019deepsigns} &\cellcolor{blue!9} 1.0000 &\cellcolor{blue!10} 0.0000 &\cellcolor{blue!10} 0.0000 & \cmark
      & \cellcolor{pink!25} 0.4453$\rightarrow$1.0000 {\color{gray!80}(0.3906)} & \cellcolor{pink!25} 0.4609$\rightarrow$1.0000 {\color{gray!80}(0.3906)} & \cellcolor{pink!25} 0.5000$\rightarrow$1.0000 {\color{gray!80}(0.4531)} & \cellcolor{pink!27} 0.5547$\rightarrow$1.0000 {\color{gray!80}(0.5156)} & \cellcolor{pink!27} 0.4531$\rightarrow$1.0000 {\color{gray!80}(0.5000)}  \\
      
    & deepipr~\cite{fan2021deepipr} &\cellcolor{blue!9} 1.0000 &\cellcolor{blue!10} 0.0000 &\cellcolor{blue!10} 0.0000 & \cmark
      & \cellcolor{pink!25} 0.4062$\rightarrow$1.0000 {\color{gray!80}(0.5156)} & \cellcolor{pink!25} 0.4062$\rightarrow$1.0000 {\color{gray!80}(0.5156)} & \cellcolor{pink!25} 0.4062$\rightarrow$1.0000 {\color{gray!80}(0.4688)} & \cellcolor{pink!27} 0.4062$\rightarrow$1.0000 {\color{gray!80}(0.5312)} & \cellcolor{pink!27} 0.4062$\rightarrow$1.0000 {\color{gray!80}(0.5938)}  \\
      
    & passport\_aware~\cite{zhang2020passport} &\cellcolor{blue!9} 1.0000 &\cellcolor{blue!10} 0.0000 &\cellcolor{blue!10} 0.0000 & \cmark
      & \cellcolor{pink!25} 0.5781$\rightarrow$1.0000 {\color{gray!80}(0.5547)} & \cellcolor{pink!25} 0.6016$\rightarrow$1.0000 {\color{gray!80}(0.5547)} & \cellcolor{pink!25} 0.5703$\rightarrow$1.0000 {\color{gray!80}(0.5234)} & \cellcolor{pink!27} 0.5703$\rightarrow$1.0000 {\color{gray!80}(0.5078)} & \cellcolor{pink!27} 0.5156$\rightarrow$1.0000 {\color{gray!80}(0.5625)} \\

    \cmidrule{1-8}

    \multirow{9}{*}{\rotatebox{90}{EfficientNet}}
    & uchida17~\cite{uchida2017embedding} &\cellcolor{blue!9} 1.0000 &\cellcolor{blue!10} 0.0000 &\cellcolor{blue!10} 0.0000 & \cmark
      & \cellcolor{pink!25} 0.5703$\rightarrow$1.0000 {\color{gray!80}(0.5078)} & \cellcolor{pink!25} 0.4922$\rightarrow$1.0000 {\color{gray!80}(0.5078)} & \cellcolor{pink!25} 0.5000$\rightarrow$1.0000 {\color{gray!80}(0.4688)} & \cellcolor{pink!27} 0.4922$\rightarrow$1.0000 {\color{gray!80}(0.5312)} & \cellcolor{pink!27} 0.5000$\rightarrow$1.0000 {\color{gray!80}(0.5000)} \\
      
    & greedy~\cite{liu2021watermarking} &\cellcolor{blue!9} 1.0000&\cellcolor{blue!10} 0.0000 &\cellcolor{blue!10} 0.0000 & \cmark
      & \cellcolor{pink!25} 0.5156$\rightarrow$1.0000 {\color{gray!80}(0.5000)}& \cellcolor{pink!25} 0.5703$\rightarrow$1.0000 {\color{gray!80}(0.5000)}& \cellcolor{pink!25}  0.4844$\rightarrow$1.0000 {\color{gray!80}(0.4453)}& \cellcolor{pink!27} 0.5469$\rightarrow$1.0000 {\color{gray!80}(0.5000)}& \cellcolor{pink!27} 0.4766$\rightarrow$1.0000 {\color{gray!80}(0.4375)} \\
      
    & lottery~\cite{chen2021you} &\cellcolor{blue!9} 1.0000&\cellcolor{blue!10} 0.0000 &\cellcolor{blue!10} 0.0000 & \cmark
      & \cellcolor{pink!25} 0.5469$\rightarrow$1.0000 {\color{gray!80}(0.4609)} & \cellcolor{pink!25} 0.4609$\rightarrow$1.0000 {\color{gray!80}(0.4609)} & \cellcolor{pink!25} 0.4453$\rightarrow$1.0000 {\color{gray!80}(0.4375)} & \cellcolor{pink!27} 0.4297$\rightarrow$1.0000 {\color{gray!80}(0.4609)} & \cellcolor{pink!27} 0.5234$\rightarrow$1.0000 {\color{gray!80}(0.4688)} \\
      
    & ipr\_gan~\cite{ong2021protecting} &\cellcolor{blue!9} 1.0000 &\cellcolor{blue!10} 0.0000 &\cellcolor{blue!10} 0.0000 & \cmark
      & \cellcolor{pink!25} 0.5312$\rightarrow$1.0000 {\color{gray!80}(0.5547)} & \cellcolor{pink!25} 0.4844$\rightarrow$1.0000 {\color{gray!80}(0.5547)} & \cellcolor{pink!25} 0.5000$\rightarrow$1.0000 {\color{gray!80}(0.5469)} & \cellcolor{pink!27} 0.5234$\rightarrow$1.0000 {\color{gray!80}(0.5469)} & \cellcolor{pink!27} 0.5234$\rightarrow$1.0000 {\color{gray!80}(0.5156)} \\
      
    & riga21~\cite{wang2021riga} & \cellcolor{blue!9} 0.9062 &\cellcolor{blue!10} 0.0000 &\cellcolor{blue!10} 0.0000 & \cmark
      & \cellcolor{pink!25} 0.6172$\rightarrow$0.9062 {\color{gray!80}(0.9062)} & \cellcolor{pink!25} 0.6172$\rightarrow$0.9062 {\color{gray!80}(0.8984)} & \cellcolor{pink!25} 0.6172$\rightarrow$0.9062 {\color{gray!80}(0.9219)} & \cellcolor{pink!27} 0.6172$\rightarrow$0.9062 {\color{gray!80}(0.9141)} & \cellcolor{pink!27} 0.6172$\rightarrow$0.9062 {\color{gray!80}(0.9062)}  \\
      
    & ipr\_ic~\cite{lim2022protect} &\cellcolor{blue!9} 1.0000 &\cellcolor{blue!10} 0.0000 &\cellcolor{blue!10} 0.0000 & \cmark
      & \cellcolor{pink!25} 0.5078$\rightarrow$1.0000 {\color{gray!80}(0.5312)} & \cellcolor{pink!25} 0.4922$\rightarrow$1.0000 {\color{gray!80}(0.5312)} & \cellcolor{pink!25} 0.4609$\rightarrow$1.0000 {\color{gray!80}(0.5391)} & \cellcolor{pink!27} 0.4922$\rightarrow$1.0000 {\color{gray!80}(0.5625)} & \cellcolor{pink!27} 0.5469$\rightarrow$1.0000 {\color{gray!80}(0.5469)}  \\
      
    & deepsigns~\cite{darvish2019deepsigns} &\cellcolor{blue!9} 1.0000 &\cellcolor{blue!10} 0.0000 &\cellcolor{blue!10} 0.0000 & \cmark
      & \cellcolor{pink!25} 0.5312$\rightarrow$1.0000 {\color{gray!80}(0.5703)} & \cellcolor{pink!25} 0.5000$\rightarrow$1.0000 {\color{gray!80}(0.5703)} & \cellcolor{pink!25} 0.4922$\rightarrow$1.0000 {\color{gray!80}(0.5625)} & \cellcolor{pink!27} 0.5234$\rightarrow$1.0000 {\color{gray!80}(0.5156)} & \cellcolor{pink!27} 0.5234$\rightarrow$1.0000 {\color{gray!80}(0.5312)}  \\
      
     & deepipr~\cite{fan2021deepipr} &\cellcolor{blue!9} 1.0000 &\cellcolor{blue!10} 0.0000 &\cellcolor{blue!10} 0.0000 & \cmark
      & \cellcolor{pink!25} 0.4062$\rightarrow$1.0000 {\color{gray!80}(0.6094)} & \cellcolor{pink!25} 0.4062$\rightarrow$1.0000 {\color{gray!80}(0.6094)} & \cellcolor{pink!25} 0.4062$\rightarrow$1.0000 {\color{gray!80}(0.3438)} & \cellcolor{pink!27} 0.4062$\rightarrow$1.0000 {\color{gray!80}(0.5469)} & \cellcolor{pink!27} 0.4062$\rightarrow$1.0000 {\color{gray!80}(0.5312)}  \\
      
    & passport\_aware~\cite{zhang2020passport} &\cellcolor{blue!9} 1.0000 &\cellcolor{blue!10} 0.0000 &\cellcolor{blue!10} 0.0000 & \cmark
      & \cellcolor{pink!25} 0.5156$\rightarrow$1.0000 {\color{gray!80}(0.4688)} & \cellcolor{pink!25} 0.5156$\rightarrow$1.0000 {\color{gray!80}(0.4688)} & \cellcolor{pink!25} 0.4844$\rightarrow$1.0000 {\color{gray!80}(0.5234)} & \cellcolor{pink!27} 0.5234$\rightarrow$1.0000 {\color{gray!80}(0.5391)} & \cellcolor{pink!27} 0.4844$\rightarrow$1.0000 {\color{gray!80}(0.5625)}  \\

    \bottomrule
    \end{tabular}

    \begin{tablenotes}
    \item $\Delta_{\mathrm{acc}}(\mathrm{clean},\mathrm{\textbf{att}acked})$ and $\Delta_{\mathrm{acc}}(\mathrm{clean},\mathrm{\textbf{rec}overed})$ report the \emph{worst-case} absolute accuracy gap over the five listed attacks for each watermark.
    \item \textbf{Tier-1 pass} indicates that the Tier-1 reference-equivalence certificate over watermarked layers passes for all runs in the suite for that model/watermark (\cmark means 100\% match rate).
    \end{tablenotes}

    \end{threeparttable}
    }
\end{table*}
\begin{table*}[t]
    \centering
    \caption{Watermark similarity restoration for \textit{InceptionV3} and \textit{DenseNet} under NSO \textcolor{magenta}{\textbf{\texttt{cat-ops}}} attacks (\textcolor{magenta}{\textbf{$0.5$}} attack ratio). Tier-1  verification passes (\cmark) in all runs and is used as the primary outcome, while Tier-2 serves as a conservative fallback.}
    \label{tab:nso_recovery_suite_cat_05}
    \vspace{5pt}
    
    \renewcommand{\arraystretch}{1.1}
    \resizebox{\linewidth}{!}{
    \begin{threeparttable}
    \begin{tabular}{c |c | ccc |c| ccccc}
    \toprule
    &
    \multirow{3}{*}{\textbf{Watermark}} &
    \multicolumn{3}{c|}{\textbf{Clean / Functional Drift}} &
    \multirow{3}{*}{\textbf{\makecell{Tier-1\\Pass}}} &
    \multicolumn{5}{c}{\textbf{Attack Sim.$\rightarrow$ Recovery Tier-1 Sim. {\color{gray!80}(Tier-2 Sim.)}}}
    \\

    \cmidrule(lr){3-5} 
    \cmidrule(lr){7-11}
     & &
     \makecell{\textbf{Clean}\\\textbf{Sim.}} &
     \makecell{$\boldsymbol{\Delta_{\mathrm{acc}}}$\\\textbf{(att.)}} &
     \makecell{$\boldsymbol{\Delta_{\mathrm{acc}}}$\\\textbf{(rec.)}} &
     &
     \makecell{\textbf{\texttt{zero}}} &
     \makecell{\textbf{\texttt{clique}}} &
     \makecell{\textbf{\texttt{split}}} &
     \makecell{\textbf{\texttt{mix-opseq}}} &
     \makecell{\textbf{\texttt{mix-opseq}}\\\textbf{\texttt{(per-merge-group)}}} 
     \\
    \midrule

    \multirow{9}{*}{\rotatebox{90}{InceptionV3}}
    & uchida17~\cite{uchida2017embedding} &\cellcolor{yellow!13} 1.0000 &\cellcolor{yellow!15} 0.0000 &\cellcolor{yellow!15} 0.0000  & \cmark 
      & \cellcolor{blue!9} 0.5000$\rightarrow$1.0000 {\color{gray}(0.4922)} & \cellcolor{blue!9} 0.5469$\rightarrow$1.0000 {\color{gray}(0.5312)} & \cellcolor{blue!9} 0.4688$\rightarrow$1.0000 {\color{gray}(0.5625)} & \cellcolor{blue!10} 0.5156$\rightarrow$1.0000 {\color{gray}(0.4688)} & \cellcolor{blue!10} 0.4844$\rightarrow$1.0000 {\color{gray}(0.5234)}  \\
      
    & greedy~\cite{liu2021watermarking} & \cellcolor{yellow!13} 1.0000&\cellcolor{yellow!15} 0.0000 &\cellcolor{yellow!15} 0.0000  & \cmark 
      & \cellcolor{blue!9} 0.4609$\rightarrow$1.0000 {\color{gray}(0.4922)}& \cellcolor{blue!9} 0.4844$\rightarrow$1.0000 {\color{gray}(0.4922)}& \cellcolor{blue!9} 0.5859$\rightarrow$1.0000 {\color{gray}(0.5078)}& \cellcolor{blue!10} 0.4922$\rightarrow$1.0000 {\color{gray}(0.5078)}& \cellcolor{blue!10} 0.5234$\rightarrow$1.0000 {\color{gray}(0.4766)}  \\
      
    & lottery~\cite{chen2021you} & \cellcolor{yellow!13} 1.0000&\cellcolor{yellow!15} 0.0000 &\cellcolor{yellow!15} 0.0000  & \cmark 
      & \cellcolor{blue!9} 0.4453$\rightarrow$1.0000 {\color{gray}(0.4922)}& \cellcolor{blue!9} 0.4531$\rightarrow$1.0000 {\color{gray}(0.4922)}& \cellcolor{blue!9} 0.5000$\rightarrow$1.0000 {\color{gray}(0.5000)}& \cellcolor{blue!10} 0.4688$\rightarrow$1.0000 {\color{gray}(0.4766)}& \cellcolor{blue!10} 0.4688$\rightarrow$1.0000 {\color{gray}(0.4922)} \\
      
    & ipr\_gan~\cite{ong2021protecting} & \cellcolor{yellow!13} 1.0000 &\cellcolor{yellow!15} 0.0000 &\cellcolor{yellow!15} 0.0000  & \cmark 
      & \cellcolor{blue!9} 0.5000$\rightarrow$1.0000 {\color{gray}(0.5156)} & \cellcolor{blue!9} 0.5000$\rightarrow$1.0000 {\color{gray}(0.5469)} & \cellcolor{blue!9} 0.5312$\rightarrow$1.0000 {\color{gray}(0.5781)} & \cellcolor{blue!10} 0.5938$\rightarrow$1.0000 {\color{gray}(0.5625)} & \cellcolor{blue!10} 0.5078$\rightarrow$1.0000 {\color{gray}(0.5078)}  \\
      
    & riga21~\cite{wang2021riga} & \cellcolor{yellow!13} 0.9688 &\cellcolor{yellow!15} 0.0000 &\cellcolor{yellow!15} 0.0000  & \cmark 
      & \cellcolor{blue!9} 0.4453$\rightarrow$0.9688{\color{gray}(0.9766)} & \cellcolor{blue!9} 0.4453$\rightarrow$0.9688{\color{gray}(0.9766)} & \cellcolor{blue!9} 0.4453$\rightarrow$0.9688{\color{gray}(0.9766)} & \cellcolor{blue!10} 0.4453$\rightarrow$0.9688{\color{gray}(0.9766)} & \cellcolor{blue!10} 0.4453$\rightarrow$0.9688{\color{gray}(0.9688)}   \\
      
    & ipr\_ic~\cite{lim2022protect} & \cellcolor{yellow!13} 1.0000 &\cellcolor{yellow!15} 0.0000 &\cellcolor{yellow!15} 0.0000  & \cmark 
      & \cellcolor{blue!9} 0.5859$\rightarrow$1.0000 {\color{gray}(0.4375)} & \cellcolor{blue!9} 0.5078$\rightarrow$1.0000 {\color{gray}(0.5000)} & \cellcolor{blue!9} 0.5156$\rightarrow$1.0000 {\color{gray}(0.5156)} & \cellcolor{blue!10} 0.5547$\rightarrow$1.0000 {\color{gray}(0.4688)} & \cellcolor{blue!10} 0.5000$\rightarrow$1.0000 {\color{gray}(0.5703)}   \\
      
    & deepsigns~\cite{darvish2019deepsigns} & \cellcolor{yellow!13} 1.0000 &\cellcolor{yellow!15} 0.0000 &\cellcolor{yellow!15} 0.0000  & \cmark 
      & \cellcolor{blue!9} 0.5000$\rightarrow$1.0000 {\color{gray}(0.5078)} & \cellcolor{blue!9} 0.5391$\rightarrow$1.0000 {\color{gray}(0.5078)} & \cellcolor{blue!9} 0.5156$\rightarrow$1.0000 {\color{gray}(0.5156)} & \cellcolor{blue!10} 0.5781$\rightarrow$1.0000 {\color{gray}(0.4688)} & \cellcolor{blue!10} 0.4531$\rightarrow$1.0000 {\color{gray}(0.5703)}   \\
      
    & deepipr~\cite{fan2021deepipr} & \cellcolor{yellow!13} 1.0000 &\cellcolor{yellow!15} 0.0000 &\cellcolor{yellow!15} 0.0000  & \cmark 
      & \cellcolor{blue!9} 0.4062$\rightarrow$1.0000 {\color{gray}(0.5156)} & \cellcolor{blue!9} 0.4062$\rightarrow$1.0000 {\color{gray}(0.4844)} & \cellcolor{blue!9} 0.4062$\rightarrow$1.0000 {\color{gray}(0.5000)} & \cellcolor{blue!10} 0.4062$\rightarrow$1.0000 {\color{gray}(0.5469)} & \cellcolor{blue!10} 0.4062$\rightarrow$1.0000 {\color{gray}(0.4219)}   \\
      
    & passport\_aware~\cite{zhang2020passport} & \cellcolor{yellow!13} 1.0000 &\cellcolor{yellow!15} 0.0000 &\cellcolor{yellow!15} 0.0000  & \cmark 
      & \cellcolor{blue!9} 0.4766$\rightarrow$1.0000 {\color{gray}(0.4922)} & \cellcolor{blue!9} 0.5156$\rightarrow$1.0000 {\color{gray}(0.5156)} & \cellcolor{blue!9} 0.4609$\rightarrow$1.0000 {\color{gray}(0.4844)} & \cellcolor{blue!10} 0.4688$\rightarrow$1.0000 {\color{gray}(0.4922)} & \cellcolor{blue!10} 0.5156$\rightarrow$1.0000 {\color{gray}(0.5469)}  \\

    \cmidrule{1-9}

    \multirow{9}{*}{\rotatebox{90}{DenseNet}}
    & uchida17~\cite{uchida2017embedding} & \cellcolor{yellow!13} 1.0000 &\cellcolor{yellow!15} 0.0000 &\cellcolor{yellow!15} 0.0000  & \cmark 
      & \cellcolor{blue!9} 0.5000$\rightarrow$1.0000 {\color{gray}(0.5547)} & \cellcolor{blue!9} 0.4688$\rightarrow$1.0000 {\color{gray}(0.5547)} & \cellcolor{blue!9} 0.4922$\rightarrow$1.0000 {\color{gray}(0.4453)} & \cellcolor{blue!10} 0.5000$\rightarrow$1.0000 {\color{gray}(0.5078)} & \cellcolor{blue!10} 0.5547$\rightarrow$1.0000 {\color{gray}(0.5156)}  \\
      
    & greedy~\cite{liu2021watermarking} & \cellcolor{yellow!13} 1.0000&\cellcolor{yellow!15} 0.0000 &\cellcolor{yellow!15} 0.0000  & \cmark 
      & \cellcolor{blue!9} 0.4609$\rightarrow$1.0000 {\color{gray}(0.5234)}& \cellcolor{blue!9} 0.4531$\rightarrow$1.0000 {\color{gray}(0.5234)}& \cellcolor{blue!9} 0.4375$\rightarrow$1.0000 {\color{gray}(0.4531)}& \cellcolor{blue!10} 0.5234$\rightarrow$1.0000 {\color{gray}(0.5625)}& \cellcolor{blue!10} 0.4844$\rightarrow$1.0000 {\color{gray}(0.4531)} \\
      
    & lottery~\cite{chen2021you} & \cellcolor{yellow!13} 1.0000&\cellcolor{yellow!15} 0.0000 &\cellcolor{yellow!15} 0.0000  & \cmark 
      & \cellcolor{blue!9} 0.4844$\rightarrow$1.0000 {\color{gray}(0.5234)} & \cellcolor{blue!9} 0.5469$\rightarrow$1.0000 {\color{gray}(0.5234)} & \cellcolor{blue!9} 0.5156$\rightarrow$1.0000 {\color{gray}(0.5000)} & \cellcolor{blue!10} 0.4922$\rightarrow$1.0000 {\color{gray}(0.5156)} & \cellcolor{blue!10} 0.4297$\rightarrow$1.0000 {\color{gray}(0.5078)}   \\
      
    & ipr\_gan~\cite{ong2021protecting} & \cellcolor{yellow!13} 1.0000 &\cellcolor{yellow!15} 0.0000 &\cellcolor{yellow!15} 0.0000  & \cmark 
      & \cellcolor{blue!9} 0.4844$\rightarrow$1.0000 {\color{gray}(0.4688)} & \cellcolor{blue!9} 0.4766$\rightarrow$1.0000 {\color{gray}(0.4688)} & \cellcolor{blue!9} 0.4766$\rightarrow$1.0000 {\color{gray}(0.4844)} & \cellcolor{blue!10} 0.5078$\rightarrow$1.0000 {\color{gray}(0.5938)} & \cellcolor{blue!10} 0.4766$\rightarrow$1.0000 {\color{gray}(0.4766)}   \\
      
    & riga21~\cite{wang2021riga} & \cellcolor{yellow!13} 1.0000 &\cellcolor{yellow!15} 0.0000 &\cellcolor{yellow!15} 0.0000  & \cmark 
      & \cellcolor{blue!9} 0.4922$\rightarrow$1.0000 {\color{gray}(0.9531)} & \cellcolor{blue!9} 0.4922$\rightarrow$1.0000 {\color{gray}(0.9531)} & \cellcolor{blue!9} 0.4922$\rightarrow$1.0000 {\color{gray}(0.9609)} & \cellcolor{blue!10} 0.4922$\rightarrow$1.0000 {\color{gray}(0.9453)} & \cellcolor{blue!10} 0.4922$\rightarrow$1.0000 {\color{gray}(0.9531)}   \\
      
    & ipr\_ic~\cite{lim2022protect} &  \cellcolor{yellow!13} 0.9922 & \cellcolor{yellow!15} 0.0000 &  \cellcolor{yellow!15} 0.0000  & \cmark 
      & \cellcolor{blue!9} 0.5469$\rightarrow$0.9922{\color{gray}(0.5312)} & \cellcolor{blue!9} 0.5703$\rightarrow$0.9922{\color{gray}(0.5312)} & \cellcolor{blue!9} 0.5234$\rightarrow$0.9922{\color{gray}(0.4609)} & \cellcolor{blue!10} 0.4688$\rightarrow$0.9922{\color{gray}(0.4141)} & \cellcolor{blue!10} 0.5703$\rightarrow$0.9922{\color{gray}(0.4766)} \\
      
    & deepsigns~\cite{darvish2019deepsigns} & \cellcolor{yellow!13} 1.0000 &\cellcolor{yellow!15} 0.0000 &\cellcolor{yellow!15} 0.0000  & \cmark 
      & \cellcolor{blue!9} 0.5547$\rightarrow$1.0000 {\color{gray}(0.5312)} & \cellcolor{blue!9} 0.5312$\rightarrow$1.0000 {\color{gray}(0.5312)} & \cellcolor{blue!9} 0.5078$\rightarrow$1.0000 {\color{gray}(0.5312)} & \cellcolor{blue!10} 0.5078$\rightarrow$1.0000 {\color{gray}(0.5781)} & \cellcolor{blue!10} 0.5391$\rightarrow$1.0000 {\color{gray}(0.5312)}  \\
      
    & deepipr~\cite{fan2021deepipr} & \cellcolor{yellow!13} 1.0000 &\cellcolor{yellow!15} 0.0000 &\cellcolor{yellow!15} 0.0000  & \cmark 
      & \cellcolor{blue!9} 0.4062$\rightarrow$1.0000 {\color{gray}(0.5312)} & \cellcolor{blue!9} 0.4062$\rightarrow$1.0000 {\color{gray}(0.5312)} & \cellcolor{blue!9} 0.4062$\rightarrow$1.0000 {\color{gray}(0.4375)} & \cellcolor{blue!10} 0.4062$\rightarrow$1.0000 {\color{gray}(0.4688)} & \cellcolor{blue!10} 0.4062$\rightarrow$1.0000 {\color{gray}(0.5156)}  \\
      
    & passport\_aware~\cite{zhang2020passport} & \cellcolor{yellow!13} 1.0000 &\cellcolor{yellow!15} 0.0000 &\cellcolor{yellow!15} 0.0000  & \cmark 
      & \cellcolor{blue!9} 0.5234$\rightarrow$1.0000 {\color{gray}(0.5469)} & \cellcolor{blue!9} 0.5000$\rightarrow$1.0000 {\color{gray}(0.5469)} & \cellcolor{blue!9} 0.5000$\rightarrow$1.0000 {\color{gray}(0.5000)} & \cellcolor{blue!10} 0.5078$\rightarrow$1.0000 {\color{gray}(0.5391)} & \cellcolor{blue!10} 0.5391$\rightarrow$1.0000 {\color{gray}(0.4844)} \\

    \bottomrule
    \end{tabular}

    \begin{tablenotes}
    \item $\Delta_{\mathrm{acc}}(\mathrm{clean},\mathrm{\textbf{att}acked})$ and $\Delta_{\mathrm{acc}}(\mathrm{clean},\mathrm{\textbf{rec}overed})$ report the \emph{worst-case} absolute accuracy gap over the five listed attacks for each watermark.
    \item \textbf{Tier-1 pass} indicates that the Tier-1 reference-equivalence certificate over watermarked layers passes for all runs in the suite for that model/watermark (\cmark means 100\% match rate).
    \end{tablenotes}

    \end{threeparttable}
    }
\end{table*}

\section{Why Can Our Recovery Accurately Detect the Presence of NSO Attacks Without Unnecessarily Modifying a Clean Model?}
\label{sec:fp-safety}

\heading{Clarifying the claim (``recognize NSO'').}
Our recovery does not attempt to \textit{classify} a model as attacked vs.\ clean using a separate detector. Instead, it is \textit{evidence-driven} and only proposes a non-identity ChannelTransform on an edge when it observes strong, consistent NSO redundancy evidence under probes.
On a clean model, those redundancy conditions rarely hold, so the induced transforms are (nearly always) identity, and the model is left unchanged. \textsc{Canon} thus ``recognizes'' NSO edits by recognizing \textit{redundancy patterns} that NSO must introduce to preserve functionality while expanding width.

\heading{Why NSO creates a recognizable signal.}
NSO primitives preserve the network function by introducing channels whose behavior is constrained:
\texttt{nso\_zero} channels are (near-)inactive, \texttt{nso\_split} channels are proportional duplicates whose aggregate equals the original, and \texttt{nso\_clique} channels are duplicates whose downstream contributions cancel.
These constructions induce repeated activation state patterns across probes and/or exact/proportional structure in weights; \textsc{Canon} targets precisely these invariants.

\heading{Conservative, multi-signal redundancy test (when we allow pruning/merging).}
For each eligible producer edge (Conv/Linear output), recovery forms candidate redundancy clusters and then filters them aggressively:
\begin{packeditemize}
  \item \textit{Activation signature clustering.}
  We record per-channel binary \textit{states} and real-valued \textit{summaries} across a probe set (BN-aware capture when applicable).
  Only channels with similar state signatures are even considered as candidates, and channels that are almost-always-on/off can be excluded by an \textit{active-ratio} gate.

  \item \textit{Proportionality refinement and scale estimation.}
  Within a signature bucket we require channels to be mutually proportional under probe summaries, using a robust median-ratio estimate and a relative error tolerance.
  This rejects accidental signature collisions that can occur in clean models.

  \item \textit{Weight-level safeguards.}
  In graph-consistent recovery, we additionally require that candidate clusters are consistent with weight proportionality (within a tight tolerance), and we explicitly augment clusters with exact weight duplicates when needed for \texttt{nso\_clique}-style recovery.

  \item \textit{Fan-out aware drop.}
  Dropping a redundant set is only permitted when its merged outgoing effect is (near-)zero \textit{for all downstream linear consumers} (intersection across \texttt{fan-out}).
  This prevents ``local'' pruning decisions that would silently change behavior along a different branch.
\end{packeditemize}
These checks make false positive compaction on clean models unlikely: a clean network needs to exhibit repeated activation signatures \texttt{and} tight proportionality across many probes \texttt{and} consistent consumer-side evidence to trigger pruning.

\heading{Graph-consistent application further prevents accidental damage.}
Even when redundancy is detected, our update rule is function-equivalent \textit{by construction}:
once an edge transform $M$ is inferred, we rewrite \textit{all} downstream linear consumers (\texttt{fan-out}) with $W \leftarrow W M$ (Eq.~\ref{eq:consumer-rewrite}). We also enforce layout compatibility at residual \texttt{add} merges (synchronize both operands to a shared layout) and propagate block-diagonal transforms through channel \texttt{cat}. Finally, we are conservative about operator coverage: e.g., we do not propose output pruning/merges for grouped/depthwise convolution \textit{producers} and treat unsupported channel-mixing modules as barriers.

\heading{Sanity checks (defensive execution).}
Recovery is run in eval mode with deterministic probes.
The implementation includes a \textit{sanity\_check} option: after applying recovery, we execute the model on the probe inputs to ensure the rewritten graph is valid (graph-consistent path) and, in the legacy pipeline, we additionally measure a pre/post output delta on a probe subset and warn if it exceeds a tolerance.
These checks are designed to catch unsafe rewrites early (crashes, obvious functional drift) rather than to encourage aggressive compaction.

\heading{Empirical evidence: “no-attack” false positive evaluation.}
We also directly measure false positives by running recovery on clean (non-attacked) watermarked models and reporting structural change, accuracy gap, and certificate outcomes, as shown in Tables~\ref{tab:fp_densenet},~\ref{tab:fp_efficientnet}, and~\ref{tab:fp_inception}.
In our current sanity run on MNIST with ResNet-18 across all considered watermark schemes, recovery showed:
(i) \texttt{nso\_zero} parameter pruning and \texttt{nso\_zero} Conv/Linear shape changes in all tested cases,
(ii) \(\Delta_{\mathrm{acc}}(\mathrm{clean},\mathrm{recovered}) = 0\) in all tested cases, and
(iii) a 100\% pass rate for the reference-equivalence certificate on watermarked layers (60/60 verified layers).
Minor fluctuations in similarity on some schemes can occur due to extractor noise, but when the certificate passes, reference-equivalence is implied on the watermarked weights (under the allowed symmetries) and similarity is treated as recorded (\S\ref{sec:system-overview}).

\begin{table}[t]
\centering
\caption{\textbf{False positive evaluation.} DenseNet recovery on clean (non-attacked) watermarked models. All test cases show zero structural false positives (no parameter pruning, no layer shape changes), perfect accuracy preservation ($\Delta_{\mathrm{acc}} = 0$), and 100\% Tier-1 reference-equivalence certificate pass rate. Watermark similarity gaps are minimal (within noise tolerance).}

\label{tab:fp_densenet}
\vspace{5pt}
\begin{threeparttable}

\resizebox{\linewidth}{!}{%
\begin{tabular}{c|cc|cc|l}
\toprule
\multirow{2}{*}{\textbf{Watermark}} & \multirow{2}{*}{\textbf{P-FPR}} & \multirow{2}{*}{\textbf{L-FP}} &  \multirow{2}{*}{\textbf{\textbf{$\Delta_{\mathrm{sim}}$}}} &  \multirow{2}{*}{\textbf{$\Delta_{\mathrm{acc}}$}} & \multicolumn{1}{c}{\textbf{ Tier-1}} \\
& &  &  &  & \multicolumn{1}{c}{\textbf{Pass}} \\
\midrule
uchida17    & \cellcolor{blue!10} $0.0000$ & \cellcolor{blue!10} $0/121$ & \cellcolor{pink!25} $0.0000$ & \cellcolor{pink!25} $0.0000$  & \cmark\,   ($2/2$) \\
greedy      & \cellcolor{blue!10} $0.0000$ & \cellcolor{blue!10} $0/121$ & \cellcolor{pink!25} $0.0000$ & \cellcolor{pink!25} $0.0000$  & \cmark\,   ($23/23$) \\
lottery     & \cellcolor{blue!10} $0.0000$ & \cellcolor{blue!10} $0/121$ & \cellcolor{pink!25} $0.0000$ & \cellcolor{pink!25} $0.0000$  & \cmark\,   ($121/121$) \\
ipr\_gan     & \cellcolor{blue!10} $0.0000$ & \cellcolor{blue!10} $0/121$ & \cellcolor{pink!25} $0.0000$ & \cellcolor{pink!25} $0.0000$  & \cmark\,   ($1/1$) \\
riga21      & \cellcolor{blue!10} $0.0000$ & \cellcolor{blue!10} $0/121$ & \cellcolor{pink!25} $0.0156$ & \cellcolor{pink!25} $0.0000$  & \cmark\,   ($2/2$) \\
ipr\_ic      & \cellcolor{blue!10} $0.0000$ & \cellcolor{blue!10} $0/121$ & \cellcolor{pink!25} $0.0000$ & \cellcolor{pink!25} $0.0000$  & \cmark\,   ($2/2$) \\
deepsigns   & \cellcolor{blue!10} $0.0000$ & \cellcolor{blue!10} $0/121$ & \cellcolor{pink!25} $0.0000$ & \cellcolor{pink!25} $0.0000$  & \cmark\,   ($8/8$) \\
deepipr     & \cellcolor{blue!10} $0.0000$ & \cellcolor{blue!10} $0/121$ & \cellcolor{pink!25} $0.0000$ & \cellcolor{pink!25} $0.0000$  & \cmark\,   ($2/2$) \\
passport\_aware & \cellcolor{blue!10} $0.0000$ & \cellcolor{blue!10} $0/121$ & \cellcolor{pink!25} $0.0156$ & \cellcolor{pink!25} $0.0000$  & \cmark\,   ($121/121$) \\
\bottomrule
\end{tabular}
}
    \begin{tablenotes}
    \scriptsize
    \item \textbf{P-FPR}: parameter false positive rate;
    \item \textbf{L-FP}: layer false positive, listed in \textit{changed / total};
    \item \textbf{$\Delta_{\mathrm{acc}}$} for accuracy gap while \textbf{$\Delta_{\mathrm{sim}}$} for watermark similarity gap.
    \end{tablenotes}
\end{threeparttable}
\end{table}

\begin{table}[t]
\centering
\caption{\textbf{False positive evaluation.} EfficientNet recovery on clean (non-attacked) watermarked models. All test cases show zero structural false positives (no parameter pruning, no layer shape changes), perfect accuracy preservation ($\Delta_{\mathrm{acc}} = 0$), and 100\% Tier-1 reference-equivalence certificate pass rate. Watermark similarity gaps are minimal (within noise tolerance).}

\label{tab:fp_efficientnet}
\vspace{5pt}
\begin{threeparttable}

\resizebox{\linewidth}{!}{%
\begin{tabular}{c|cc|cc|l}
\toprule
\multirow{2}{*}{\textbf{Watermark}} & \multirow{2}{*}{\textbf{P-FPR}} & \multirow{2}{*}{\textbf{L-FP}} &  \multirow{2}{*}{\textbf{\textbf{$\Delta_{\mathrm{sim}}$}}} &  \multirow{2}{*}{\textbf{$\Delta_{\mathrm{acc}}$}} & \multicolumn{1}{c}{\textbf{ Tier-1}} \\
& &  &  &  & \multicolumn{1}{c}{\textbf{Pass}} \\
\midrule
uchida17    & \cellcolor{blue!10} $0.0000$ & \cellcolor{blue!10} $0/82$ & \cellcolor{pink!25} $0.0000$ & \cellcolor{pink!25} $0.0000$  & \cmark\,   ($6/6$) \\
greedy      & \cellcolor{blue!10} $0.0000$ & \cellcolor{blue!10} $0/82$ & \cellcolor{pink!25} $0.0000$ & \cellcolor{pink!25} $0.0000$  & \cmark\,   ($14/14$) \\
lottery     & \cellcolor{blue!10} $0.0000$ & \cellcolor{blue!10} $0/82$ & \cellcolor{pink!25} $0.0391$ & \cellcolor{pink!25} $0.0000$  & \cmark\,   ($33/33$) \\
ipr\_gan     & \cellcolor{blue!10} $0.0000$ & \cellcolor{blue!10} $0/82$ & \cellcolor{pink!25} $0.0000$ & \cellcolor{pink!25} $0.0000$  & \cmark\,   ($1/1$) \\
riga21      & \cellcolor{blue!10} $0.0000$ & \cellcolor{blue!10} $0/82$ & \cellcolor{pink!25} $0.0000$ & \cellcolor{pink!25} $0.0000$  & \cmark\,   ($6/6$) \\
ipr\_ic      & \cellcolor{blue!10} $0.0000$ & \cellcolor{blue!10} $0/82$ & \cellcolor{pink!25} $0.0000$ & \cellcolor{pink!25} $0.0000$  & \cmark\,   ($6/6$) \\
deepsigns   & \cellcolor{blue!10} $0.0000$ & \cellcolor{blue!10} $0/82$ & \cellcolor{pink!25} $0.0000$ & \cellcolor{pink!25} $0.0000$  & \cmark\,   ($8/8$) \\
deepipr     & \cellcolor{blue!10} $0.0000$ & \cellcolor{blue!10} $0/82$ & \cellcolor{pink!25} $0.0000$ & \cellcolor{pink!25} $0.0000$  & \cmark\,   ($6/6$) \\
passport\_aware & \cellcolor{blue!10} $0.0000$ & \cellcolor{blue!10} $0/82$ & \cellcolor{pink!25} $0.0000$ & \cellcolor{pink!25} $0.0000$  & \cmark\,   ($82/82$) \\
\bottomrule
\end{tabular}
}
    \begin{tablenotes}
    \scriptsize
    \item \textbf{P-FPR}: parameter false positive rate;
    \item \textbf{L-FP}: layer false positive, listed in \textit{changed / total};
    \item \textbf{$\Delta_{\mathrm{acc}}$} for accuracy gap while \textbf{$\Delta_{\mathrm{sim}}$} for watermark similarity gap.
    \end{tablenotes}
\end{threeparttable}
\end{table}

\begin{table}[t]
\centering
\caption{\textbf{False positive evaluation.} Inception recovery on clean (non-attacked) watermarked models. All test cases show zero structural false positives (no parameter pruning, no layer shape changes), perfect accuracy preservation ($\Delta_{\mathrm{acc}} = 0$), and 100\% Tier-1 reference-equivalence certificate pass rate. Watermark similarity gaps are minimal (within noise tolerance).}

\label{tab:fp_inception}
\vspace{5pt}
\begin{threeparttable}

\resizebox{\linewidth}{!}{%
\begin{tabular}{c|cc|cc|l}
\toprule
\multirow{2}{*}{\textbf{Watermark}} & \multirow{2}{*}{\textbf{P-FPR}} & \multirow{2}{*}{\textbf{L-FP}} &  \multirow{2}{*}{\textbf{\textbf{$\Delta_{\mathrm{sim}}$}}} &  \multirow{2}{*}{\textbf{$\Delta_{\mathrm{acc}}$}} & \multicolumn{1}{c}{\textbf{ Tier-1}} \\
& &  &  &  & \multicolumn{1}{c}{\textbf{Pass}} \\
\midrule
uchida17    & \cellcolor{blue!10} $0.0000$ & \cellcolor{blue!10} $0/95$ & \cellcolor{pink!25} $0.0000$ & \cellcolor{pink!25} $0.0000$  & \cmark\,   ($3/3$) \\
greedy      & \cellcolor{blue!10} $0.0000$ & \cellcolor{blue!10} $0/95$ & \cellcolor{pink!25} $0.0000$ & \cellcolor{pink!25} $0.0000$  & \cmark\,   ($29/29$) \\
lottery     & \cellcolor{blue!10} $0.0000$ & \cellcolor{blue!10} $0/95$ & \cellcolor{pink!25} $0.0000$ & \cellcolor{pink!25} $0.0000$  & \cmark\,   ($95/95$) \\
ipr\_gan     & \cellcolor{blue!10} $0.0000$ & \cellcolor{blue!10} $0/95$ & \cellcolor{pink!25} $0.0000$ & \cellcolor{pink!25} $0.0000$  & \cmark\,   ($1/1$) \\
riga21      & \cellcolor{blue!10} $0.0000$ & \cellcolor{blue!10} $0/95$ & \cellcolor{pink!25} $0.0000$ & \cellcolor{pink!25} $0.0000$  & \cmark\,   ($3/3$) \\
ipr\_ic      & \cellcolor{blue!10} $0.0000$ & \cellcolor{blue!10} $0/95$ & \cellcolor{pink!25} $0.0000$ & \cellcolor{pink!25} $0.0000$  & \cmark\,   ($3/3$) \\
deepsigns   & \cellcolor{blue!10} $0.0000$ & \cellcolor{blue!10} $0/95$ & \cellcolor{pink!25} $0.0000$ & \cellcolor{pink!25} $0.0000$  & \cmark\,   ($8/8$) \\
deepipr     & \cellcolor{blue!10} $0.0000$ & \cellcolor{blue!10} $0/95$ & \cellcolor{pink!25} $0.0000$ & \cellcolor{pink!25} $0.0000$  & \cmark\,   ($3/3$) \\
passport\_aware & \cellcolor{blue!10} $0.0000$ & \cellcolor{blue!10} $0/95$ & \cellcolor{pink!25} $0.0000$ & \cellcolor{pink!25} $0.0000$  & \cmark\,   ($95/95$) \\
\bottomrule
\end{tabular}
}
    \begin{tablenotes}
    \scriptsize
    \item \textbf{P-FPR}: parameter false positive rate;
    \item \textbf{L-FP}: layer false positive, listed in \textit{changed / total};
    \item \textbf{$\Delta_{\mathrm{acc}}$} for accuracy gap while \textbf{$\Delta_{\mathrm{sim}}$} for watermark similarity gap.
    \end{tablenotes}
\end{threeparttable}
\end{table}

\section{Additional Discussion on Original NSO's Defense Settings}
\label{app:nso_defense_settings}

This section expands the comparison with Yan et al.~\cite{yan2023} by detailing their preliminary defense settings (\S7.4) and clarifying why those settings do not directly provide a reliable pathway to restore watermark-verifiable parameterizations under modern, graph-consistent NSO attacks.

\heading{NSO context: function preservation versus structural indices.}
Yan et al.~\cite{yan2023} formalize NSO as a \textit{function-equivalent} obfuscation: the attacker injects redundancy (dummy neurons/channels) and rewires internal representations while maintaining the input--output function.
Their attack design incorporates camouflaging operations that exploit invariances of common architectures, including (i) permuting channels/neuron orderings and (ii) applying compensating per-channel scaling that preserves the composed mapping but blurs statistical fingerprints.
They also discuss structural edits such as kernel expansion, which can further mask where redundancy is injected.
These transformations preserve task utility, yet they break a verifier premise of many white-box watermark schemes: the extractor typically reads a message from a set of indices/locations that are assumed to remain stable across deployment edits.

\heading{Defense settings in \S7.4 of \cite{yan2023}.}
The ``preliminary defense'' (\S7.4 of \cite{yan2023}) is evaluated under three defender knowledge levels:
\begin{packeditemize}
    \item \textit{Partially knowledgeable.} The defender has limited data samples and attempts to detect and remove suspicious neurons. This setting captures practical constraints but offers weak guidance for restoring an aligned layout.
    \item \textit{Skilled.} The defender has access to full training data distribution (or equivalent in-distribution samples), improving anomaly signals but lacking a structural reference for mapping post-attack channels back to the expected indices.
    \item \textit{Fully knowledgeable.} The defender additionally has the original (pre-attack) model, enabling direct alignment and recovery guided by a reference parameterization.
\end{packeditemize}
A central observation is the gap between ``removing suspicious units'' and ``recovering a watermark-verifiable parameterization'':
without a reference model, elimination-based procedures can remain essentially non-recovering for extraction (e.g., extraction error close to chance), while stronger recovery is mainly feasible in the fully knowledgeable setting.
Yan et al.\ explicitly leave more effective dummy-neuron elimination and de-obfuscation as future work.

\begin{table*}[!t]
    \centering
    \caption{Watermark similarity restoration for \textit{ResNet-18} under NSO \textcolor{magenta}{\textbf{\texttt{add-ops}}} attacks (\textcolor{magenta}{\textbf{$0.2$}} attack ratio) \textcolor{brown!80}{without permutation or scaling}. 
    Tier-1 reference-equivalence verification passes (\cmark) in all runs. Tier-1 watermark similarity of the recovered models is the primary outcome, while Tier-2 similarity is the conservative fallback under the assumption that Tier-1 failed.
    }
    \label{tab:nso_recovery_without_scale_02}
    \vspace{5pt}
    \renewcommand{\arraystretch}{1.1}
    \resizebox{\linewidth}{!}{
    \begin{threeparttable}
    \begin{tabular}{c | c | c | c | ccccc }
    \toprule
    & \multirow{3}{*}{\textbf{Watermark}} &
    \multirow{3}{*}{\makecell{\textbf{Clean}\\ \textbf{Sim.}}}  &
    \multirow{3}{*}{\textbf{\makecell{Tier-1\\Pass}}} &
    \multicolumn{5}{c}{\textbf{Attack Sim.$\rightarrow$ Recovery Tier-1 Sim. {\color{brown!80}(Tier-2 Sim.)}}}\\

    \cmidrule(lr){5-9}
     & & & &
     \makecell{\textbf{\texttt{nso\_zero}}} &
     \makecell{\textbf{\texttt{nso\_clique}}} &
     \multicolumn{1}{c}{\textbf{\texttt{nso\_split}}} &
     \makecell{\textbf{\texttt{mix-opseq}}} &
     \makecell{\textbf{\texttt{mix-opseq}}\\\textbf{\texttt{(per-merge-group)}}} 
     \\
    \midrule

    \multirow{9}{*}{\rotatebox{90}{\makecell{ResNet-18}}}
    & uchida17~\cite{uchida2017embedding} &\cellcolor{blue!9} 1.0000 & \cmark
      & \cellcolor{pink!25} 0.5703$\rightarrow$1.0000 {\color{brown!80}(1.0000)} & \cellcolor{pink!25} 0.5312$\rightarrow$1.0000 {\color{brown!80}(1.0000)} & \cellcolor{pink!25} 0.4844$\rightarrow$1.0000 {\color{brown!80}(0.5781)} & \cellcolor{pink!27} 0.4531$\rightarrow$1.0000 {\color{brown!80}(0.6859)} & \cellcolor{pink!27} 0.6094$\rightarrow$1.0000 {\color{brown!80}(1.0000)}  \\
      
    & greedy~\cite{liu2021watermarking} &\cellcolor{blue!9} 1.0000& \cmark
      & \cellcolor{pink!25} 0.5000$\rightarrow$1.0000 {\color{brown!80}(0.5078)}& \cellcolor{pink!25} 0.4844$\rightarrow$1.0000 {\color{brown!80}(0.5078)}& \cellcolor{pink!25} 0.4609$\rightarrow$1.0000 {\color{brown!80}(0.5453)} & \cellcolor{pink!27}  0.5703$\rightarrow$1.0000 {\color{brown!80}(0.5688)} & \cellcolor{pink!27} 0.5000$\rightarrow$1.0000 {\color{brown!80}(0.6078)}\\
      
    & lottery~\cite{chen2021you} &\cellcolor{blue!9} 1.0000& \cmark
      & \cellcolor{pink!25} 0.5547$\rightarrow$1.0000 {\color{brown!80}(0.5708)} & \cellcolor{pink!25} 0.5000$\rightarrow$1.0000 {\color{brown!80}(0.5078)} & \cellcolor{pink!25} 0.5312$\rightarrow$1.0000 {\color{brown!80}(0.5516)} & \cellcolor{pink!27} 0.4688$\rightarrow$1.0000 {\color{brown!80}(0.5078)} & \cellcolor{pink!27} 0.4609$\rightarrow$1.0000 {\color{brown!80}(0.4844)} \\
      
    & ipr\_gan~\cite{ong2021protecting} &\cellcolor{blue!9} 1.0000 & \cmark
      & \cellcolor{pink!25} 0.5703$\rightarrow$1.0000 {\color{brown!80}(1.0000)} & \cellcolor{pink!25} 0.5703$\rightarrow$1.0000 {\color{brown!80}(1.0000)} & \cellcolor{pink!25} 0.5625$\rightarrow$1.0000 {\color{brown!80}(0.6250)} & \cellcolor{pink!27} 0.5234$\rightarrow$1.0000 {\color{brown!80}(0.7031)} & \cellcolor{pink!27} 0.5469$\rightarrow$1.0000 {\color{brown!80}(1.0000)} \\
      
    & riga21~\cite{wang2021riga} & \cellcolor{blue!9} 0.9766 & \cmark
      & \cellcolor{pink!25} 0.4531$\rightarrow$0.9766 {\color{brown!80}(0.9688)} & \cellcolor{pink!25} 0.4531$\rightarrow$0.9766 {\color{brown!80}(0.9688)} & \cellcolor{pink!25} 0.4531$\rightarrow$0.9766 {\color{brown!80}(0.9688)} & \cellcolor{pink!27} 0.4531$\rightarrow$0.9766 {\color{brown!80}(0.9688)} & \cellcolor{pink!27} 0.4531$\rightarrow$0.9766 {\color{brown!80}(0.9688)} \\
      
    & ipr\_ic~\cite{lim2022protect} &\cellcolor{blue!9} 1.0000 & \cmark
      & \cellcolor{pink!25} 0.5156$\rightarrow$1.0000 {\color{brown!80}(0.9297)} & \cellcolor{pink!25} 0.4766$\rightarrow$1.0000 {\color{brown!80}(0.9297)} & \cellcolor{pink!25} 0.5156$\rightarrow$1.0000 {\color{brown!80}(0.5547)} & \cellcolor{pink!27} 0.5234$\rightarrow$1.0000 {\color{brown!80}(0.6484)} & \cellcolor{pink!27} 0.4141$\rightarrow$1.0000 {\color{brown!80}(0.5547)} \\
      
    & deepsigns~\cite{darvish2019deepsigns} &\cellcolor{blue!9} 1.0000 &  \cmark
      & \cellcolor{pink!25} 0.5547$\rightarrow$1.0000 {\color{brown!80}(1.0000)} & \cellcolor{pink!25} 0.5547$\rightarrow$1.0000 {\color{brown!80}(1.0000)} & \cellcolor{pink!25} 0.5547$\rightarrow$1.0000 {\color{brown!80}(0.5946)} & \cellcolor{pink!27} 0.4844$\rightarrow$1.0000 {\color{brown!80}(0.6250)} & \cellcolor{pink!27} 0.5391$\rightarrow$1.0000 {\color{brown!80}(1.0000)} \\
      
    & deepipr~\cite{fan2021deepipr} &\cellcolor{blue!9} 1.0000 & \cmark
      & \cellcolor{pink!25} 0.4062$\rightarrow$1.0000 {\color{brown!80}(1.0000)} & \cellcolor{pink!25} 0.4062$\rightarrow$1.0000 {\color{brown!80}(1.0000)} & \cellcolor{pink!25} 0.4062$\rightarrow$1.0000 {\color{brown!80}(0.5781)} & \cellcolor{pink!27} 0.4062$\rightarrow$1.0000 {\color{brown!80}(0.6406)} & \cellcolor{pink!27} 0.4062$\rightarrow$1.0000 {\color{brown!80}(0.7812)} \\
      
    & passport\_aware~\cite{zhang2020passport} &\cellcolor{blue!9} 1.0000 &  \cmark
      & \cellcolor{pink!25} 0.5703$\rightarrow$1.0000 {\color{brown!80}(1.0000)} & \cellcolor{pink!25} 0.5703$\rightarrow$1.0000 {\color{brown!80}(1.0000)} & \cellcolor{pink!25} 0.5625$\rightarrow$1.0000 {\color{brown!80}(0.5870)} & \cellcolor{pink!27} 0.5312$\rightarrow$1.0000 {\color{brown!80}(0.5938)} & \cellcolor{pink!27} 0.5000$\rightarrow$1.0000 {\color{brown}(0.8359)} \\

    \bottomrule
    \end{tabular}

    \begin{tablenotes}
    \item \textbf{Tier-1 pass} indicates that the Tier-1 reference-equivalence certificate over watermarked layers passes for all runs in the suite for that model/watermark. We use \cmark to denote a 100\% match rate.
    \end{tablenotes}

    \end{threeparttable}
    }
\end{table*}

\heading{Single primitive evaluation v.s. compositional NSO.}
In \S7.4 of~\cite{yan2023}, dummy neurons generated by \texttt{nso\_zero}, \texttt{nso\_clique}, and \texttt{nso\_split} are injected \textit{independently}, i.e., defenses are evaluated against individual primitive families in isolation. This is informative but under-approximates a realistic adaptive attacker who composes primitives to maximize ambiguity while remaining function-equivalent.

Our efforts include \texttt{mix-opseq}, where a primitive sequence is sampled and applied sequentially within eligible layers, and \texttt{mix-opseq (per-merge-group)}, which reflects graph constraints while maximizing heterogeneity across residual/\texttt{cat} structure.
Compositional NSOs cover redundancy patterns (e.g., clique membership and split duplicates) within a layer and across graph boundaries, undermining defenses that implicitly assume independently identifiable primitives.

\heading{Why the elimination of~\cite{yan2023} is not sufficient for compatible recovery.}
Eliminating anomalous channels does not by itself restore a canonical layout compatible with extractors, for three reasons.
First, permutation/scaling camouflage breaks index semantics: after NSO, the correspondence between post-attack indices and the extractor’s assumed indices may be arbitrary (up to permutation), and values may be rescaled in compensating ways.
Second, recovery must be \textit{global}: any change to a producer’s channel basis must be propagated to \textit{all} downstream consumers; otherwise the defense either fails to remove obfuscation or breaks functional equivalence.
Third, modern CNN graphs impose merge constraints: residual \texttt{add} requires aligned channel semantics across branches, while channel \texttt{cat} allows independent layouts across branches but produces mixed consumers downstream.
Therefore, an NSO defense must couple redundancy identification with graph-consistent rewriting.



\section{End-to-end Recovery Time under Larger Attack Ratio (i.e., $\rho=0.5$)}
\label{sec:tab_nso_0.5}

To avoid redundancy in the main text, we report the corresponding end-to-end results under a larger injection ratio $\rho=0.5$ in Tables~\ref{tab:nso_recovery_suite_add_05} and~\ref{tab:nso_recovery_suite_cat_05}.
The results mirror the $\rho=0.2$ suite: NSO remains function-equivalent, and \textsc{Canon} continues to restore verifiability across architectures and watermarking schemes under stronger structural obfuscation.

\section{Supplementary Experiments: Cifar100}
\label{sec:wxperiments_cifar100}

\begin{figure}[!t]
    \centering
    \subfigure[\shortstack{\small MNIST}]{
    \begin{minipage}[t]{0.225\textwidth}
    \centering
    \includegraphics[width=1.8in]{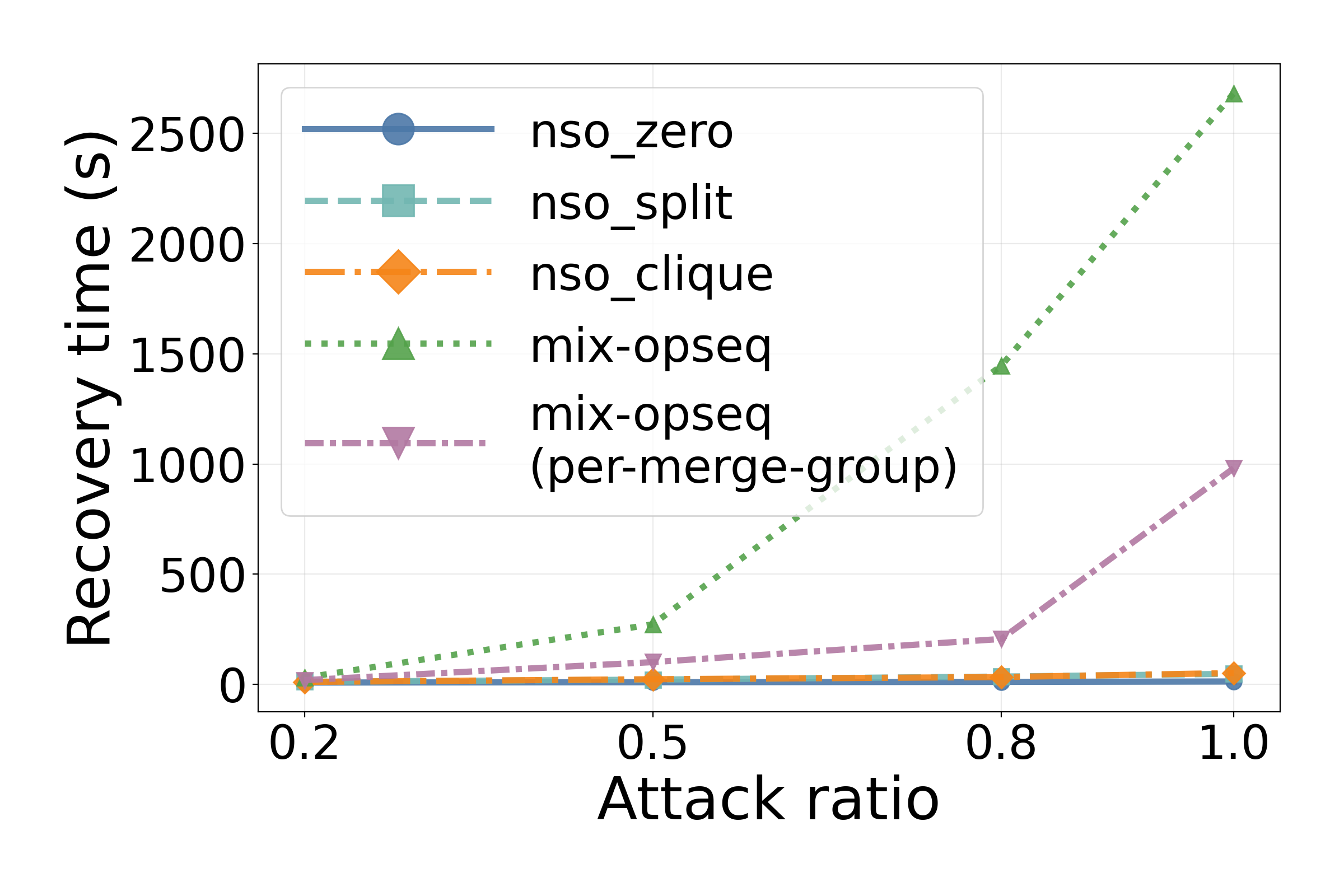}
    \end{minipage}
    \label{recover_time_mnist}
    }
    \subfigure[\shortstack{\small CIFAR-100}]{
    \begin{minipage}[t]{0.225\textwidth}
    \centering
    \includegraphics[width=1.8in]{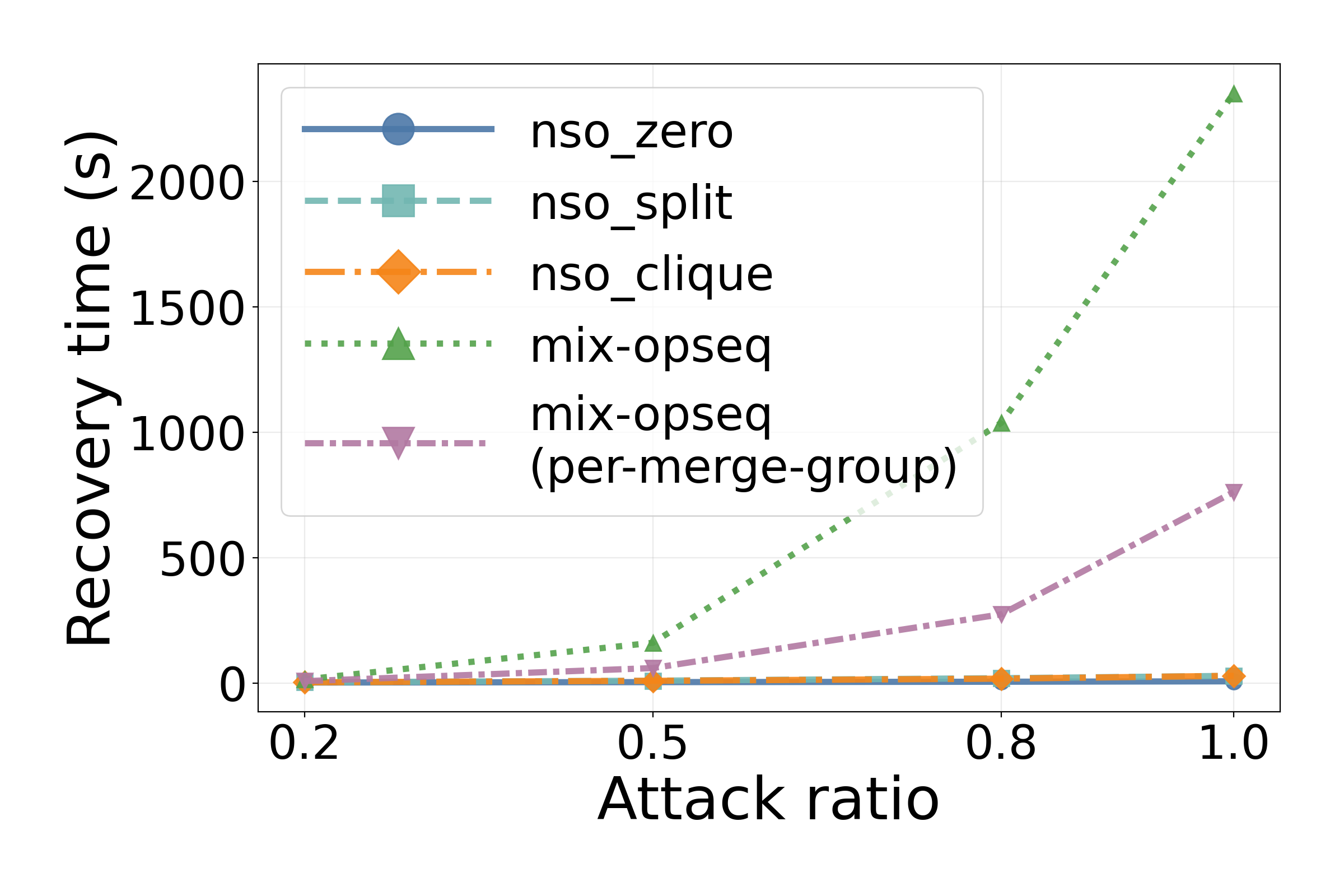}
    \end{minipage}
    \label{recover_time_cifar100}
    }
    \caption{\textsc{Canon} recovery time on ResNet-18 under MNIST and CIFAR-100 dataset as the NSO injection ratio $\rho$ increases.
    Curves correspond to five attack variants (\texttt{nso\_zero}, \texttt{nso\_split}, \texttt{nso\_clique}, \texttt{mix-opseq}, \texttt{mix-opseq (per-merge-group)}).
    Higher $\rho$ increases the number of injected channels and thus widens the attacked network, which raises probe-time activation collection and amplifies the cost of redundancy clustering and graph-consistent consumer rewrites; the effect is most pronounced for compositional \texttt{mix-opseq} attacks.
    }
    \label{fig:runtime_resnet18_ratio}
\end{figure}

\begin{table*}[t]
    \centering
    \caption{Watermark similarity restoration for \textit{ResNet-18} under \textbf{CIFAR-100} dataset. Tier-1  verification passes (\cmark) in all runs and is used as the primary outcome, while Tier-2 serves as a conservative fallback.
    }
    \label{tab:nso_recovery_suite_add_05}
    \vspace{5pt}
    
    \renewcommand{\arraystretch}{1.1}
    \resizebox{\linewidth}{!}{
    \begin{threeparttable}
    \begin{tabular}{c |c |c | ccc |c| ccccc}
    \toprule
    & \multirow{3}{*}{\textbf{\makecell{Attack\\Ratio}}} &
    \multirow{3}{*}{\textbf{Watermark}} &
    \multicolumn{3}{c|}{\textbf{Clean / Functional Drift}} &
    \multirow{3}{*}{\textbf{\makecell{Tier-1\\Pass}}} &
    \multicolumn{5}{c}{\textbf{Attack Sim.$\rightarrow$ Recovery Tier-1 Sim. {\color{gray!80}(Tier-2 Sim.)}}}
    \\

    \cmidrule(lr){4-6} 
    \cmidrule(lr){8-12}
     & & &
     \makecell{\textbf{Clean}\\\textbf{Sim.}} &
     \makecell{$\boldsymbol{\Delta_{\mathrm{acc}}}$\\\textbf{(att.)}} &
     \makecell{$\boldsymbol{\Delta_{\mathrm{acc}}}$\\\textbf{(rec.)}} &
     &
     \makecell{\textbf{\texttt{zero}}} &
     \makecell{\textbf{\texttt{clique}}} &
     \makecell{\textbf{\texttt{split}}} &
     \makecell{\textbf{\texttt{mix-opseq}}} &
     \makecell{\textbf{\texttt{mix-opseq}}\\\textbf{\texttt{(per-merge-group)}}} 
     \\
    \midrule

        \multirow{36}{*}{\rotatebox{90}{\makecell{ResNet-18 + CIFAR-100}}}
    & \multirow{9}{*}{0.2} & uchida17~\cite{uchida2017embedding} &\cellcolor{blue!9} 1.0000 &\cellcolor{blue!10} 0.0000 &\cellcolor{blue!10} 0.0000 & \cmark
      & \cellcolor{pink!25} 0.4844$\rightarrow$1.0000 {\color{gray!80}(0.5391)} & \cellcolor{pink!25} 0.4375$\rightarrow$1.0000 {\color{gray!80}(0.5391)} & \cellcolor{pink!25} 0.4766$\rightarrow$1.0000 {\color{gray!80}(0.4766)} & \cellcolor{pink!27} 0.4219$\rightarrow$1.0000 {\color{gray!80}(0.5156)} & \cellcolor{pink!27} 0.4297$\rightarrow$1.0000 {\color{gray!80}(0.5312)} \\
      
   & & greedy~\cite{liu2021watermarking} &\cellcolor{blue!9} 1.0000&\cellcolor{blue!10} 0.0000 &\cellcolor{blue!10} 0.0000 & \cmark
      & \cellcolor{pink!25} 0.4453$\rightarrow$1.0000 {\color{gray!80}(0.3828)}& \cellcolor{pink!25} 0.4219$\rightarrow$1.0000 {\color{gray!80}(0.3828)}& \cellcolor{pink!25} 0.4688$\rightarrow$1.0000 {\color{gray!80}(0.4297)}& \cellcolor{pink!27} 0.4297$\rightarrow$1.0000 {\color{gray!80}(0.5400)}& \cellcolor{pink!27} 0.4375$\rightarrow$1.0000 {\color{gray!80}(0.4844)} \\
      
   & & lottery~\cite{chen2021you} &\cellcolor{blue!9} 1.0000&\cellcolor{blue!10} 0.0000 &\cellcolor{blue!10} 0.0000 & \cmark
      & \cellcolor{pink!25} 0.5000$\rightarrow$1.0000 {\color{gray!80}(0.5703)} & \cellcolor{pink!25} 0.4922$\rightarrow$1.0000 {\color{gray!80}(0.5703)} & \cellcolor{pink!25} 0.5000$\rightarrow$1.0000 {\color{gray!80}(0.5156)} & \cellcolor{pink!27} 0.4922$\rightarrow$1.0000 {\color{gray!80}(0.5078)} & \cellcolor{pink!27} 0.5234$\rightarrow$1.0000 {\color{gray!80}(0.5312)}   \\
      
   &  & ipr\_gan~\cite{ong2021protecting} &\cellcolor{blue!9} 1.0000 &\cellcolor{blue!10} 0.0000 &\cellcolor{blue!10} 0.0000 & \cmark
      & \cellcolor{pink!25} 0.4922$\rightarrow$1.0000 {\color{gray!80}(0.5000)} & \cellcolor{pink!25} 0.5859$\rightarrow$1.0000 {\color{gray!80}(0.5000)} & \cellcolor{pink!25} 0.5391$\rightarrow$1.0000 {\color{gray!80}(0.4688)} & \cellcolor{pink!27} 0.5625$\rightarrow$1.0000 {\color{gray!80}(0.5938)} & \cellcolor{pink!27} 0.4609$\rightarrow$1.0000 {\color{gray!80}(0.5312)}   \\
      
   & & riga21~\cite{wang2021riga} & \cellcolor{blue!9} 0.9688 &\cellcolor{blue!10} 0.0000 &\cellcolor{blue!10} 0.0000 & \cmark
      & \cellcolor{pink!25} 0.5547$\rightarrow$0.9688 {\color{gray!80}(0.9844)} & \cellcolor{pink!25} 0.5547$\rightarrow$0.9688 {\color{gray!80}(0.9844)} & \cellcolor{pink!25} 0.5547$\rightarrow$0.9688 {\color{gray!80}(0.9922)} & \cellcolor{pink!27} 0.5547$\rightarrow$0.9688 {\color{gray!80}(0.9766)} & \cellcolor{pink!27} 0.5547$\rightarrow$0.9688 {\color{gray!80}(0.9844)} \\
      
  &  & ipr\_ic~\cite{lim2022protect} &\cellcolor{blue!9} 1.0000 &\cellcolor{blue!10} 0.0000 &\cellcolor{blue!10} 0.0000 & \cmark
      & \cellcolor{pink!25} 0.5234$\rightarrow$1.0000 {\color{gray!80}(0.4844)} & \cellcolor{pink!25} 0.4531$\rightarrow$1.0000 {\color{gray!80}(0.4844)} & \cellcolor{pink!25} 0.5078$\rightarrow$1.0000 {\color{gray!80}(0.4688)} & \cellcolor{pink!27} 0.5156$\rightarrow$1.0000 {\color{gray!80}(0.4609)} & \cellcolor{pink!27} 0.4766$\rightarrow$1.0000 {\color{gray!80}(0.5078)}  \\
      
   & & deepsigns~\cite{darvish2019deepsigns} &\cellcolor{blue!9} 1.0000 &\cellcolor{blue!10} 0.0000 &\cellcolor{blue!10} 0.0000 & \cmark
      & \cellcolor{pink!25} 0.4844$\rightarrow$1.0000 {\color{gray!80}(0.4531)} & \cellcolor{pink!25} 0.4375$\rightarrow$1.0000 {\color{gray!80}(0.4531)} & \cellcolor{pink!25} 0.3750$\rightarrow$1.0000 {\color{gray!80}(0.4688)} & \cellcolor{pink!27} 0.4844$\rightarrow$1.0000 {\color{gray!80}(0.5156)} & \cellcolor{pink!27} 0.4531$\rightarrow$1.0000 {\color{gray!80}(0.4375)}  \\
      
   & & deepipr~\cite{fan2021deepipr} &\cellcolor{blue!9} 1.0000 &\cellcolor{blue!10} 0.0000 &\cellcolor{blue!10} 0.0000 & \cmark
      & \cellcolor{pink!25} 0.4062$\rightarrow$1.0000 {\color{gray!80}(0.5625)} & \cellcolor{pink!25} 0.4062$\rightarrow$1.0000 {\color{gray!80}(0.5625)} & \cellcolor{pink!25} 0.4062$\rightarrow$1.0000 {\color{gray!80}(0.6250)} & \cellcolor{pink!27} 0.4062$\rightarrow$1.0000 {\color{gray!80}(0.5312)} & \cellcolor{pink!27} 0.4062$\rightarrow$1.0000 {\color{gray!80}(0.5469)}  \\
      
  &  & passport\_aware~\cite{zhang2020passport} &\cellcolor{blue!9} 1.0000 &\cellcolor{blue!10} 0.0000 &\cellcolor{blue!10} 0.0000 & \cmark
      & \cellcolor{pink!25} 0.5078$\rightarrow$1.0000 {\color{gray!80}(0.5625)} & \cellcolor{pink!25} 0.4844$\rightarrow$1.0000 {\color{gray!80}(0.5625)} & \cellcolor{pink!25} 0.5391$\rightarrow$1.0000 {\color{gray!80}(0.5078)} & \cellcolor{pink!27} 0.4922$\rightarrow$1.0000 {\color{gray!80}(0.5000)} & \cellcolor{pink!27} 0.4609$\rightarrow$1.0000 {\color{gray!80}(0.4922)} \\

      \cmidrule(lr){2-12}
      
    & \multirow{9}{*}{0.5} & uchida17~\cite{uchida2017embedding} &\cellcolor{blue!9} 1.0000 &\cellcolor{blue!10} 0.0000 &\cellcolor{blue!10} 0.0000 & \cmark
      & \cellcolor{pink!25} 0.5156$\rightarrow$1.0000 {\color{gray!80}(0.5156)} & \cellcolor{pink!25} 0.5156$\rightarrow$1.0000 {\color{gray!80}(0.5156)} & \cellcolor{pink!25} 0.5234$\rightarrow$1.0000 {\color{gray!80}(0.4766)} & \cellcolor{pink!27} 0.4453$\rightarrow$1.0000 {\color{gray!80}(0.4609)} & \cellcolor{pink!27} 0.4531$\rightarrow$1.0000 {\color{gray!80}(0.5469)} \\
      
   & & greedy~\cite{liu2021watermarking} &\cellcolor{blue!9} 1.0000&\cellcolor{blue!10} 0.0000 &\cellcolor{blue!10} 0.0000 & \cmark
      & \cellcolor{pink!25} 0.5469$\rightarrow$1.0000 {\color{gray!80}(0.4688)}& \cellcolor{pink!25} 0.5547$\rightarrow$1.0000 {\color{gray!80}(0.4688)}& \cellcolor{pink!25} 0.5703$\rightarrow$1.0000 {\color{gray!80}(0.5156)}& \cellcolor{pink!27} 0.5391$\rightarrow$1.0000 {\color{gray!80}(0.5703)}& \cellcolor{pink!27} 0.5703$\rightarrow$1.0000 {\color{gray!80}(0.4688)} \\
      
   & & lottery~\cite{chen2021you} &\cellcolor{blue!9} 1.0000&\cellcolor{blue!10} 0.0000 &\cellcolor{blue!10} 0.0000 & \cmark
      & \cellcolor{pink!25} 0.5469$\rightarrow$1.0000 {\color{gray!80}(0.5078)} & \cellcolor{pink!25} 0.4766$\rightarrow$1.0000 {\color{gray!80}(0.5078)} & \cellcolor{pink!25} 0.5156$\rightarrow$1.0000 {\color{gray!80}(0.5156)} & \cellcolor{pink!27} 0.5547$\rightarrow$1.0000 {\color{gray!80}(0.5547)} & \cellcolor{pink!27} 0.5547$\rightarrow$1.0000 {\color{gray!80}(0.5078)}   \\
      
   &  & ipr\_gan~\cite{ong2021protecting} &\cellcolor{blue!9} 1.0000 &\cellcolor{blue!10} 0.0000 &\cellcolor{blue!10} 0.0000 & \cmark
      & \cellcolor{pink!25} 0.5859$\rightarrow$1.0000 {\color{gray!80}(0.5625)} & \cellcolor{pink!25} 0.5328$\rightarrow$1.0000 {\color{gray!80}(0.5625)} & \cellcolor{pink!25} 0.5547$\rightarrow$1.0000 {\color{gray!80}(0.5312)} & \cellcolor{pink!27} 0.4844$\rightarrow$1.0000 {\color{gray!80}(0.5312)} & \cellcolor{pink!27} 0.5781$\rightarrow$1.0000 {\color{gray!80}(0.5156)}   \\
      
   & & riga21~\cite{wang2021riga} & \cellcolor{blue!9} 0.9688 &\cellcolor{blue!10} 0.0000 &\cellcolor{blue!10} 0.0000 & \cmark
      & \cellcolor{pink!25} 0.5547$\rightarrow$0.9688 {\color{gray!80}(0.9844)} & \cellcolor{pink!25} 0.5547$\rightarrow$0.9688 {\color{gray!80}(0.9844)} & \cellcolor{pink!25} 0.5547$\rightarrow$0.9688 {\color{gray!80}(0.9844)} & \cellcolor{pink!27} 0.5547$\rightarrow$0.9688 {\color{gray!80}(0.9844)} & \cellcolor{pink!27} 0.5547$\rightarrow$0.9688 {\color{gray!80}(0.9766)} \\
      
  &  & ipr\_ic~\cite{lim2022protect} &\cellcolor{blue!9} 1.0000 &\cellcolor{blue!10} 0.0000 &\cellcolor{blue!10} 0.0000 & \cmark
      & \cellcolor{pink!25} 0.4844$\rightarrow$1.0000 {\color{gray!80}(0.4531)} & \cellcolor{pink!25} 0.4922$\rightarrow$1.0000 {\color{gray!80}(0.4531)} & \cellcolor{pink!25} 0.4219$\rightarrow$1.0000 {\color{gray!80}(0.5234)} & \cellcolor{pink!27} 0.5000$\rightarrow$1.0000 {\color{gray!80}(0.5312)} & \cellcolor{pink!27} 0.4688$\rightarrow$1.0000 {\color{gray!80}(0.5078)}  \\
      
   & & deepsigns~\cite{darvish2019deepsigns} &\cellcolor{blue!9} 1.0000 &\cellcolor{blue!10} 0.0000 &\cellcolor{blue!10} 0.0000 & \cmark
      & \cellcolor{pink!25} 0.5234$\rightarrow$1.0000 {\color{gray!80}(0.4844)} & \cellcolor{pink!25} 0.5391$\rightarrow$1.0000 {\color{gray!80}(0.4844)} & \cellcolor{pink!25} 0.4688$\rightarrow$1.0000 {\color{gray!80}(0.5312)} & \cellcolor{pink!27} 0.4922$\rightarrow$1.0000 {\color{gray!80}(0.4844)} & \cellcolor{pink!27} 0.5156$\rightarrow$1.0000 {\color{gray!80}(0.4531)}  \\
      
   & & deepipr~\cite{fan2021deepipr} &\cellcolor{blue!9} 1.0000 &\cellcolor{blue!10} 0.0000 &\cellcolor{blue!10} 0.0000 & \cmark
      & \cellcolor{pink!25} 0.4062$\rightarrow$1.0000 {\color{gray!80}(0.5625)} & \cellcolor{pink!25} 0.4062$\rightarrow$1.0000 {\color{gray!80}(0.5625)} & \cellcolor{pink!25} 0.4062$\rightarrow$1.0000 {\color{gray!80}(0.5625)} & \cellcolor{pink!27} 0.4062$\rightarrow$1.0000 {\color{gray!80}(0.6562)} & \cellcolor{pink!27} 0.4062$\rightarrow$1.0000 {\color{gray!80}(0.5469)}  \\
      
  &  & passport\_aware~\cite{zhang2020passport} &\cellcolor{blue!9} 1.0000 &\cellcolor{blue!10} 0.0000 &\cellcolor{blue!10} 0.0000 & \cmark
      & \cellcolor{pink!25} 0.5781$\rightarrow$1.0000 {\color{gray!80}(0.4844)} & \cellcolor{pink!25} 0.5859$\rightarrow$1.0000 {\color{gray!80}(0.4844)} & \cellcolor{pink!25} 0.5156$\rightarrow$1.0000 {\color{gray!80}(0.5078)} & \cellcolor{pink!27} 0.4922$\rightarrow$1.0000 {\color{gray!80}(0.5000)} & \cellcolor{pink!27} 0.5547$\rightarrow$1.0000 {\color{gray!80}(0.4766)} \\

      \cmidrule(lr){2-12}

      & \multirow{9}{*}{0.8} & uchida17~\cite{uchida2017embedding} &\cellcolor{blue!9} 1.0000 &\cellcolor{blue!10} 0.0000 &\cellcolor{blue!10} 0.0000 & \cmark
      & \cellcolor{pink!25} 0.5625$\rightarrow$1.0000 {\color{gray!80}(0.5391)} & \cellcolor{pink!25} 0.5078$\rightarrow$1.0000 {\color{gray!80}(0.5391)} & \cellcolor{pink!25} 0.4922$\rightarrow$1.0000 {\color{gray!80}(0.5156)} & \cellcolor{pink!27} 0.4375$\rightarrow$1.0000 {\color{gray!80}(0.5469)} & \cellcolor{pink!27} 0.5625$\rightarrow$1.0000 {\color{gray!80}(0.5312)} \\
      
   & & greedy~\cite{liu2021watermarking} &\cellcolor{blue!9} 1.0000&\cellcolor{blue!10} 0.0000 &\cellcolor{blue!10} 0.0000 & \cmark
      & \cellcolor{pink!25} 0.4453$\rightarrow$1.0000 {\color{gray!80}(0.5547)}& \cellcolor{pink!25} 0.4531$\rightarrow$1.0000 {\color{gray!80}(0.5547)}& \cellcolor{pink!25} 0.5625$\rightarrow$1.0000 {\color{gray!80}(0.4922)}& \cellcolor{pink!27} 0.5625$\rightarrow$1.0000 {\color{gray!80}(0.4062)}& \cellcolor{pink!27} 0.4609$\rightarrow$1.0000 {\color{gray!80}(0.5078)} \\
      
   & & lottery~\cite{chen2021you} &\cellcolor{blue!9} 1.0000&\cellcolor{blue!10} 0.0000 &\cellcolor{blue!10} 0.0000 & \cmark
      & \cellcolor{pink!25} 0.4531$\rightarrow$1.0000 {\color{gray!80}(0.5156)} & \cellcolor{pink!25} 0.5078$\rightarrow$1.0000 {\color{gray!80}(0.5156)} & \cellcolor{pink!25} 0.5391$\rightarrow$1.0000 {\color{gray!80}(0.5234)} & \cellcolor{pink!27} 0.5234$\rightarrow$1.0000 {\color{gray!80}(0.4844)} & \cellcolor{pink!27} 0.5312$\rightarrow$1.0000 {\color{gray!80}(0.4922)}   \\
      
   &  & ipr\_gan~\cite{ong2021protecting} &\cellcolor{blue!9} 1.0000 &\cellcolor{blue!10} 0.0000 &\cellcolor{blue!10} 0.0000 & \cmark
              & \cellcolor{pink!25} 0.4609$\rightarrow$1.0000 {\color{gray!80}(0.5312)} & \cellcolor{pink!25} 0.5859$\rightarrow$1.0000 {\color{gray!80}(0.5312)} & \cellcolor{pink!25} 0.5625$\rightarrow$1.0000 {\color{gray!80}(0.5938)} & \cellcolor{pink!27} 0.5312$\rightarrow$1.0000 {\color{gray!80}(0.5781)} & \cellcolor{pink!27} 0.5547$\rightarrow$1.0000 {\color{gray!80}(0.5156)}   \\
      
   & & riga21~\cite{wang2021riga} & \cellcolor{blue!9} 0.9688 &\cellcolor{blue!10} 0.0000 &\cellcolor{blue!10} 0.0000 & \cmark
      & \cellcolor{pink!25} 0.5547$\rightarrow$0.9688 {\color{gray!80}(0.9766)} & \cellcolor{pink!25} 0.5547$\rightarrow$0.9688 {\color{gray!80}(0.9766)} & \cellcolor{pink!25} 0.5547$\rightarrow$0.9688 {\color{gray!80}(0.9688)} & \cellcolor{pink!27} 0.5547$\rightarrow$0.9688 {\color{gray!80}(0.9844)} & \cellcolor{pink!27} 0.5547$\rightarrow$0.9688 {\color{gray!80}(0.9844)} \\
      
  &  & ipr\_ic~\cite{lim2022protect} &\cellcolor{blue!9} 1.0000 &\cellcolor{blue!10} 0.0000 &\cellcolor{blue!10} 0.0000 & \cmark
      & \cellcolor{pink!25} 0.4922$\rightarrow$1.0000 {\color{gray!80}(0.5156)} & \cellcolor{pink!25} 0.5000$\rightarrow$1.0000 {\color{gray!80}(0.5156)} & \cellcolor{pink!25} 0.5234$\rightarrow$1.0000 {\color{gray!80}(0.5156)} & \cellcolor{pink!27} 0.5469$\rightarrow$1.0000 {\color{gray!80}(0.5000)} & \cellcolor{pink!27} 0.4766$\rightarrow$1.0000 {\color{gray!80}(0.4531)}  \\
      
   & & deepsigns~\cite{darvish2019deepsigns} &\cellcolor{blue!9} 1.0000 &\cellcolor{blue!10} 0.0000 &\cellcolor{blue!10} 0.0000 & \cmark
      & \cellcolor{pink!25} 0.5625$\rightarrow$1.0000 {\color{gray!80}(0.5312)} & \cellcolor{pink!25} 0.4688$\rightarrow$1.0000 {\color{gray!80}(0.5312)} & \cellcolor{pink!25} 0.4531$\rightarrow$1.0000 {\color{gray!80}(0.5469)} & \cellcolor{pink!27} 0.4922$\rightarrow$1.0000 {\color{gray!80}(0.4844)} & \cellcolor{pink!27} 0.5625$\rightarrow$1.0000 {\color{gray!80}(0.4531)}  \\
      
   & & deepipr~\cite{fan2021deepipr} &\cellcolor{blue!9} 1.0000 &\cellcolor{blue!10} 0.0000 &\cellcolor{blue!10} 0.0000 & \cmark
      & \cellcolor{pink!25} 0.4062$\rightarrow$1.0000 {\color{gray!80}(0.5312)} & \cellcolor{pink!25} 0.4062$\rightarrow$1.0000 {\color{gray!80}(0.5312)} & \cellcolor{pink!25} 0.4062$\rightarrow$1.0000 {\color{gray!80}(0.5469)} & \cellcolor{pink!27} 0.4062$\rightarrow$1.0000 {\color{gray!80}(0.6250)} & \cellcolor{pink!27} 0.4062$\rightarrow$1.0000 {\color{gray!80}(0.5781)}  \\
      
  &  & passport\_aware~\cite{zhang2020passport} &\cellcolor{blue!9} 1.0000 &\cellcolor{blue!10} 0.0000 &\cellcolor{blue!10} 0.0000 & \cmark
      & \cellcolor{pink!25} 0.5234$\rightarrow$1.0000 {\color{gray!80}(0.5156)} & \cellcolor{pink!25} 0.5391$\rightarrow$1.0000 {\color{gray!80}(0.5156)} & \cellcolor{pink!25} 0.5156$\rightarrow$1.0000 {\color{gray!80}(0.4688)} & \cellcolor{pink!27} 0.5391$\rightarrow$1.0000 {\color{gray!80}(0.4922)} & \cellcolor{pink!27} 0.5312$\rightarrow$1.0000 {\color{gray!80}(0.5234)} \\

      \cmidrule(lr){2-12}

      & \multirow{9}{*}{1.0} & uchida17~\cite{uchida2017embedding} &\cellcolor{blue!9} 1.0000 &\cellcolor{blue!10} 0.0000 &\cellcolor{blue!10} 0.0000 & \cmark
      & \cellcolor{pink!25} 0.5000$\rightarrow$1.0000 {\color{gray!80}(0.4531)} & \cellcolor{pink!25} 0.5078$\rightarrow$1.0000 {\color{gray!80}(0.4531)} & \cellcolor{pink!25} 0.5156$\rightarrow$1.0000 {\color{gray!80}(0.4922)} & \cellcolor{pink!27} 0.4531$\rightarrow$1.0000 {\color{gray!80}(0.5078)} & \cellcolor{pink!27} 0.4844$\rightarrow$1.0000 {\color{gray!80}(0.5234)} \\
      
   & & greedy~\cite{liu2021watermarking} &\cellcolor{blue!9} 1.0000&\cellcolor{blue!10} 0.0000 &\cellcolor{blue!10} 0.0000 & \cmark
      & \cellcolor{pink!25} 0.5547$\rightarrow$1.0000 {\color{gray!80}(0.6016)}& \cellcolor{pink!25} 0.5391$\rightarrow$1.0000 {\color{gray!80}(0.6016)}& \cellcolor{pink!25} 0.5859$\rightarrow$1.0000 {\color{gray!80}(0.5547)}& \cellcolor{pink!27} 0.5574$\rightarrow$1.0000 {\color{gray!80}(0.5469)}& \cellcolor{pink!27} 0.5391$\rightarrow$1.0000 {\color{gray!80}(0.6328)} \\
      
   & & lottery~\cite{chen2021you} &\cellcolor{blue!9} 1.0000&\cellcolor{blue!10} 0.0000 &\cellcolor{blue!10} 0.0000 & \cmark
      & \cellcolor{pink!25} 0.4688$\rightarrow$1.0000 {\color{gray!80}(0.5000)} & \cellcolor{pink!25} 0.5078$\rightarrow$1.0000 {\color{gray!80}(0.5000)} & \cellcolor{pink!25} 0.5625$\rightarrow$1.0000 {\color{gray!80}(0.5703)} & \cellcolor{pink!27} 0.5547$\rightarrow$1.0000 {\color{gray!80}(0.4922)} & \cellcolor{pink!27} 0.5156$\rightarrow$1.0000 {\color{gray!80}(0.4922)}   \\
      
   &  & ipr\_gan~\cite{ong2021protecting} &\cellcolor{blue!9} 1.0000 &\cellcolor{blue!10} 0.0000 &\cellcolor{blue!10} 0.0000 & \cmark
      & \cellcolor{pink!25} 0.5156$\rightarrow$1.0000 {\color{gray!80}(0.5312)} & \cellcolor{pink!25} 0.6328$\rightarrow$1.0000 {\color{gray!80}(0.5312)} & \cellcolor{pink!25} 0.5391$\rightarrow$1.0000 {\color{gray!80}(0.4688)} & \cellcolor{pink!27} 0.4688$\rightarrow$1.0000 {\color{gray!80}(0.4844)} & \cellcolor{pink!27} 0.5703$\rightarrow$1.0000 {\color{gray!80}(0.5312)} \\
      
   & & riga21~\cite{wang2021riga} & \cellcolor{blue!9} 0.9688 &\cellcolor{blue!10} 0.0000 &\cellcolor{blue!10} 0.0000 & \cmark
      & \cellcolor{pink!25} 0.5547$\rightarrow$0.9688 {\color{gray!80}(0.9922)} & \cellcolor{pink!25} 0.5547$\rightarrow$0.9688 {\color{gray!80}(0.4922)} & \cellcolor{pink!25} 0.5547$\rightarrow$0.9688 {\color{gray!80}(0.9844)} & \cellcolor{pink!27} 0.5547$\rightarrow$0.9688 {\color{gray!80}(0.9766)} & \cellcolor{pink!27} 0.5547$\rightarrow$0.9688 {\color{gray!80}(0.9844)} \\
      
  &  & ipr\_ic~\cite{lim2022protect} &\cellcolor{blue!9} 1.0000 &\cellcolor{blue!10} 0.0000 &\cellcolor{blue!10} 0.0000 & \cmark
      & \cellcolor{pink!25} 0.5703$\rightarrow$1.0000 {\color{gray!80}(0.5078)} & \cellcolor{pink!25} 0.5938$\rightarrow$1.0000 {\color{gray!80}(0.5078)} & \cellcolor{pink!25} 0.5000$\rightarrow$1.0000 {\color{gray!80}(0.5078)} & \cellcolor{pink!27} 0.5078$\rightarrow$1.0000 {\color{gray!80}(0.4688)} & \cellcolor{pink!27} 0.4688$\rightarrow$1.0000 {\color{gray!80}(0.4922)}  \\
      
   & & deepsigns~\cite{darvish2019deepsigns} &\cellcolor{blue!9} 1.0000 &\cellcolor{blue!10} 0.0000 &\cellcolor{blue!10} 0.0000 & \cmark
      & \cellcolor{pink!25} 0.5156$\rightarrow$1.0000 {\color{gray!80}(0.5312)} & \cellcolor{pink!25} 0.5703$\rightarrow$1.0000 {\color{gray!80}(0.5312)} & \cellcolor{pink!25} 0.3750$\rightarrow$1.0000 {\color{gray!80}(0.5469)} & \cellcolor{pink!27} 0.4844$\rightarrow$1.0000 {\color{gray!80}(0.5312)} & \cellcolor{pink!27} 0.4453$\rightarrow$1.0000 {\color{gray!80}(0.5469)}  \\
      
   & & deepipr~\cite{fan2021deepipr} &\cellcolor{blue!9} 1.0000 &\cellcolor{blue!10} 0.0000 &\cellcolor{blue!10} 0.0000 & \cmark
      & \cellcolor{pink!25} 0.4062$\rightarrow$1.0000 {\color{gray!80}(0.5938)} & \cellcolor{pink!25} 0.4062$\rightarrow$1.0000 {\color{gray!80}(0.5938)} & \cellcolor{pink!25} 0.4602$\rightarrow$1.0000 {\color{gray!80}(0.5938)} & \cellcolor{pink!27} 0.4062$\rightarrow$1.0000 {\color{gray!80}(0.5652)} & \cellcolor{pink!27} 0.4062$\rightarrow$1.0000 {\color{gray!80}(0.6406)}  \\
      
  &  & passport\_aware~\cite{zhang2020passport} &\cellcolor{blue!9} 1.0000 &\cellcolor{blue!10} 0.0000 &\cellcolor{blue!10} 0.0000 & \cmark
      & \cellcolor{pink!25} 0.5859$\rightarrow$1.0000 {\color{gray!80}(0.5234)} & \cellcolor{pink!25} 0.5391$\rightarrow$1.0000 {\color{gray!80}(0.5234)} & \cellcolor{pink!25} 0.5156$\rightarrow$1.0000 {\color{gray!80}(0.5312)} & \cellcolor{pink!27} 0.4609$\rightarrow$1.0000 {\color{gray!80}(0.5234)} & \cellcolor{pink!27} 0.5156$\rightarrow$1.0000 {\color{gray!80}(0.4925)} \\
    \bottomrule
    \end{tabular}
    \begin{tablenotes}
    \item $\Delta_{\mathrm{acc}}(\mathrm{clean},\mathrm{\textbf{att}acked})$ and $\Delta_{\mathrm{acc}}(\mathrm{clean},\mathrm{\textbf{rec}overed})$ report the \emph{worst-case} absolute accuracy gap over the five listed attacks for each watermark.
    \item \textbf{Tier-1 pass} indicates that the Tier-1 reference-equivalence certificate over watermarked layers passes for all runs in the suite for that model/watermark (\cmark means 100\% match rate).
    \end{tablenotes}
    \end{threeparttable}
    }
\end{table*}

To further demonstrate that \textsc{Canon} is not tied to a particular data distribution, we repeat our end-to-end evaluation on CIFAR-100 using the same ResNet-18 backbone, watermarking suite, and NSO stressors as in the main experiments. The recovery procedure remains unchanged: \textsc{Canon} operates on the traced computation graph and probe-time activation signatures, and does not assume access to the training data distribution beyond the probes used at verification time.

Table~\ref{tab:nso_recovery_suite_add_05} summarizes watermark restoration under CIFAR-100 across four injection ratios ($\rho\in\{0.2,0.5,0.8,1.0\}$), five NSO attack variants, and nine white-box watermarking schemes. The key observation is that task utility is fully preserved: for every watermark and ratio, both $\Delta_{\mathrm{acc}}(\mathrm{clean},\mathrm{attacked})$ and $\Delta_{\mathrm{acc}}(\mathrm{clean},\mathrm{recovered})$ remain $0$, confirming that the evaluated NSO settings operate in the intended \emph{function-preserving} regime and that \textsc{Canon} does not introduce accuracy regressions while undoing structural obfuscation.

More importantly, the primary outcome in our evaluation, Tier-1 reference-equivalence certification, passes (\cmark) for all runs across all ratios and watermarking schemes, indicating that recovered watermark-bearing tensors match the clean reference up to the allowed channel symmetries. We therefore treat the recovered model as successfully canonicalized back into an extractor-compatible parameterization. 
Tier-2 similarity is reported only as an auxiliary, conservative reference: it reflects each scheme's native extractor score on the (potentially re-indexed) recovered weights and is not used for acceptance when Tier-1 passes.

Figure~\ref{fig:runtime_resnet18_ratio} reports recovery runtime as the NSO injection ratio increases, for both MNIST and CIFAR-100. The overall trend is consistent across datasets: higher $\rho$ increases the number of injected channels and widens the attacked network, which raises probe-time activation collection cost and amplifies redundancy clustering and graph-consistent consumer rewrites. This confirms that \textsc{Canon}'s runtime scales primarily with the \emph{structural complexity} introduced by NSO rather than dataset-specific factors.

The curves also highlight that runtime is strongly \emph{attack-dependent}. Single-primitive attacks (\texttt{nso\_zero}, \texttt{nso\_split}, \texttt{nso\_clique}) remain comparatively inexpensive and grow moderately with $\rho$, whereas compositional attacks are the dominant cost drivers: \texttt{mix-opseq} exhibits the steepest growth as $\rho$ increases, and \texttt{mix-opseq (per-merge-group)} sits in between. 

Taken together with the Tier-1 success in Table~\ref{tab:nso_recovery_suite_add_05}, these results show that \textsc{Canon} maintains full recovery while incurring a predictable, attack- and ratio-driven verification-time cost profile on CIFAR-100.

\begin{table*}[!htbp]
    \centering
    \caption{Tier-1 reference-equivalence sensitivity for \textit{ResNet-18} under NSO \textcolor{magenta}{\textbf{\texttt{add-ops}}} attacks (\textcolor{magenta}{\textbf{$0.2$}} attack ratio) \textcolor{brown!80}{with permutation or scaling}. Each cell reports the Tier-1 certificate statistics across permutation tolerance (\textbf{Perm Tol.}): maximum relative error (\textbf{max\_rel\_err}) and greedy match fraction (\textbf{match\_frac}).}
    \label{tab:perm_tol_sweep_resnet18_02}
    \vspace{5pt}
    \renewcommand{\arraystretch}{1.1}
    \resizebox{\linewidth}{!}{
    \begin{threeparttable}
    \renewcommand{\arraystretch}{1.35}
\begin{tabular}{c | c | cc | cc | cc | cc | cc}
\toprule
\textbf{Watermark} & \textbf{Perm Tol.}
& \multicolumn{2}{c|}{\textbf{nso\_zero}}
& \multicolumn{2}{c|}{\textbf{nso\_clique}}
& \multicolumn{2}{c|}{\textbf{nso\_split}}
& \multicolumn{2}{c|}{\textbf{mix-opseq}}
& \multicolumn{2}{c}{\textbf{mix-opseq (per-merge-group)}} \\
\cmidrule(lr){3-4}\cmidrule(lr){5-6}\cmidrule(lr){7-8}\cmidrule(lr){9-10}\cmidrule(lr){11-12}
& & \textbf{max\_rel\_err} & \textbf{match\_frac}
  & \textbf{max\_rel\_err} & \textbf{match\_frac}
  & \textbf{max\_rel\_err} & \textbf{match\_frac}
  & \textbf{max\_rel\_err} & \textbf{match\_frac}
  & \textbf{max\_rel\_err} & \textbf{match\_frac} \\
\midrule

\multirow{7}{*}{\makecell[l]{%
uchida17~\cite{uchida2017embedding}\\
greedy~\cite{liu2021watermarking}\\
lottery~\cite{chen2021you}\\
ipr\_gan~\cite{ong2021protecting}\\
riga21~\cite{wang2021riga}\\
ipr\_ic~\cite{lim2022protect}\\
deepsigns~\cite{darvish2019deepsigns}\\
deepipr~\cite{fan2021deepipr}\\
passport\_aware~\cite{zhang2020passport}}}
& 0.001 & 0.0000 & 1.0000 & 0.0000 & 1.0000 & 0.0000 & 1.0000 & 0.0000 & 1.0000 & 0.0000 & 1.0000 \\
& 0.002 & 0.0000 & 1.0000 & 0.0000 & 1.0000 & 0.0000 & 1.0000 & 0.0000 & 1.0000 & 0.0000 & 1.0000 \\
& 0.005 & 0.0000 & 1.0000 & 0.0000 & 1.0000 & 0.0000 & 1.0000 & 0.0000 & 1.0000 & 0.0000 & 1.0000 \\
& 0.010 & 0.0000 & 1.0000 & 0.0000 & 1.0000 & 0.0000 & 1.0000 & 0.0000 & 1.0000 & 0.0000 & 1.0000 \\
& 0.020 & 0.0000 & 1.0000 & 0.0000 & 1.0000 & 0.0000 & 1.0000 & 0.0000 & 1.0000 & 0.0000 & 1.0000 \\
& 0.050 & 0.0000 & 1.0000 & 0.0000 & 1.0000 & 0.0000 & 1.0000 & 0.0000 & 1.0000 & 0.0000 & 1.0000 \\
& 0.100 & 0.0000 & 1.0000 & 0.0000 & 1.0000 & 0.0000 & 1.0000 & 0.0000 & 1.0000 & 0.0000 & 1.0000 \\

\bottomrule
\end{tabular}

    \begin{tablenotes}
    \item \textbf{Cell format:} (\textbf{max\_rel\_err}, \textbf{match\_frac}) for the Tier-1 certificate at the given \textbf{Perm Tol.}. Here, all entries satisfy max\_rel\_err$=0.0000$ and match\_frac$=1.0000$.
    \end{tablenotes}

    \end{threeparttable}
    }
    \label{tab:perm_tol}
\end{table*}

\section{Supplementary Attack Setting: Disabling Permutation and Scaling}
\label{sec:tab_nso_without_permute_scale}

We provide a complementary NSO stressor that \textit{disables ``explicit'' channel permutation and per-channel scaling} to help interpret the Tier-2 similarity behavior. 
\S\ref{sec:exp} evaluates a stronger attack setting that includes explicitly applied permutation and scaling, i.e., symmetry-breaking operations chosen by the attacker that preserve task utility while intentionally scrambling channel identity and thus aggressively disrupting index-bound extractors.
When these explicit symmetry-breaking knobs are disabled, exact index stability is still \textit{not} guaranteed. This is because NSO primitives first inject redundancy, and the subsequent compaction step (drop/merge) can induce ``implicit'' channel relabeling, and any producer--consumer alignment procedure may introduce arbitrary tie-breaking among redundant channels.
Nevertheless, under this milder setting, the same recovery pipeline yields substantially higher Tier-2 watermark similarity, while Tier-1 certification still passes across all runs, as shown in Table~\ref{tab:nso_recovery_without_scale_02}.

\heading{Tier-1 vs.\ Tier-2 measure different invariants.}
In our two-tier evaluation, Tier-1 searches for an alignment between recovered and clean model weights up to the allowed symmetries (output-channel permutation and, when enabled, BN-consistent positive scaling) under a tolerance. So, it can certify recovery whenever the recovered tensors lie in the same symmetry-equivalence class as the reference, even if their raw indices are not identical.
In contrast, Tier-2 reports each scheme's native extraction similarity by running the extractor directly on the unaligned recovered tensors. Since most white-box extractors are index/position-based and not permutation invariant, Tier-2 is not guaranteed to increase even when recovery is correct in the sense of canonicalization.

\heading{Disabling permutation/scaling makes Tier-2 similarity visibly improve.}
When permutation and scaling are disabled, the stressor is dominated by redundancy injection and graph-consistent rewrites, and the recovered compact layout is much closer to the clean channel basis in a way that index-bound extractors can directly benefit from. Table~\ref{tab:nso_recovery_without_scale_02} reflects that the reported Tier-2 similarities rebound toward the clean baseline. For instance, several schemes recover Tier-2 similarity close to 1.0 under \texttt{nso\_zero}/\texttt{nso\_clique}, and the overall extractor-native evidence is markedly stronger than under the stronger attack configuration in \S\ref{sec:exp}.

Most importantly, this supplementary setting confirms that the low Tier-2 similarity observed under the strong NSO configuration is primarily driven by explicit symmetry-breaking edits (permutation/scaling) that target index-bound extractors, rather than by a failure of \textsc{Canon}. 
Even in this attack setting without permutation or scaling, Tier-2 remains a conservative indicator, whereas Tier-1 provides the definitive confirmation of full recovery, as all runs pass Tier-1 reference-equivalence certification.

\section{Tight reference-equivalence certificates under strong NSO}\label{sec:cert_strength}

Table~\ref{tab:perm_tol} underscores a central strength of our recovery--verification pipeline: \textsc{Canon} enables a \textit{tight} Tier-1 reference-equivalence certificate that is stable across attack variants and tolerance settings.
Under NSO \texttt{add-ops} at $0.2$ attack ratio \textit{with deliberate channel permutation and/or positive scaling enabled}, every evaluated watermark scheme admits an exact one-to-one correspondence between recovered and clean reference \textit{output channels} in watermark-bearing layers.

Concretely, \texttt{Perm Tol.} is an $\ell_2$-based \textit{relative} residual tolerance applied \textit{after} the best one-to-one channel matching is found.
For each recovered output channel $i$ and reference output channel $j$, incoming weights are flattened into vectors $w^{(\mathrm{rec})}_i$ and $w^{(\mathrm{ref})}_j$, and a pairwise residual matrix is formed as
\[
R_{i,j}\;=\;\frac{\bigl\lVert w^{(\mathrm{rec})}_i - w^{(\mathrm{ref})}_j \bigr\rVert_2}{\bigl\lVert w^{(\mathrm{ref})}_j \bigr\rVert_2 + \epsilon}\,.
\]
An $\ell_2$-greedy one-to-one matching then repeatedly selects the smallest remaining $R_{i,j}$ while enforcing uniqueness, yielding a permutation $\pi$.
The certificate passes at \texttt{Perm Tol.} if the worst matched residual is within tolerance,
\[
\max_i \; R_{i,\pi(i)} \;\le\; \texttt{Perm Tol.}\,.
\]

Across all watermarks and all attacks in the suite, the matched residuals are numerically zero, giving \textbf{exactly} $\texttt{max\_rel\_err}=0.0000$ and \textbf{full coverage} $\texttt{match\_frac}=1.0000$, consistently from $\texttt{Perm Tol.}=0.001$ up to $0.1$.
This tolerance-insensitivity indicates that recovery does not merely yield approximate alignment, but returns watermark-bearing tensors to the same symmetry-equivalence class as the clean reference (up to the explicitly permitted channel permutation and BN-consistent positive scaling).
As a result, ownership checks behave as if performed on the clean reference modulo these allowed symmetries, even under NSO-induced structural edits that preserve task utility while severely disrupting index-bound extractors.

\end{document}